\documentclass[aps,pra,preprint,11pt,amsmath,amssymb,showpacs,floats]{revtex4}
\usepackage{psfrag}
\usepackage{graphicx}

\begin{document} 




\def\integer{{\mathbb{Z}}}
\def\pinteger{{\mathbb{N}}}
\def\spinteger{{\mathbb{N}}^\star}
\def\real{{\mathbb{R}}}
\def\preal{{\mathbb{R}}_+}
\def\complex{{\mathbb{C}}}
\def\rational{{\mathbb{Q}}}

\newcommand{\I}{{\rm{i}}}
\newcommand{\D}{{\rm{d}}}
\newcommand{\E}{{\rm{e}}}

\newcommand{\tr}{\operatorname{tr}}
\newcommand{\Span}{\operatorname{span}}
\newcommand{\sinc}{\operatorname{sinc}}
\newcommand{\card}{\operatorname{card}}
\newcommand{\supp}{\operatorname{supp}}
\newcommand{\vp}{\operatorname{vp}}
\newcommand{\cartprod}{\operatornamewithlimits{\times}}
\newcommand{\union}{\operatornamewithlimits{\cup}}
\newcommand{\intersection}{\operatornamewithlimits{\cap}}

\newcommand{\Sc}{{\cal S}}
\newcommand{\Pc}{{\cal P}}
\newcommand{\Bc}{{\cal B}}
\newcommand{\Ac}{{\cal A}}
\newcommand{\timeord}{{\cal T}}
\newcommand{\oforder}{{\cal O}}
\newcommand{\ket}[1]{| #1 \rangle}
\newcommand{\bra}[1]{\langle #1 |}
\newcommand{\braket}[2]{\langle #1 | #2 \rangle}
\newcommand{\ketbra}[2]{| #1 \rangle \langle #2 |}
\newcommand{\meanB}[1]{\langle #1 \rangle }
\newcommand{\meanBx}[2]{\langle #1 \rangle_{#2 } }
\newcommand{\lamb}{\lambda}
\newcommand{\HS}{{H}_{\cal{S}} }
\newcommand{\HB}{{H}_{\cal{B}} }
\newcommand{\OB}{{O}_{\cal{B}} }
\newcommand{\HP}{{H}_{\cal{P}} }
\newcommand{\HPS}{{H}_{\cal{PS}} }
\newcommand{\HtildePS}{\widetilde{H}_{\cal{PS}} }
\newcommand{\HPB}{{H}_{\cal{PB}} }
\newcommand{\HtildePB}{\widetilde{H}_{\cal{PB}} }
\newcommand{\HPSB}{{H} }
\newcommand{\rhoS}{ {\rho}_{\cal{S}} }
\newcommand{\rhoP}{ {\rho}_{\cal{P}} }
\newcommand{\rhoPeq}{ {\rho}_{\,\cal{P}}^{(\rm{eq})} }
\newcommand{\rhoB}{ {\rho}_{\,\cal{B}}^{(\rm{eq})} }
\newcommand{\rhoPS}{ {\rho}_{\cal{PS}} }
\newcommand{\rhotildePS}{ \widetilde{\rho}_{\cal{PS}} }
\newcommand{\rhotildeS}{ \widetilde{\rho}_{\cal{S}} }
\newcommand{\rhoPB}{ {\rho}_{\,\cal{PB}}^{(\rm{eq})} }
\newcommand{\rhoPBeq}{ {\rho}_{\,\cal{PB}}^{(\rm{eq})} }
\newcommand{\rhoPSB}{ {\rho} }
\newcommand{\UtildePSB}{ \widetilde{U} }
\newcommand{\trP}{\operatorname{tr}_{\cal P}}
\newcommand{\trS}{\operatorname{tr}_{\cal S}}
\newcommand{\trB}{\operatorname{tr}_{\cal B}}
\newcommand{\TS}{T_{\cal{S}} }
\newcommand{\TP}{T_{\cal{P}} }
\newcommand{\TBmax}{T_{\cal{B}} }
\newcommand{\TBmin}{t_{\cal{B}} }
\newcommand{\ZB}{Z_{\cal{B}} }
\newcommand{\ZBx}[1]{Z_{#1} }
\newcommand{\ZPB}{Z_{\cal{P B}} }
\newcommand{\ds}{\delta s }
\newcommand{\dectime}{t_{\rm dec} }
\newcommand{\dectimemax}{t_{\rm dec}^{(\rm max )}}
\newcommand{\enttime}{t_{\rm ent} }
\newcommand{\club}{${\clubsuit}$}
\newcommand{\newparagraph}{\vspace{0mm}}




\title{Quantum measurements without macroscopic superpositions}
\author{Dominique Spehner$^{1}\,$\footnote{e-mail: spehner@ujf-grenoble.fr}
and Fritz Haake$^{2}$}
\affiliation{$^1$Institut Fourier, 100 rue des Maths, 38402 Saint-Martin 
d'H\`eres, France\\
$^2$Fachbereich Physik, Universit{\"a}t Duisburg-Essen,
  47048 Duisburg, Germany}

\date{\today}

\renewcommand{\baselinestretch}{2}

\begin{abstract}
  We study a class of quantum measurement models. A microscopic object
  is entangled with a macroscopic pointer such that each eigenvalue of
  the measured object observable is tied up with a specific pointer
  deflection. Different pointer positions mutually decohere under the
  influence of a bath.  Object-pointer entanglement and
  decoherence of distinct pointer readouts proceed simultaneously.
  Mixtures of macroscopically distinct object-pointer states may then
  arise without intervening macroscopic superpositions.  Initially,
  object and apparatus are statistically independent while the latter
  has pointer and bath correlated according to a metastable local thermal
  equilibrium.  We obtain explicit results
  for the object-pointer dynamics with temporal coherence decay in
  general neither exponential nor Gaussian.  The decoherence time does
  not depend on details of the pointer-bath coupling if it is smaller
  than the bath correlation time, whereas in the opposite Markov
  regime the decay depends strongly on whether that coupling is Ohmic
  or super-Ohmic.
\end{abstract}


\maketitle



\renewcommand{\baselinestretch}{1}

\section{Introduction}

The interpretation and theoretical description of measurements on
quantum systems have been under intense debate since the birth of
quantum mechanics~\cite{Wheeler,vonNeumann}.  In the last three decades the major
role played by environment-induced decoherence in a measurement
process has been fully acknowledged thanks to the works of Zeh, Zurek,
and others~(see~\cite{Giulini,Zurek81,Zurek91,Zurek03} and references
therein).  A renewal of interest for quantum detection and decoherence
came in the last decade with new developments in quantum information.
It is desirable to better understand the relation between quantum and
classical information and how one can convert one into another.
Moreover, good control over all sources of decoherence is required
for quantum information processing.  On the experimental side,
measurements can be used either to extract information on quantum
states or to monitor quantum systems (quantum
trajectories~\cite{Knight98}, quantum Zeno effect~\cite{Fisher01,Toschek01}).
Experimental data are now available for the decoherence time in
microwave cavities~\cite{Brune96}, in trapped ions~\cite{Myatt00}, in solid
state devices like quantum dots~\cite{Borri} and superconducting tunnel junction
nanocircuits~\cite{Vion02,Buisson06}, for fullerene molecules decohered by
collisions with a background gas~\cite{Hornberger03},
and for beams of electrons
decohered by Coulomb interaction with a semiconducting
plate~\cite{Sonnentag}.  These and other experiments call for studies
of concrete models for quantum measurements.  Various models
have been investigated so far (see, e.g.,~\cite{HaakeZuk} and the
interesting statistical physics models of Refs.\ \cite{Balian01,Balian03}) but a
satisfactory treatment of decoherence resulting from the many-body
interactions in the measurement apparatus is still lacking.

\newparagraph

A measurement on a quantum system consists in letting this
system (called ``object'' in the following) interact with a
measurement apparatus, in such a way that some information about the
state of the object is transferred to the apparatus.  As already
recognized by Bohr, even though the composite (object
and apparatus) system has to be described by quantum theory, some part
of the apparatus (called the ``pointer'' in the following) must be
capable of classical behavior.  The interaction builds up a one-to-one
correspondence between the eigenvalues $s$ of the measured observable
$S$ (supposed here to have a discrete spectrum) and macroscopically
distinguishable pointer states (characterized e.g. by sharply defined
pointer positions separated by macroscopic distances).  In addition to
this object-apparatus coupling, the measurement must involve some
``superselection rules'' destroying the coherences between the pointer
states~\cite{Giulini,Zurek81,Zurek91,Wigner63}.  Most previous
discussions in the literature consider these two processes separately:
A first step (``premeasurement'') exclusively treats the unitary
evolution entangling object and pointer.  For an object initially
uncorrelated to the apparatus and in a linear superposition
$\ket{\psi_{\cal S}}=\sum_s c_s \ket{s}$ of eigenstates of $S$, this
entanglement produces a superposition of macroscopically
distinguishable object-pointer states
$ 
\ket{\Psi_{\rm ent}} = \sum_s c_s \ket{s} \otimes
\ket{\psi^{\,s}_{\cal P}}$,
where $\ket{\psi^{\,s}_{\cal P}}$ is 
the  pointer state tied up with the eigenvalue $s$  
(for simplicity we provisionally assume  that this state is
pure). 
The ``Schr{\"o}dinger cat'' state $\ket{\Psi_{\rm ent}}$  is taken as the initial
state for a second dynamical process, decoherence, which leads to
superselection rules;  there, the quantum
correlations between object and apparatus are transformed into
classical correlations, as the superposition of object-pointer states
is degraded to a statistical mixture of the same states
according to 
$\ketbra{\Psi_{\rm ent}}{\Psi_{\rm ent}}  
\rightarrow 
\sum_{s} | c_s |^2 \, \ketbra{s}{s} \otimes
\ketbra{\psi^{\,s}_{\cal P}}{\psi^{\, s}_{\cal P}}
$.
For such a sequential treatment to make physical sense, the duration
of the entanglement process  must be short
compared with the decoherence time $\dectime$ associated with the
latter transformation.  However, it is known that
$\dectime$ is extremely short for macroscopic superpositions.  The
present paper is devoted to the more realistic situation where
entanglement and decoherence proceed simultaneously.
If the
characteristic time for entanglement is larger than $\dectime$,
macroscopic superpositions decohere to mixtures faster than
entanglement can create them. The measurement process
 then yields the
final mixture without involving
a Schr{\"o}dinger cat state at any previous moment, 
the object-pointer initial product state
being directly transformed as    
\begin{equation}\label{eq-simultaneous}
\ketbra{\psi_{\cal S}}{\psi_{\cal S}} \otimes 
 \ketbra{\psi^{\,0}_{\cal P}}{\psi^{\,0}_{\cal P}}
\longrightarrow 
 \sum_{s} | c_s |^2 \, \ketbra{s}{s} \otimes
\ketbra{\psi^{\,s}_{\cal P}}{\psi^{\, s}_{\cal P}}\;.
\end{equation}
Here  $\ket{\psi_{\cal S}}=\sum_s c_s \ket{s}$, 
$\ket{\psi^{\,0}_{\cal P}}$, and $\ket{\psi^{\,s}_{\cal P}}$ refer 
to the  object initial state, the  pointer initial state, 
and the pointer state tied up with $s$, respectively.

\newparagraph

Our model is a three-partite model and consists of the quantum object to be
measured, a single ``pointer'' degree of freedom of the apparatus
singled out by its strong coupling to the object and by affording a
macroscopic range of ``read-out'' values, and a ``bath''
comprising all other degrees of freedom of the apparatus.  A
pointer-bath coupling is responsible for decoherence.  Overcoming
limitations of many previous approaches, we (i) allow for
object-pointer entanglement and decoherence of distinct pointer
readouts to proceed simultaneously, (ii) cope with initial
correlations between pointer and bath by considering them initially in
a metastable local thermal equilibrium, and (iii) go beyond the Markovian treatment of
decoherence.  The physical relevance of point (i) has been
discussed above. This simultaneity of entanglement and
decoherence and the possibility of having decoherence
much faster than  entanglement  
have been considered in~\cite{Balian01,Balian03}.
Let us now comment on (ii) and (iii).  Most models
studied so far (in particular in~\cite{Balian01,Balian03}) 
are based on the assumption that the pointer and bath
are initially statistically independent.  Taking instead the whole
apparatus to be initially in a  local thermal equilibrium seems more realistic.
The Markov approximation mentioned in (iii) consists in neglecting
memory effects for the (reduced) object-pointer dynamics. It assumes
the decoherence time $\dectime$ to be larger than the bath correlation
time, a condition not satisfied in some
experiments~\cite{Borri,Vion02,Buisson06}. Since decoherence for
macroscopic and even mesoscopic
superpositions is faster than bath relaxation~\cite{Haake01},
this approximation is clearly unjustified if such superpositions
arise during the object-pointer evolution.

\newparagraph

We shall assume a certain ordering of time scales.  One
of these, denoted by $\TS$, is the characteristic time for
the evolution of the measured observable $S$ under the Hamiltonian of
the object.  A second (classical) time scale $\TP$ characterizes
significant changes in position of the pointer under its proper
Hamiltonian (i.e., in the absence of coupling with the object).  The
initial temperature $T$ of the apparatus sets a time scale $\hbar
\beta = \hbar/(k_B T)$, referred to below as the thermal time ($k_B$
is the Boltzmann constant).  Finally, the object and pointer are put
in contact during a time $t_{\rm int}$.  For a macroscopic pointer,
the limit $\hbar \beta \ll \TP$ seems difficult to avoid. Similarly,
the decoherence and object-pointer interaction times $\dectime$ and
$t_{\rm int}$ are small compared with $\TP$.  During an ideal
measurement, the measured observable $S$ 
may change  but
weakly under the full (object + apparatus) Hamiltonian $H$,
i.e., $\E^{\I t H/\hbar} S \E^{-\I t H/\hbar} \simeq S$  for $0\leq t \leq t_{\rm
  int},t_{\rm dec}$. 
Only under this condition can an eigenstate of
the measured observable $S$ be left almost unchanged by the measurement.
One has to require that
(i)~the
object-pointer interaction Hamiltonian producing the entanglement
commutes with $S$ (see~\cite{Giulini,Zurek81,Zurek91}) and
(ii)~$\TS$ be much
larger than $t_{\rm int}$ and $\dectime$.  It is thus legitimate to
assume
\begin{equation}\label{timescales}
t_{\rm int} , t_{\rm dec} \ll \TS 
\quad , \quad t_{\rm int} , t_{\rm dec} , \hbar\beta \ll \TP\,.
\end{equation}
As far as we are aware, this separation of time scales in ideal
measurements has not been
fully exploited in previous works except in Ref.\ \cite{Balian01}. 
Unlike in the latter reference, given (\ref{timescales}) 
 we shall not need a further hypothesis on the bath
correlation time $\TBmax$ and its relation with $t_{\rm int}$ and
$t_{\rm dec}$.  

\newparagraph

A further key input in what follows is the quantum
central limit theorem (QCLT)~\cite{QCTL,Verbeure} which implies
Gaussian statistics (Wick theorem) for the bath coupling agent in the
pointer-bath interaction.  This will allow us to study a broad class
of pointers and baths, following the approach of Ref.~\cite{Haake01}.
The harmonic oscillator bath linearly coupled to the
pointer~\cite{Caldeira-Leggett83,Weiss} is one member of this class,
but more general (non-harmonic) baths as well as nonlinear couplings
in the position $X$ of the pointer will be also considered.  It turns
out that the decoherence time $\dectime$ may be considerably reduced
by allowing such non-linear couplings.

\newparagraph

The paper is organized as follows. The model and its different time
scales are introduced in Sec.~\ref{sec-model}.  We discuss the
separation of the time scales (\ref{timescales}) and the ensuing simplification of the
object-pointer dynamics in Sec.~\ref{sec-sep_time_scales}.
Sec.~\ref{sec-entanglementvsdecoherence} contains a separate study
of the two dynamical processes producing the entanglement of the
object with the pointer and the loss of coherences between
well-separated pointer readings.  That section is pedagogical in
character and serves to fix the notation and to introduce the relevant
time scales; readers familiar with the theory of quantum
measurement might want to skip the section save for
Secs.~\ref{sec-summary_ent_dec} and~\ref{sec-unstable_V}.  Our
principal results are presented in Sec.~\ref{simul_results},
discussed in Sec.~\ref{sec_dec_times}, and finally derived in
Sec.~\ref{sec_derivation}.  Our conclusions are drawn in
Sec.~\ref{sec-conclusion}.  Appendices~\ref{app-noninteracting_gas}
and \ref{app-total_equilibrium} are devoted to an example for a
measurement apparatus and to a technical derivation of an
approximation for the pointer-bath thermal state. We review in
Appendix~\ref{app-correl_functions} the general properties of
two-point correlation functions used in
Secs.~\ref{simul_results}-\ref{sec_derivation}.  Finally, we discuss
the QCLT and its consequences (Wick theorem for the bath correlation
functions) in Appendix~\ref{app-characteristic-functional}.
Let us point out that a short report of our results can be found in~\cite{Letter}.

\newparagraph

Before going on, some remarks may be permitted to put our paper
in perspective.  Remaining within the frame of quantum mechanics and
its probabilistic interpretation, we are concerned with unitary
evolution of the composite system object + apparatus. 
We discard information about 
the (dynamics of) the microscopic degrees of
freedom of the apparatus (``bath'')
and their entanglement with pointer and object by tracing out the
bath  (see Secs.~\ref{sec-decoherence}
and~\cite{Giulini,Zurek03,ZurekPRL03}).
We so obtain a reduced object-pointer density operator with an
irreversible evolution.  Pointer and object end up in the mixed state
(\ref{eq-simultaneous}) wherein the different pointer states
$|\psi_{\cal P}^{\,s} \rangle$ correspond to macroscopically
distinguishable positions.
 Such states have quantum uncertainties in
position and momentum much smaller than the scales of macroscopic
readings. 
Therefore, the irreversible process (\ref{eq-simultaneous}) leaves a
pointer position revealing an eigenvalue of the measured object
variable.  Over many runs of the measurement, the outcome
$|\psi^{\,s}_{\cal P}\rangle$ arises with probability
$|c_s|^2$. Similar behavior arises for all processes where
initially microscopic fluctuations evolve towards macroscopically
distinct outcomes. A nice example is provided by superfluorescence
where light pulses with substantial shot-to-shot fluctuations grow
from initial quantum uncertainties
\cite{supfluor,supfluDutch,supfluEssen}. --- Let us also recall that
 quantum mechanics is not compatible with the
idea that the specific outcome of a single run is predetermined by
some unknown but ``real'' property of the object (such a property being
independent of the measurement apparatus)~\cite{Peres}. 
Competing hidden-variable theories which indulge in such
more ``intuitive'' notions of reality than quantum mechanics have been
experimentally falsified \cite{Freedman72,Aspect_Modestus} in ever larger
classes, most recently even in non-local variants
\cite{Zeilipompus_Maximus}, while the respective quantum predictions
were invariably confirmed.

\section{Model} \label{sec-model}

As many authors~\cite{Giulini,Zurek81,Zurek91,Zurek03,HaakeZuk,
Balian01,Balian03}
we consider a three-partite system: the object of measurement is some
microscopic system ($\Sc$); a single-degree-of-freedom macroscopic
pointer ($\Pc$) will allow readouts; finally, a bath ($\Bc$) with
many ($N\gg 1$) degrees of freedom serves to decohere distinct pointer
readings. We shall have to deal with the following dynamical
variables: for $\Sc$, the observable ${S}$ to be measured; for $\Pc$
the position ${X}$ and momentum ${P}$; and for $\Bc$, a certain
coupling agent $B$ given by a sum of $N$ self-adjoint operators
$B_\nu$ acting on single degrees of freedom of the bath.  The pointer
is coupled to $\Sc$ and $\Bc$ via the Hamiltonians 
\begin{equation} \label{eq-int_Hamiltonian}
\HPS = \epsilon  {S} {P}
\quad , \quad
\HPB = {B} {X}^\alpha
\quad , \quad 
{B}=N^{-1/2} \sum_{\nu=1}^N B_\nu
\end{equation}
where $\epsilon$ is a coupling constant and $\alpha$ a positive
integer. 
The object-pointer coupling $\HPS$ is chosen so as to (i) not change
the  measured observable $S$ (i.e., $[\HPS,S]=0$); (ii)~be
capable of shifting the pointer position by an amount proportional to
 ${S}$, in such a way that each eigenvalue $s$ of
$S$ becomes tied up with a specific pointer reading; (iii)~be a strong
coupling ($\epsilon$ is large), so that different eigenvalues $s\neq
s'$ eventually become associated with pointer readings separated by
large distances.  The pointer-bath interaction $\HPB$ is
chosen for the most efficient decoherence of distinct pointer
positions~\cite{Haake01}.  Depending on the value of $\alpha$,
nonlinear ($\alpha >1$) as well as linear ($\alpha=1$) couplings will
be considered~\cite{footnote-generalization_f(X)}.
The additivity of the bath coupling agent $B$ in single-degree-of-freedom contributions
$B_\nu$ having zero
mean and positive variance with respect to the bath thermal state 
will allow us to invoke the quantum central limit theorem  when
taking the limit $N\rightarrow \infty$. 
The factor $N^{-1/2}$ in front of the sum in (\ref{eq-int_Hamiltonian}) 
is introduced for convergence purposes (the same scaling with $N$ is
familiar to the classical CLT); note that the pointer-bath
coupling constants are incorporated within the operators $B_\nu$.

\newparagraph

The free evolutions of $\Sc$, $\Pc$, and $\Bc$ are generated by
the respective Hamiltonians $\HS$, $\HP$, and $\HB$. We do not have to
specify $\HS$. The pointer Hamiltonian $\HP=P^2/2M+V(X)$ has a
potential $V(x)$ with a local minimum at $x=0$, so that $V'(0)=0$ and
$V''(0)>0$.  The bath Hamiltonian $\HB$ is like $B$ a sum of
Hamiltonians acting on single degrees of freedom, $\HB = \sum_{\nu}
H_{{\cal B},\nu}$.  We thus disregard couplings between different
degrees of freedom of the bath.  The Hamiltonian of the full system
$\Sc+\Pc+\Bc$ is $\HPSB = \HS + \HP + \HB + \HPS + \HPB$.
An example of a physical system realizing the apparatus $\Pc+\Bc$ is
given in Appendix~\ref{app-noninteracting_gas}.

\newparagraph

We now proceed to describing the initial states allowed for.  It is
appropriate to require initial statistical independence between object
and apparatus. The initial density operator $\rhoS$ of the object may
represent a pure or a mixed state.  Two types of initial conditions
for the apparatus will be considered. The first one, to be referred to
as partial equilibrium, is a product state in which $\Pc$ has some
density operator $\rhoP$ and $\Bc$ is at thermal equilibrium with the
Gibbs density operator $\rhoB = Z_{\cal B}^{-1} \exp(-\beta \HB)$,
wherein $\beta=(k_B T)^{-1}$ is the inverse temperature.  For this
first initial state all three subsystems are statistically
independent.  In the second (more realistic) initial state, the
apparatus is in thermal equilibrium according to the density operator
$\rhoPBeq = Z_{\cal{PB}}^{-1} \,\E^{-\beta (\HP + \HB + \HPB)} $.
The two initial states of $\Sc+\Pc+\Bc$ are
\begin{subequations}
\label{eq-init_state}
\begin{eqnarray} 
\rhoPSB (0) &= &\rhoS  \otimes \rhoP \otimes \rhoB \qquad  
\mbox{partial-equilibrium apparatus}\label{parteqapp}\\
\rhoPSB (0) &= & \rhoS  \otimes \rhoPBeq \quad \quad\qquad \;\,
\mbox{equilibrium apparatus}\,.\label{eqapp}
\end{eqnarray}
\end{subequations}
We further specify the partial-equilibrium state (\ref{parteqapp}) by
requiring that the probability density $\bra{x}\rhoP\ket{x}$ to find
the pointer at position $x$ has a single peak of width $\Delta x =
\Delta$ centered at $x=0$. A momentum uncertainty $\Delta p = 2\pi
\hbar/\lamb$ defines a 
second length scale $\lamb$.  A macroscopic pointer has both $\Delta$
and $\lamb$ negligibly small against any macroscopic read-out scale
$\Delta_{\rm class}$,
\begin{equation} \label{eq-lenghtscales}
\lamb
\leq 4 \pi \Delta \ll \Delta_{\rm class}\;,
\end{equation}
where the first inequality is the
uncertainty principle. We shall also  require that 
\begin{equation} \label{eq-no_squeezing}
\frac{\lamb \Delta}{2 \pi \hbar} = \frac{\Delta x}{\Delta p} 
\approx ( M V''(0) )^{-1/2}
\end{equation}
which means that the state $\rhoP$ is not highly
squeezed in momentum or in position.
As a concrete example we may
consider a Gaussian pointer density matrix 
\begin{equation} \label{eq-rhoP} 
\bra{x} \rhoP \ket{x'} =\frac{1}{\sqrt{2\pi\Delta^2 }} 
\, \E^{-(x+x')^2/(8 \Delta^2)}\, 
\E^{-2 \pi^2 (x-x')^2/\lamb^2} \;.
\end{equation}
If $\Pc$ is initially in a pure state then $\tr_{\Pc}\rhoP^2=\int \D x
\, \D x'\bra{x} \rhoP \ket{x'}^2 = 1$, which implies that this state
has the minimum uncertainty product $\Delta x\Delta p=\hbar/2$, i.e., $\lamb =4
\pi \Delta$.  

\newparagraph

The Gaussian density (\ref{eq-rhoP}) also arises if $\Pc$ is in a
Gibbs state $\rhoPeq= Z_{\cal{P}}^{-1} \E^{-\beta \HP}$ provided that
the potential $V(x)$ is confining and $\beta$ is small enough.  To see
this, we note that the pointer observables $X$ and $P$ evolve
noticeably under the Hamiltonian $\HP$ on a {\it classical time scale}
$\TP$, which is much larger than all other (quantum) time scales in
the model.  In particular, $\TP$ is much larger than the thermal time,
$\TP \gg \hbar \beta$.  As a result, the matrix elements $\bra{x}
\rhoPeq \ket{x'}$ of $\rhoPeq$ can be approximated by
$Z_{\cal{P}}^{-1} \E^{-\beta( V(x) + V(x'))/2} \E^{-2 \pi^2
  (x-x')^2/\lambda_{\rm th}^2}$, wherein $\lambda_{\rm th}= 2 \pi
\hbar (\beta/M)^{1/2}$ is the thermal de Broglie wavelength. The
reader may recognize in this expression the short-time behavior of the
quantum propagator $\bra{x} \E^{-\I t \HP/\hbar} \ket{x'}$ for $t=-\I
\hbar \beta$ (see e.g.~\cite{Schulmann}).  Since the potential $V(x)$
has a local minimum at $x=0$, it can be approximated near the origin
by a quadratic potential, $V(x) \simeq V(0) + x^2 V''(0)/2$.
Therefore, for small $x$ and $x'$, $\bra{x} \rhoPeq \ket{x'}$ has the
Gaussian form (\ref{eq-rhoP}) with $\Delta = \Delta_{\rm th} = (\beta
V''(0) )^{-1/2}$ and $\lamb \simeq \lambda_{\rm th}$.  It is important
to bear in mind the separation of length scales $ \lambda_{\rm th} \ll
\Delta_{\rm th} \ll \Delta_{\rm class} $.  Inasmuch as the pointer
classical time scale $\TP$ may be defined as $\TP=( M/V''(0))^{1/2}$,
the fact that $\lambda_{\rm th}$ is much smaller than $\Delta_{\rm
  th}$ is equivalent to $\TP \gg \hbar \beta$.  To fix ideas, for $\TP
=$ 1 s, $M=$ 1 g, $\Delta_{\rm class}=1$ cm, and a temperature of 1 K
the above-mentioned length scales differ by more than eight orders of
magnitude. Hence (\ref{eq-lenghtscales}) and (\ref{eq-no_squeezing})
are well satisfied if $\rhoP = \rhoP^{(\rm eq)}$.

\newparagraph

All of these illustrations, including the Gaussian (\ref{eq-rhoP}),
are meant to give an intuitive picture. What we shall need in actual
fact is the quasi-classical nature of the pointer initial state, as
implied by (\ref{eq-lenghtscales}) and (\ref{eq-no_squeezing}),
together with the single-peak character of the initial density of
pointer positions.

\newparagraph 

Let us point out an essential difference between our model 
and the interacting
spin model of~\cite{Balian03}. Unlike in this reference,
$\Sc$ is strongly coupled to a single degree of freedom 
(the pointer $\Pc$) of the apparatus, e.g. with its total momentum
$P$ in a given direction (see Appendix~\ref{app-noninteracting_gas}). The coupling of
$\Sc$ with the other apparatus degrees of freedom (the bath $\Bc$, for us)      
is assumed to be much weaker and can therefore be neglected, as will
be seen in Sec.~\ref{sec-Markov}. Given the separation of time
scales (\ref{timescales}) and our choice of a quasi-classical pointer
initial state,  the 
pointer Hamiltonian $\HP$ only plays  a role in providing
an amplification mechanism, as we shall see in 
Secs.~\ref{sec-simplified_dynamics} and \ref{sec-unstable_V}. 
Hence allowing $\Pc$ to have two or three 
degrees of freedom, instead of one, would make the notation  more cumbersome 
without changing significantly the results.   
Our results below should also remain valid if the bath consists of
interacting degrees of freedom (like in a spin chain) 
 provided that the (spin-spin) correlations $\langle B_\mu B_\nu \rangle$ 
in  the bath thermal state decay more rapidly
than $1/|\mu-\nu|$ as $|\mu-\nu |\rightarrow \infty$. In fact, the
validity of the QCLT can be extended in this
context~\cite{Verbeure}. Decoherence via coupling with a bath of interacting spins and
random matrix models for the coupling and bath have been considered in
\cite{Balian03,Seligman07,Lutz01}.

\newparagraph

We shall study the dynamics of the reduced state of
$\Sc+\Pc$ (object and pointer).  That state is defined by a density
operator $\rhoPS(t)$ obtained by tracing out the bath degrees of
freedom in the state of $\Sc+\Pc+\Bc$, 
\begin{equation} \label{eq-rhoPS}
\rhoPS (t) 
 = 
\tr_{\cal{B}} \left( \E^{-\I t \HPSB/\hbar} \rhoPSB(0) \, \E^{\I t \HPSB/\hbar} \right)
\;.
\end{equation}
Here and in what follows, $\tr_j$ refers to the partial trace over the
Hilbert space of $j=\Sc,\Pc$ or $\Bc$.  When tracing
out the bath we admit the inability to acquire
information about it~\cite{ZurekPRL03}.

\section{Separation of time scales}
\label{sec-sep_time_scales}

\subsection{Time scales of  object, pointer, and bath}
\label{sec-time scales}

Let us denote by $\widetilde{S} (t)$ the time-evolved observable $S$
in the absence of the coupling $\HPS$, i.e., for the
dynamics implemented by the ``free Hamiltonian'' $\HS$. Similarly, let
$\widetilde{X} (t)$ and $\widetilde{B} (t)$ be the time-evolved
observables $X$ and $B$ when both couplings $\HPS$ and $\HPB$ are
turned off: namely,
\begin{equation} \label{eq-def_int_picture}
\widetilde{O}_j (t) = \E^{\I t {H}_j/\hbar}\, {O}_j \, e^{-\I t
  {H}_j/\hbar}
\quad , \quad 
O_j = S, X \; {\rm{or}} \; B \quad , 
 \quad  j=\Sc,\Pc \; {\rm{or}}\;  \Bc\;.
\end{equation}
One may associate with the time evolution of $\widetilde{X} (t)$,
$\widetilde{B} (t)$, and $\widetilde{S} (t)$ four distinct time
scales.  The time scale $\TP= ( M/V''(0))^{1/2}$ has been already
introduced in Sec.~\ref{sec-model}; it is the time scale for
significant evolution of $\widetilde{X}(t)$ (or, equivalently, of
$\widetilde{P} (t) = M \D \widetilde{X}/\D t\,$) when the pointer is
in the initial state $\rhoP$. The Gaussian form (\ref{eq-rhoP}) for
$\rhoP$ and the no-squeezing condition (\ref{eq-no_squeezing}) make
sure that $\TP$ is indeed a classical time.

\newparagraph

The {\it bath correlation time} $\TBmax$ is defined with the help of
the $n$-point correlation functions
\begin{equation} \label{eq-bath_correl_function}
h_n (t_1, \ldots, t_n) 
 = 
 \tr_{\Bc} 
  \bigl( \widetilde{B}(t_1) \ldots \widetilde{B}(t_n)  \rhoB
  \bigr)\;.
\end{equation}
For simplicity we assume
\begin{equation} \label{eq-B=0}
\tr_{\Bc}(B \rhoB)=0
\;.
\end{equation}
Since the bath has infinitely many degrees of freedom, $h_n(t_1, \ldots, t_n)$
decays to zero as $|t_m-t_l|$ goes to infinity.  We
define $\TBmax$ (respectively $\TBmin$) as the largest (smallest) time
constant characterizing the variations of $h_n$.  It follows from the
QCLT of Ref.~\cite{QCTL} that for a
bath coupling agent $B$ and Hamiltonian $\HB$ which are sums of $N$
independent contributions coming from single degrees of freedom, the $n$-point
functions (\ref{eq-bath_correl_function}) satisfy the bosonic Wick
theorem in the limit $N \gg 1$. This means that $h_n$ vanishes if $n$
is odd and is given if $n$ is even by sums of products of 
two-point functions,
\begin{equation} \label{eq-Wick}
h_n (t_1,\ldots ,t_n) 
 = \sum_{\text{pairing of $\{1,\ldots,n \}$}} 
  h_2 (t_{i_1}, t_{j_1} ) 
   \ldots 
    h_2  (t_{i_{n/2}}, t_{j_{n/2} } )\;.
\end{equation}
That manifestation of the QCLT amounts to
Gaussian statistics for the bath correlation functions. It follows
from (\ref{eq-Wick}) that
$\TBmax$ ($\TBmin$) can be defined more simply as
the largest (smallest) time 
scale associated with the variations of $h_2 (t_1,t_2)=h_2 (t_1-t_2)$ as function of
$t=t_1-t_2$.  More precisely, $h_2(t) \simeq 0$ whenever $|t|\gg
\TBmax$ and $h_2(t) \simeq h_2(0)$ whenever $|t|\ll \TBmin$.  Note
that with $\Bc$ in thermal equilibrium, the thermal time $\hbar \beta$
figures among the decay rates of $h_2$ and thus $\TBmin \leq \hbar
\beta \leq \TBmax$.

\newparagraph

The time scale $\TS$ is defined in analogy to $\TBmin$, so as
to signal significant variation of the object $n$-point functions
$\tr_{\Sc} (\widetilde{S}(t_1) \ldots \widetilde{S}(t_n)\rhoS)$.  Let
us stress that $\TS$ can be larger than the typical inverse 
Bohr frequency $\hbar/|E -E'|$ of $\Sc$ (here $E$ and $E'$ are two
eigenvalues of $\HS$).  For instance, if $S$ is (or
commutes with) the energy $\HS$, then $\TS=\infty$.

\subsection{Simplified dynamics and initial state}
\label{sec-simplified_dynamics}

We assume that the object and pointer
observables $\widetilde{S}(t)$, $\widetilde{X}(t)$, and
$\widetilde{P}(t)$ do not evolve noticeably under the ``free''
Hamiltonian $\HS+\HP$ during the time span of the measurement, so that
$t_{\rm int},\dectime \ll \TS,\TP$.  We show now that thanks to this
separation of time scales, the impact of $\HS$ and $\HP$ on the
dynamics can be fully accounted for at times $t\ll \TS, \TP$ by
modifying the initial states (\ref{eq-init_state}) according to
\begin{equation} \label{eq-slippage_init_cond}
\rho(0) 
 \longrightarrow  
  \E^{- {\I} t (\HS + \HP)/{\hbar}} \rho(0)\, \E^{ {\I} t (\HS + \HP) / {\hbar}}\;.
\end{equation}
 With that slippage of the initial condition accounted for, one
makes a small error by otherwise dropping $\HS$ and $\HP$ from the
total Hamiltonian $H$ in the object-pointer state (\ref{eq-rhoPS}).

\newparagraph

Actually, for times $t$ short compared with $\TS$ and $\TP$, the full
evolution operator can be approximated by
\begin{equation} \label{eq-approx_total_evol_op}
\E^{- \I t \HPSB/\hbar} 
 \simeq
    \E^{- {\I} t (  \HB + \HPS + \HPB)/{\hbar}}\,\E^{- \I t ( \HS + \HP )/ \hbar }
\quad , \quad |t| \ll \TS , \TP\;.
\end{equation}
To justify that simplification we express 
this evolution operator in the interaction picture
with respect to $H_0= \HS+\HP$ as 
\begin{equation} \label{eq-object-pointer_evol_op}
\E^{ {\I} t H_0/{\hbar}} \E^{- {\I} t \HPSB/{\hbar} } 
 = 
   {\cal T} \exp \biggl\{ -\frac{\I}{\hbar} \int_0^t \D \tau \, 
    \Bigl(  \HB + \epsilon \widetilde{S} (\tau)\widetilde{P} (\tau)
     +  B \widetilde{X} (\tau)^\alpha \Bigr) \biggr\}\,;
\end{equation}
here $\timeord$ denotes the time ordering and $ \widetilde{S}(\tau)$,
$\widetilde{P} (\tau)$, and $\widetilde{X} (\tau)$ are given by
(\ref{eq-def_int_picture}).  Note that for $|t| \ll \TS, \TP$ these
operators are almost constant in time between $\tau=0$ and $\tau=t$
and may thus be replaced in (\ref{eq-object-pointer_evol_op}) by $S$,
$P$, and $X$. In other words, the right-hand side of
(\ref{eq-object-pointer_evol_op}) can be approximated by $\exp \{ - \I
t (\HB + \HPS+\HPB) /\hbar\}$, whereupon
(\ref{eq-approx_total_evol_op}) is obtained by taking the adjoint and
by setting $t \to -t$.

\newparagraph

More specific remarks are in order for each of our two initial states
(\ref{eq-init_state}). We first comment on the partial-equilibrium
(\ref{parteqapp}). Due to its assumed quasi-classical nature, the
pointer state $\rhoP$ is weakly modified by the substitution
(\ref{eq-slippage_init_cond}) in the range of time under study.
Deferring the justification of that statement to
Appendix~\ref{app-total_equilibrium} we shall use $\E^{-\I t
  \HP/\hbar} \rhoP \,\E^{\I t \HP/\hbar} \simeq \rhoP$ for $t\ll\TP$.
As regards the object of measurement $\Sc$, the free time evolution of
the object in
\begin{equation} \label{eq-rhoS(0)}
\rhoS^{0}(t) = \E^{-\I t \HS/\hbar} \rhoS \,\E^{\I t \HS/\hbar}
\end{equation}
cannot be neglected, even for $t\ll \TS$.  (For instance, for $S= \HS$
one has $\TS = \infty$ and $\bra{s}\rhoS^{0}(t) \ket{s'}= \E^{-\I t
  (s-s')/\hbar} \bra{s}\rhoS \ket{s'}$ is not close to
$\bra{s}\rhoS\ket{s'}$ for all finite times $t$ if $s\not=s'$).
However, it suffices for our purposes to notice that 
 the diagonal elements $\bra{s} \rhoS^{0}(t) \ket{s} \simeq \bra{s}
\rhoS\ket{s}$ remain nearly unaffected by the free evolution when 
$t\ll \TS$.
Actually,
$\tr (\widetilde{S} (t) \rhoS ) = \sum_s s \bra{s}
\rhoS^{0} (t)\ket{s}$ has to approximate $\tr ( S \rhoS ) = \sum_s s
\bra{s} \rhoS \ket{s}$ in this limit by definition of $\TS$.  

\newparagraph

When allowing the apparatus to start out from thermal equilibrium
according to (\ref{eqapp}), we shall take advantage of the
``high-temperature'' condition $\hbar \beta \ll \TP$ discussed in the
Introduction.  We argue in Appendix~\ref{app-total_equilibrium} that
under this condition (which to violate for a macroscopic pointer would
be a nearly impossible task) and for a weak enough pointer-bath
coupling satisfying $\eta_{\rm th} = h_2(0)^{1/2}
\Delta_{\rm th}^\alpha \beta \lesssim 1$, the Gibbs state of the
apparatus can be approximated by 
\begin{equation} \label{eq-approx_total_equilibrium}
\rhoPBeq   
 \simeq  \frac{1}{\ZPB} \E^{- \beta \HP/2} \E^{-\beta ( \HB + \HPB
  )}\E^{- \beta \HP/2}  \quad , \quad  \hbar \beta \ll \TP\; . 
\end{equation}
Moreover, $\E^{-\I t \HP/\hbar} \rhoPBeq \,\E^{\I t \HP/\hbar}\simeq
\rhoPBeq$ as long as $t , \hbar \beta \ll \TP$.

\newparagraph

In conclusion, for both  initial states (\ref{parteqapp}) and  (\ref{eqapp}),
the substitution (\ref{eq-slippage_init_cond}) amounts to
replacing $\rhoS$ by $\rhoS^{0}(t)$ in the object-pointer initial state.

\begin{figure}
\centering
\psfrag{D}{$\Delta$}
\psfrag{E}{$\epsilon t \delta s$}
\psfrag{(a)}{(a) $t=0$}
\psfrag{(b)}{(b) $t \approx t_{\rm ent}$}
\includegraphics[width=9.5cm]{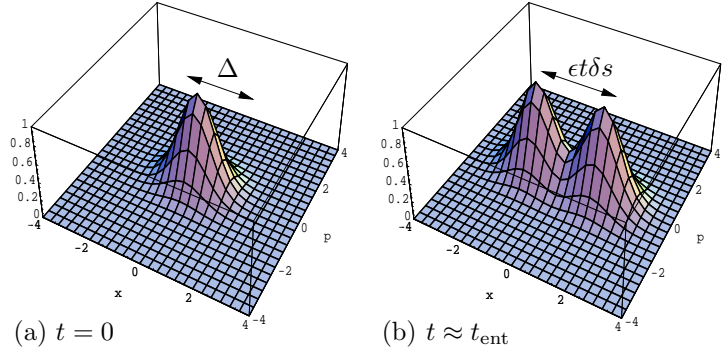}
\caption{ \label{fig-2}
Wigner function $W_{\cal P} (x,p;t)$ of the pointer reduced state $\rhoP (t)$;  $S$ is a
  spin one-half with two eigenvalues $\pm \ds/2$ and $\langle \pm | \rhoS | \pm \rangle = 1/2$. 
Note the absence of
  ripples: the reduced state is a mixture, not a superposition of single-peak
  states. In the horizontal axis, position and  momentum are measured in units of $\Delta/2$
  and $\Delta p$; in the vertical axis, units are 
  such that $W_{\cal P} (x,p;t)$ has maximum value $1$. }
\end{figure}

\section{Entanglement and decoherence separated}
\label{sec-entanglementvsdecoherence}
\subsection{Entanglement of object and pointer} 
\label{sec-entanglement}

Before studying the dynamics generated by the total Hamiltonian $H$,
it is instructive to discuss what happens if we discard the free
dynamics of $\Sc$, $\Pc$, and $\Bc$ as well as the pointer-bath
interaction. For the initial state (\ref{parteqapp}), the bath can
then be ignored and the object-pointer entanglement produced by the
interaction $\HPS= \epsilon {S} {P}$ becomes particularly easy to
describe.  Recalling that ${P}$ is the generator of space translations
we have $\E^{\I\epsilon {S} {P}t /\hbar}\ket{s,x}=\ket{s,x-t\epsilon
  s}$, where $\ket{s,x}$ is the joint eigenstate of ${S}$ and ${X}$
with eigenvalues $s$ and $x$, normalized as
$\braket{s,x}{s',x'}=\delta_{ss'}\delta(x-x')$.  Hence an initial
product state of $\Sc+\Pc$ becomes at time~$t$
\begin{eqnarray} \label{eq-Qsuperposition}
\rhoPS (t)&=& \E^{-\I t\HPS / {\hbar}} \rhoS \otimes 
\rhoP \,\E^{\I t\HPS / {\hbar}} \nonumber\\
&=& \sum_{s,s'}\bra{s} \rhoS \ket{s'} \ketbra{s}{s'} 
 \int \D x \, \D  x'\, \bra{x_s(t)} \rhoP \ket{x_{s'}' (t)} 
 \ketbra{x}{x'}
\end{eqnarray}
with
\begin{equation} \label{eq_x_s(t)}
x_s(t) = x - t\epsilon s\,,\quad x_{s'}' (t)=x'-t \epsilon s'\,,
\end{equation}  
and, for the Gaussian initial state (\ref{eq-rhoP}), 
\begin{eqnarray} \label{pointercoherence}
\bra{x_s(t)} \rhoP \ket{x_{s'}' (t)} 
=  
\frac{1}{\sqrt{2\pi\Delta^2 }}\, 
\E^{-(x+x' - t \epsilon (s+s') )^2/(8 \Delta^2)}\, 
\E^{-2 \pi^2 (x-x' - t \epsilon(s - s') )^2 /\lamb^2}
\,.
\end{eqnarray} 
It is now well to put forth a specification: throughout the present
paper we assume for simplicity that $S$ has a {\it discrete and
  non-degenerate spectrum}.  Moreover, if the Hilbert space of $\Sc$
has infinite dimension we restrict ourselves to initial states of the
object satisfying $\bra{s} \rhoS \ket{s'} =0$ if $s$ and $s'$ belong to
a part of the spectrum containing arbitrarily close eigenvalues (near
an accumulation point).
\newparagraph

In the state (\ref{eq-Qsuperposition}), the diagonal ($s=s'$) matrix
elements of the object state $\rhoS$ are multiplied by the pointer
density matrix $\rhoP$ shifted by $t \epsilon s$ in position space, as
given by (\ref{pointercoherence}) for $s=s'$.  The interaction has
thus tied up each eigenstate $\ket{s}$ of $S$ with a pointer state
which has position $x \simeq t \epsilon s $ with uncertainty $\Delta$
and momentum $p \simeq 0$ with uncertainty $2 \pi \hbar / \lamb$. 
In position representation, each of these pointer states has a peak 
at $x=t \epsilon s$. The different peaks   
 are separated  by at least by the distance $t \epsilon \ds$, where
$\ds$ is the minimum of $|s-s'|$ over all pairs $(s,s')$ of
non-degenerate eigenvalues such that $\bra{s} \rhoS \ket{s'} \not= 0$.
In order to be able to infer the value of $s$ from the position of the
pointer, one must wait until all peaks are well resolved.  That
resolvability begins at the {\it entanglement time}
\begin{equation} \label{eq-t_ent}
t_{\text{ent}}= \frac{\Delta}{\epsilon\, \ds} \;.
\end{equation}
At that time, the reduced pointer density operator $\rhoP (t)=
\tr_{\cal S} (\rhoPS (t))$ has a Wigner function as represented in
Fig.\ \ref{fig-2}(b).  Much later yet, the separation between the peaks
reaches a macroscopic value $\Delta_{\rm class}$ at the time
\begin{equation} \label{eq-t_class}
t_{\text{class}}= \frac{\Delta_{\rm class}}{\epsilon\,\ds} \gg
t_{\text{ent}} \; ,
\end{equation}
allowing for a ``reading'' of the result by a classical
observer. 

\newparagraph

The entanglement in the state (\ref{eq-Qsuperposition}) comes from the
off-diagonal ($s\not= s'$) contributions in $\rhoS$. Due to the
peak structure of the pointer matrix elements
(\ref{pointercoherence}), for fixed $s \not=s'$, $| \bra{s,x}
\rhoPS(t) \ket{s',x'}|$ reaches its maximal value when $x=\epsilon st$
and $x'=\epsilon st$. For those values of $x$ and $x'$,
\begin{equation} \label{eq-no_decay_of_coherences}
\langle s , x=\epsilon t s | \rhoPS (t) | s', x'=\epsilon t s' \rangle
 = \bra{s} \rhoS   \ket{s'}\, \bra{0} \rhoP \ket{0} 
\end{equation}
is time-independent and proportional to $\bra{s} \rhoS \ket{s'}$.
Hence all coherences between different eigenstates of $S$ present in
the initial state of the object are still alive, no matter how large
the time $t$ is.  At times $t \gtrsim t_{\rm class}$, $\rhoPS (t)$
resembles a Schr{\"o}dinger cat state, i.e., has nonzero matrix
elements between macroscopically distinguishable pointer position
eigenstates.  For such an object-pointer state, no classical
probabilistic interpretation is possible: one cannot assign a
probability to the pointer being located e.g. in the vicinity of
$x=\epsilon t s$, henceforth implying that $S$ has the value $s$.  In
a quantum measurement, the entanglement process must be completed by a
decoherence process suppressing the coherences
(\ref{eq-no_decay_of_coherences}) for $s\not= s'$.

\subsection{Decoherence  and ``disentanglement'' of object and pointer} 
\label{sec-decoherence}

We now turn to the decoherence brought about by the pointer-bath
interaction $\HPB= B X^\alpha$, momentarily disregarding all other terms in the
full Hamiltonian $H$.  As shown in \cite{Haake01} for a similar model,
a quantum superposition of coherent states of $\Pc$ with
well-separated peaks in position evolves under $\HPB$ to a statistical
mixture of these coherent states. In the situation under study here,
$\Sc$ and $\Pc$ are entangled, and then decoherence also modifies 
$\Sc$.  The present subsection highlights the
 fundamental role of this decoherence in a measurement
(for more details, see~\cite{Giulini,Zurek91,Zurek03}).

\newparagraph

Assume object and pointer at time $t_0$ entangled, with $\rhoPS =
\rhoPS^{\;\rm ent}$ given by (\ref{eq-Qsuperposition}); the time $t_0$
should be chosen larger than $t_{\text{ent}}$, possibly as large as
the classical time scale introduced above, $\enttime\ll t_0 \approx
t_{\rm class}$. At time $t_0$, the state of $\Sc+\Pc+\Bc$ is $\rhoPSB
(t_0) = \rhoPS^{\;\rm ent} \otimes \rho_{\cal B}$ and the pointer-bath
coupling $\HPB$ is switched on. To simplify the discussion, let us
take for $\rho_{\cal B}$ a pure state $\rho_{\cal B} =
\ketbra{\Psi_0}{\Psi_0}$, where $\ket{\Psi_0}=\otimes_\nu
\ket{\psi_\nu}$ is a product of $N$ single-degree-of-freedom
wavefunctions.  (All arguments below can be easily extended to a bath
in an initial mixed state like $\rho_{\cal B}=\rhoB$.)  Moreover, let
us suppose that $\bra{\psi_\nu} B_\nu \ket{\psi_\nu} = 0$ and that
higher moments $\meanB{ B_\nu^q} = \bra{\psi_\nu} B_\nu^q
\ket{\psi_\nu}$ ($q=2,3,\ldots$) are bounded uniformly in $\nu$.  The
eigenstates $\ket{s,x}$ are entangled at time $t > t_0$ with the bath
states $\ket{\Psi_{x}(t)} = \E^{-\I (t-t_0) x^\alpha B/ \hbar}
\ket{\Psi_0}$. The density operator of $\Sc+\Pc+\Bc$ reads
\begin{eqnarray} \label{eq-Qsuperposition_is_gone}
\nonumber
\rhoPSB (t) 
&  = & 
\E^{-{\I} \HPB (t-t_0)/{\hbar}} \rhoPS^{\;\rm ent} \otimes \rho_{\cal
  B} \,
 \E^{{\I} \HPB (t-t_0)/{\hbar}}
\\
& = & 
  \sum_{s,s'} \int \D x \D  x'\, \bra{s,x}\rhoPS^{\;\rm ent} \ket{s',x'}\;
 \ketbra{s}{s'} \otimes \ketbra{x}{x'} \otimes  \ketbra{\Psi_x(t)}{\Psi_{x'} (t)} \;.
\end{eqnarray}
We now argue that for $x\not= x'$, the scalar product
$\braket{\Psi_x(t)}{\Psi_{x'}(t)}$ is vanishingly small when the time
span $t-t_0$ is larger than a certain {\it decoherence time} $\dectime
(x,x')$. Due to the additivity (\ref{eq-int_Hamiltonian}) of
$B$, 
\begin{eqnarray} \label{eq-scalar_prod}
\nonumber
\braket{\Psi_x(t)}{\Psi_{x'}(t)}
& = &  
  \prod_{\nu=1}^N \bra{\psi_\nu} 
   \E^{-\frac{\I(t-t_0)}{\hbar \sqrt{N}} 
  ({x'}^\alpha-x^\alpha) B_\nu } \ket{\psi_\nu}
\\
& = &\prod_{\nu=1}^N 
 \Bigl( 1 - 
  \frac{(t-t_0)^2 ({x '}^\alpha-x^\alpha)^2 \bra{\psi_\nu} B_\nu^2 \ket{\psi_\nu}}{2 N \hbar^2}
   + \oforder ( N^{-3/2} )
 \Bigr) \;.
\end{eqnarray} 
Taking the limit $N \to \infty$ for fixed values of $t$, $t_0$,
$x$, and $x'$ we obtain 
\begin{eqnarray} 
\label{eq-ln_scalar_prod}
\braket{\Psi_x(t)}{\Psi_{x'}(t)}& = & \E^{-D_t(x,x')}
\;\; = \;\;
\displaystyle
\exp \left\{ - \frac{(t-t_0)^2}{\dectime (x,x')^2}+ \oforder ( N^{-1/2} ) \right\} \\  
 \dectime(x,x') & = & 
\label{eq-decohe_time}
\displaystyle
\frac{\sqrt{2} \,\hbar}{| x'^\alpha-x^\alpha| 
\sqrt{\meanB{B^2}}}
\end{eqnarray}
with $\meanB{B^2} = \bra{\Psi_0} B^2 \ket{\Psi_0} = N^{-1} \sum_{\nu}
\bra{\psi_\nu} B^2_\nu \ket{\psi_\nu}$. We have so far retraced the
proof of the (classical) central limit theorem.

\newparagraph

Taking the partial trace of (\ref{eq-Qsuperposition_is_gone}) over the
bath Hilbert space yields
\begin{equation} \label{eq-intermediate_step}
\rhoPS(t) 
 =
   \sum_{s,s'} \int \D x \D  x'\, \bra{s} \rhoS \ket{s'} 
     \bra{x_s(t_0)} \rhoP \ket{x_{s'}'( t_0 )} 
       \E^{-D_t(x,x')}\, \ketbra{s}{s'} \otimes \ketbra{x}{x'}       
\;.
\end{equation}
Due to the coupling with the bath, each matrix element of $\rho_{\cal
  P S}^{\, {\rm ent}}$ is now multiplied by the scalar product
(\ref{eq-ln_scalar_prod}).  Let us consider a particular term $s \not=
s'$ in the sum in the right-hand side of (\ref{eq-intermediate_step}).  To simplify the
forthcoming discussion, we assume that $\alpha=1$.  As follows from
the peak structure of the pointer coherences (\ref{pointercoherence}),
only the terms satisfying $x\simeq x_{s 0} = t_0 \epsilon s$ and $x'
\simeq x_{s' 0} = t_0 \epsilon s'$ with uncertainty $\Delta$
contribute significantly to the integral over $x$ and $x'$.  For those terms, 
$D_t (x,x') = D_t ( x_{s 0},x_{s' 0}
) ( 1 +\oforder (\enttime / t_0 )) $, see
(\ref{eq-t_ent}), (\ref{eq-ln_scalar_prod}),
and (\ref{eq-decohe_time}).  Therefore, if $t-t_0$ is large compared
with $\dectime (x_{s 0},x_{s' 0})$ and $t_0 \gg \enttime$, the product $\bra{x_s(t_0)} \rhoP
\ket{x_{s'}'( t_0 )}\,\E^{-D_t (x,x')}$ is vanishingly small for all
values of $(x,x')$.  The off-diagonal terms corresponding to
$s\not=s'$ then become negligible in the object-pointer state
(\ref{eq-intermediate_step}).  It is worth emphasizing that $\dectime
(x_{s 0},x_{s' 0})$ can be much smaller than the dissipation time
scale on which the pointer-bath coupling irreversibly changes the
pointer position.

\newparagraph

We would like to point out that the aforementioned damping of the
coherences is related to a lack of information about the bath in a
more subtle way than what is suggested by the partial trace in
(\ref{eq-intermediate_step}).  In fact, some partial knowledge of
the bath state would not inhibit this decoherence.  More precisely, in
order to obtain some information on the coherences in the full density
operator (terms proportional to $\ketbra{s}{s'}$ with $s\not= s'$ in
(\ref{eq-Qsuperposition_is_gone})) at times $t-t_0\gg \dectime (x_{s
  0}, x_{s' 0})$, it is necessary to perform a measurement on some
bath observable $O_B$ satisfying $\bra{\Psi_{x_{s 0}}(t)} O_B
\ket{\Psi_{x_{s' 0}}(t)} \not= 0$ at such time.  It can be shown by
repeating the arguments yielding to (\ref{eq-ln_scalar_prod}) that
such an observable must be non-local, i.e., it must act non-trivially
on {\it all} bath degrees of freedom except for a finite number of
them.  Considering that measuring such an observable is
``unrealistic'', everything happens as if 
the $s\not= s'$ terms have disappeared
in (\ref{eq-Qsuperposition_is_gone}).  The crucial point
is that the object-pointer state is entangled by $\HPB$ with {\it a
  very large number} $N$ of bath variables, so that information about
the coherences is spread out between these many variables after some
time.  Macroscopically distinguishable object-pointer states are then
entangled with bath states which are almost orthogonal in many
subspaces of the bath Hilbert space.  This makes the situation quite
different from the entanglement discussed in
Sec.~\ref{sec-entanglement}: there, the object state was
entangled with a single pointer variable $x$ and it was implicitly
assumed that any pointer observable (in particular, its position $X$)
could be ``observed'' at some ultimate stage of the measurement.
We refer the reader to \cite{Bohm51} (Sec. 22.11), \cite{Wigner63},
\cite{Giulini} (Chapter 2), and~\cite{Zurek03}   
for related discussions on this very important conceptual point. 

\newparagraph

Let us define the decoherence time $\dectime$ as the largest of the
times $\dectime(x_{s 0},x_{s' 0})$ for all distinct eigenvalues $s$ and
$s'$ such that $\bra{s} \rhoS \ket{s'} \not= 0$. For $t-t_0 \gg
\dectime$, the object-pointer state has shed all terms $s\neq s'$ in
the double sum in the density operator (\ref{eq-Qsuperposition}),
\begin{equation} \label{eq-post-meas-1}
\rhoPS (t)\simeq \sum_{s} \bra{s} \rhoS \ket{s} \,\ketbra{s}{s} \otimes \rhoP^{\,s} (t) 
\end{equation}
wherein 
it has been assumed that $\dectime \ll t-t_0 \ll \TBmin, \TS,\TP$ 
(so that $\HB$, $\HS$, and $\HP$ can be neglected) and
$t_{\rm{ent}} \ll t_0$ and we have set
\begin{equation} \label{eq-post-meas-3}
\rhoP^{\,s} (t)
=
 \int \D x \,\D  x'\,\bra{x_s (t_0)} \rhoP \ket{x_{s}' (t_0)} \,\E^{-(t-t_0)^2 /\dectime (x,x')^2} 
\ketbra{x}{x'}\;.
 \end{equation}
While $\rhoPS (t)$ is not (and actually never can become) strictly
diagonal in the position basis of the pointer, 
the matrix elements of the pointer state
(\ref{eq-post-meas-3}) almost vanish if $|x-x'|$ is larger than either the
uncertainty $\lamb$ (see~(\ref{pointercoherence})) or the decoherence length $\sqrt{2} \hbar/(
\alpha\, \Delta^{\alpha-1} (t-t_0) \sqrt{\meanB{B^2}})$ (see the second factor
inside the integral in (\ref{eq-post-meas-3})).

\newparagraph

It is worth noting that the object-pointer states appearing in the sum
over $s$ in (\ref{eq-post-meas-1}) are product states; $\rhoPS(t)$ is
a statistical mixture of these states with probabilities $p_s = \bra{s} \rhoS
\ket{s}$.  Hence the decoherence {\it disentangles} $\Sc$ and $\Pc$.
This implies that in the time regime indicated after 
(\ref{eq-post-meas-1}), $\Sc$ and $\Pc$ can be given independent
states $\rhoS(t)$ and $\rhoP (t)$,
\begin{equation} \label{eq-post-meas-2}
\rhoS(t) 
 = 
\displaystyle
  \sum_{s} p_s
   \ketbra{s}{s} 
\quad , \quad 
\rhoP (t) 
  =  
\displaystyle
  \sum_{s} p_s \, \rhoP^{\,s} (t) \;.
\end{equation}
The object $\Sc$ is in one of the
eigenstates 
$\ket{s}$ with probability $p_s$, in
agreement with von Neumann's postulate.  
The pointer $\Pc$ is in the quasi-classical state $\rhoP^{\,s}(t)$, 
with the same probability.

\subsection{Summary}
\label{sec-summary_ent_dec}

Let us sum up the discussion of the two previous subsections about the
object-pointer entanglement produced by the interaction $\HPS$ and the
decoherence arising from the coupling with the bath $\HPB$.  The
dynamics implemented by $\HPS$ uniquely ties up after the entanglement
time $\enttime$ each eigenvalue $s$ of $S$ with a characteristic
pointer position $x_{s}(t)$.  Such neighboring pointer positions
differ by more than the uncertainty $\Delta$ then. 
Note that arbitrarily close eigenvalues cannot be resolved within
a finite time: in fact, $\enttime$ tends to be 
large for an object initially in a superposition of eigenstates
$\ket{s}$ with closely lying eigenvalues $s$
(e.g. near an accumulation point of the spectrum), 
i.e., for small values of $\delta s$ in (\ref{eq-t_ent}); this 
limits in practice the precision of the measurement of $S$. In the absence of
any other interaction, after a time $t_{\rm class}\gg \enttime$ the
initial product state of the object and pointer has evolved into a
Schr{\"o}dinger cat state.
Nothing irreversible is brought about by
the dynamics: the entanglement can be as easily undone as done, by
applying the Hamiltonian $\HPS$ with the parameter reset $\epsilon \to
-\epsilon $.

\newparagraph

The dynamics generated by $\HPB$ brings about
decoherence.  After the decoherence time $\dectime$, any pair of
object-pointer states corresponding to macroscopically distinguishable
pointer positions are entangled with almost orthogonal bath states. 
 After averaging the object-apparatus state over the bath
variables, one obtains an object-pointer state $\rhoPS (t)$ in which
all information about the coherences between such 
states is missing, i.e., all coherences for pairs $(s,s')$
of distinct eigenvalues are suppressed.  The irretrievable loss of
information about the bath goes hand in hand with the irreversibility
of the object-pointer dynamics.  

\newparagraph

An object-pointer state $\rhoPS(t)$ describes an accomplished
measurement under two conditions:
\begin{itemize}
\item[(i)] All coherences $\bra{s,x} \rhoPS (t) \ket{s',x'}$
  corresponding to $s \not= s'$ have disappeared, so that $\Sc+\Pc$ is
  in a statistical mixture of separable states like in
  (\ref{eq-post-meas-1}); this occurs at time $t \gg \dectime$.
\item[(ii)]
The separation between the peaks of the distinguished pointer states
$\rhoP^{\,s}(t)$ reaches a macroscopic value $\Delta_{\rm class}$;
this occurs at time $t_0 \gtrsim t_{\rm class}$,
see~(\ref{eq-t_class}).
\end{itemize}
Only for $t_0 \gtrsim t_{\rm class}$ can a classical observer infer a
measured value $s$ by looking at the position of the pointer.  Such a
``reading'' of the pointer, while still a physical process in
principle perturbing $\Pc$, surely cannot blur the distinction of the
peaks. Rather, the pointer will behave classically under a reading,
i.e., it will not noticeably react.

\begin{center}
\begin{figure}
\centering
\includegraphics[width=7.5cm]{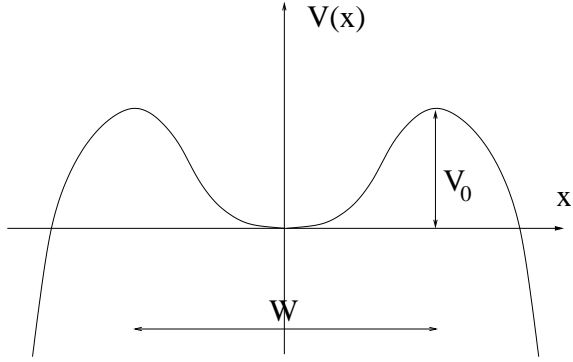}
\caption{\label{fig-0} 
 Sketch of a candidate for the pointer potential. The height $V_0$ of the
potential barriers around $x=0$  and the width $W$ of the potential wall 
 are much larger than $k_B
T$ and the
thermal fluctuation
$\Delta_{\rm th}$.
}
\end{figure}
\end{center}

\vspace*{-1.5cm}

\subsection{Unstable pointer potentials and amplification}
\label{sec-unstable_V}

It is clear from (\ref{eq-t_class}) that condition (ii) can hardly
arise unless the object-pointer coupling constant $\epsilon$ is very
large.  This is related to the well-known
amplification problem in quantum measurements~\cite{Bohm51}.  In order
to get rid of this unrealistic condition on $\epsilon$, one may
consider a different situation than that described in
Sec.~\ref{sec-entanglement}. Let us take a non-confining pointer
potential $V(x)$ with two potential barriers separated by a distance
$W$, see Fig.\ \ref{fig-0}.  The height of these barriers is large
compared with the thermal energy $1/\beta$. We now replace the initial
states $\rhoP=\rhoP^{\rm (eq)}$ in (\ref{parteqapp}) and $\rhoPB$ in
(\ref{eqapp}) by {\it local equilibria} within the potential well.
(This local equilibrium for the apparatus can be achieved by first
preparing $\Pc$ in some state localized near $x=0$ at time
$t=-t_{\rm i}$, with $t_{\rm i}$ larger than the relaxation time but
small compared with the tunneling escape time, and then letting $\Pc$
interact with $\Bc$ until $t=0$.) 
Our previous statements about the distinct peaks in the pointer
density produced by the object-pointer interaction remain valid
for such initial states. 
The interaction $\HPS$ is switched
off at some time $t_{\rm int}$.  
If $t_{\rm int}$ is larger than $W/(\epsilon \delta s)$,
the separation between the peaks in the pointer density at time
$t_{\rm int}$ will be subsequently amplified by the pointer dynamics.
Assuming also that $\Delta \ll W \ll \epsilon \, \delta s \,\TP,
\Delta_{\rm class}$, one has $t_{\rm int} \approx W/(\epsilon \delta
s) \ll \TP, t_{\rm class}$.  In this situation, the small quantum
system $\Sc$ must be able to perturb the pointer strongly enough in
order to produce in it a ``mesoscopic change'' (i.e., a distance $W$
between the peaks in its density), instead of a macroscopic change as
required in Sec.~\ref{sec-entanglement}.  In particular, if
$\rhoP$ is a Gibbs state with position uncertainty $\Delta=\Delta_{\rm
  th}= (\beta V''(0))^{-1/2}$, this arises when the height $V_0
\approx W^2 V''(0)$ of the two potential barriers satisfies
$\beta^{-1} \ll V_0 \ll M (\epsilon\, \delta s )^2$ and $ V_0 \ll
\Delta_{\rm class}^2 V''(0)$ (recall that $\TP^2 = M/V''(0)$).  Then
condition~(ii) of the preceding subsection will be fulfilled after the
object-pointer interaction has been turned off, at time $t_0 \approx
\TP$.  If moreover $V_0 \ll V''(0) ( \epsilon\,\delta s \,\TS )^2$,
 the simplification of the dynamics
discussed in Sec.~\ref{sec-sep_time_scales} can be used 
since the object-pointer interaction time satisfies $t_{\rm int} \ll \TS, \TP$.

\vspace*{5mm}

\section{Simultaneous entanglement and decoherence}  
\label{simul_results}

We now present and discuss the main results of this work, before
deriving them in Sec.~\ref{sec_derivation}.  We are interested in
the object-pointer dynamics when, unlike in the situation just
described, $\Sc$ and $\Pc$ evolve under the {\it simultaneous action}
of $\HPS$ and $\HPB$.  Furthermore, in contrast to
Sec.~\ref{sec-decoherence}, we do not neglect the bath
Hamiltonian $\HB$.

\subsection{Partial-equilibrium initial state}
\label{sec-parteqapp}

Let us first consider the evolution of the initial state
(\ref{parteqapp}). Due to both the initial statistical independence
and our special choice of the interactions, the density matrix of
$\Sc+\Pc$ retains at ``short'' times $t \ll \TS,\TP$ a remarkably
simple product structure (see the discussion in
Sec.~\ref{sec-sep_time_scales}),
\begin{equation} \label{eq-result-S}
\langle s , x | \rhoPS (t) | s', x' \rangle
 = 
  \bra{s} \rhoS^{0} (t)  \ket{s'}\;
   \bra{x_{s} (t)} \rhoP  \ket{x_{s'}' (t)} \;
    \exp
     \left\{
       - D_t \bigl( x_s(t),x_{s'}'(t);s,s' \bigr) 
        - \I \phi_t
     \right\}
        \,.
\end{equation}
Here $\rhoS^{0}(t)$ is given by
(\ref{eq-rhoS(0)}), $x_s(t)=x-t \epsilon s$, $x_{s'}'(t)=x'-t
\epsilon s'$, and $\phi_t$ is a certain real
phase (depending on $t$, $x$, $x'$, $s$, and $s'$) which we do not
specify here since it is irrelevant for decoherence.  We shall derive in
Sec.~\ref{sec_derivation} the decoherence exponent $D_t$, 
\begin{eqnarray} \label{eq-D(i)}
\nonumber
D_t (x,x';s,s')   
& = &  
\frac{1}{2 \hbar^2} \int_0^t  \! \D \tau_1 \! \int_0^t \! \D \tau_2
   \bigl(  
    ( x'+ \tau_1 \epsilon s' )^\alpha
    - ( x+ \tau_1 \epsilon s )^\alpha
   \bigr) \times
\\
& & \bigl(  
    ( x'+ \tau_2 \epsilon s' )^\alpha
    - ( x+ \tau_2 \epsilon s )^\alpha
   \bigr)  h (\tau_1-\tau_2)
\end{eqnarray}
 where
$h(\tau_1-\tau_2)=h_2(\tau_1,\tau_2)$ is the bath two-point function defined
in~(\ref{eq-bath_correl_function}).  The first factor in
(\ref{eq-result-S}) accounts for free evolution of the object initial
state $\rhoS$, as generated by $\HS$, see (\ref{eq-rhoS(0)}).  It is
equal to $p_s = \bra{s} \rhoS \ket{s}$ if $s=s'$ and $t \ll \TS$. 
 The second factor is nothing
but the matrix element (\ref{pointercoherence}) of the shifted pointer initial
state. Here, the Hamiltonian $\HP$
does not show up because of our assumption $t \ll \TP$ and our choice
of a quasi-classical initial state $\rhoP$.  Most important is now the
third factor in (\ref{eq-result-S}); it accounts for decoherence,
i.e., for the suppression of coherences with respect to pointer
displacements associated with different eigenvalues $s\neq s'$.

\newparagraph

The exponent $D_t$ has the following properties:
\begin{itemize}
\item[(a)] $D_t(x,x';s,s') \geq 0$ for all values of $x,x',s$, and
  $s'$.
\item[(b)] $D_t(x,x';s,s') = 0$ initially (for $t=0$)
for all matrix elements and at all later times for the diagonal
matrix elements ($x=x'$ and $s=s'$).
\item[(c)] 
$D_t(x_s(t),x_{s'}'(t);s,s')= D_{-t} (x,x'; s,s')$.
\end{itemize}
The non-negativity (a) is a consequence of the fact
that the correlation function $h(t)$ and its real part $\Re h (t)$
are of positive type, i.e., they have
nonnegative Fourier transforms $\widehat{h} (\omega)$ and
$\widehat{(\Re h)}(\omega)$. 
Actually, one may rewrite (\ref{eq-D(i)}) as   
\begin{equation} \label{eq-D(i)_with_FT}
D_t (x,x';s,s') 
 =  
  \frac{1}{2 \hbar^2} \int_{0}^{\infty} 
  \frac{\D \omega}{\pi} (\widehat{\Re h}) (\omega)
    \left|
    \int_0^t \D \tau 
     \bigl(  
      ( x'+ \tau \epsilon s' )^\alpha - ( x+ \tau \epsilon s )^\alpha
     \bigr) \E^{-\I \, \omega \tau}
   \right|^2 \;\geq \; 0  
\end{equation}
where we have used $\widehat{(\Re h)}(\omega)=\widehat{(\Re  h)}(-\omega)$ and 
$\widehat{(\Im h)}(\omega)=-\widehat{(\Im  h)}(-\omega)$.
Property (c) is easily
checked by a change of the time integration variable in
(\ref{eq-D(i)_with_FT}). Let us recall from
Sec.~\ref{sec-entanglement}
that the
dynamics generated by $\HPS$ maps   
 the object-pointer coordinate $(x,s)$ to 
$(x_s(t),s)$  after time $t$ and, similarly, $(x',s')$ is mapped to
$(x_{s'}' (t),s')$. Hence one may 
interpret (c) as the invariance of $D_t$ under time reversal, i.e., 
under  $t \to -t$ and the exchange of the initial and final coordinates.

\subsection{Equilibrium apparatus initial state}
\label{sec-tot_eq_init_state}

Our result for the initial state (\ref{eqapp}) looks quite similar to
that for the initial state (\ref{parteqapp}). Before stating it, let
us introduce the  effective pointer potential
\begin{equation} \label{eq-effective_pot}
V_{\rm eff} (x) = V(x) -\hbar^{-1} \gamma_0 \, x^{2 \alpha} \;,
\end{equation} 
wherein $\gamma_0$ is given in terms of the imaginary part of the bath
correlation function $h(t)$ by
\begin{equation} \label{eq-gamma0}
\gamma_0= \int_{-\infty}^{0}  \D \tau \,\Im h(\tau) 
 \;.
\end{equation}
We write $ Z_{{\cal P},{\rm eff}} = \int \D x \, \E^{-\beta V_{\rm eff}(
  x)}$ for the partition function associated with $V_{\rm eff}$. It
follows from the general properties of $h(t)$ that $0 \leq \gamma_0 \leq \hbar \beta
h(0)/2$ (see Appendix~\ref{app-correl_functions}). Considering e.g. a
linear pointer-bath coupling, $2\gamma_0/\hbar$ is the mean force per unit
length exerted by the bath on the pointer.  Note that $V_{\rm eff}
(x)$ is a non-confining potential if $V(x)= o(x^{2 \alpha})$ at large
distances. For instance, if $\Pc$ is a harmonic oscillator ($V(x) \propto
x^2$ for all $x$) and $\alpha>1$ then $V_{\rm eff} (x)$ looks like in
Fig.\ \ref{fig-0}.  This means that an initial pointer density
localized around $x=0$ will tunnel away and eventually spread over the
whole real line once the pointer-bath coupling is switched on.
In such a case the
apparatus equilibrium state $\rhoPBeq$ must be replaced by a local
thermal 
equilibrium (see Sec.~\ref{sec-unstable_V}).
This local equilibrium  exists under certain conditions on the pointer-bath coupling 
to be discussed below.

\newparagraph

 As we shall show in Sec.~\ref{sec-derivation-eqapp}, the
object-pointer density operator is given at times $t \ll \TS, \TP$ by
\begin{equation} \label{eq-result-S_totaleq}
\bra{s, x} \rhoPS (t) \ket{s',x'} 
 =  
\bra{s}  \rhoS^{0} (t)   \ket{s'}\,
R_t  \bigl( x_s(t), x_{s'}'(t) ; s,s' \bigr)
\exp \left\{  - D_t(x_{s}(t) ,x_{s'}'(t) ; s,s') -\I \phi_t \right\}
\end{equation}
with the same decoherence exponent  $D_t$ and phase $\phi_t$ as above. 
The only difference between  (\ref{eq-result-S_totaleq}) and
 the formula (\ref{eq-result-S}) for the partial-equilibrium
initial state lies in the replacement of the initial pointer density
$\bra{x}\rhoP \ket{x'}$  
by the function  $R_t(x,x';s,s')$.   
For a time $t$  
short enough so that $D_t(x,x';s,s')\lesssim 1$, 
this function is given by the Gibbs-type density 
\begin{equation} \label{eq-R_tsimeqR_0}
R_t (x,x';s,s') 
 \simeq 
  R_0 (x,x') 
  =  Z_{{\cal P}, {\rm    eff}}^{-1} \,
\E^{
   -\beta 
    ( V_{\rm eff}( x) + V_{\rm eff}(x' ) )/2} \, \E^{- 2 \pi^2 ( x'-x)^2/\lambda_{\rm th}^2}\;.
\end{equation}
For larger times $t$ (with the proviso $t \ll \TS, \TP$), $R_t$ is
given by the more complicated integral (\ref{eq-R}) or, in the special
case $\alpha=2$, by the formula (\ref{eq-result-S_totaleq_alpha=2})
below.  Let us only mention here that for $\alpha=1$,
(\ref{eq-R_tsimeqR_0}) gives the correct answer up to a phase factor
for all times $t \ll \TS,\TP$.  Interestingly,
(\ref{eq-result-S_totaleq}) entails the following result on the
reduced pointer initial state
\begin{equation} \label{eq-R0} 
\bra{x} \tr_{\Bc} \bigl( \rhoPB \bigr) \ket{x'}  = R_0 (x,x')  \; .
\end{equation}
Comparing (\ref{eq-R_tsimeqR_0}) with the expression of $\bra{x}
\rhoP^{(\rm eq)} \ket{x'}$ at high temperatures given in
Sec.~\ref{sec-model}, we see that the coupling between $\Pc$ and
$\Bc$ can be fully accounted for by the effective potential
(\ref{eq-effective_pot}).  Furthermore, for a linear coupling
$\alpha=1$ the matrix elements (\ref{eq-R_tsimeqR_0}) can be approximated for
small $x$ and $x'$ by the Gaussian (\ref{eq-rhoP}) with an almost
unchanged uncertainty in momentum, $\Delta p \simeq 2 \pi
\hbar/\lambda_{\rm th}$, and a renormalized uncertainty in position
$\Delta_{\rm eff} \geq \Delta_{\rm th}$ given by $\Delta_{\rm
  eff}^{-2} = \beta V_{\rm eff}''(0)=\beta ( V''(0) - 2 \gamma_0/\hbar
)$ (see Sec.~\ref{sec-model}).

\newparagraph

Our results (\ref{eq-result-S_totaleq}-\ref{eq-R0}) rely, in addition
to $t \ll \TP, \TS$, on two additional hypotheses: {(a)} the
separation of time scales $\hbar \beta \ll \TP$ or, equivalently, the
separation of length scales $\lambda_{\rm th} \ll \Delta_{\rm th}$
(see Sec.~\ref{sec-simplified_dynamics}); {(b)} a weak enough
pointer-bath coupling satisfying
\begin{equation} \label{eq-stability_cond} 
\begin{cases}
\eta_{\rm th}  < 1/\sqrt{2}
& {\text{ if }} \; \alpha=1 
\\
\eta_{\rm th}    
\ll 1   
&  {\text{ if }} \; \alpha >1 
\end{cases}
\qquad {\rm{ with }} \qquad
\eta_{\rm th}= \langle B^2 \rangle^{1/2} \Delta^{\alpha}_{\rm th}\,
\beta \;.
\end{equation}
Here $\langle B^2\rangle=\tr_\Bc ( B^2 \rhoB)=h(0)$ is the thermal
variance of the bath coupling agent.  Condition
(\ref{eq-stability_cond}) is motivated by the following requirement:
The effective potential (\ref{eq-effective_pot}) must have a local
minimum at $x=0$ and the height of the potential barriers surrounding
the origin must be large compared with the thermal energy $1/\beta$.
Only under that  condition can pointer and bath be prepared in a
local thermal state in which the pointer reduced state has a
single peak at the origin like in Fig.\ \ref{fig-2}(a). If  the coupling $\HPB$
induces an instability in the pointer-bath dynamics, 
we must replace the Gibbs state $\rhoPB$ in
(\ref{eqapp}) by that local thermal state, as explained in
Sec.~\ref{sec-unstable_V}.  Note that we exclude here
pointers being at a critical point of a phase transition considered
in~\cite{Balian01,Balian03}.

\newparagraph

We first consider the case $\alpha=1$. If $V(x)=V''(0) x^2/2$, the
aforementioned requirement is met whenever $V_{\rm eff}''(0)>0$,
i.e., $\gamma_0/\hbar <V''(0)/2$. This stability condition is
well known for a harmonic oscillator interacting linearly with a bath
of harmonic oscillators~\cite{Haake-Reibold85}.  For a potential
$V(x)$ which is non-quadratic at large distances $|x| \gtrsim W$, we
must stipulate a bit more, e.g.  $\gamma_0 /\hbar <
V''(0)/4$, in order that the height of the two potential barriers
be large compared with $1/\beta$. Bearing in mind that $\gamma_0
\leq \hbar \beta \langle B^2\rangle /2$, the latter condition is
satisfied under our hypothesis (\ref{eq-stability_cond}).  Most
importantly, it implies $\Delta_{\rm th} \leq \Delta_{\rm eff} \leq
\sqrt{2} \Delta_{\rm th}$, so that the various length scales are
ordered as $\lambda_{\rm th} \ll \Delta_{\rm th} \approx \Delta_{\rm
  eff} \ll W$.

\newparagraph

Now turning to the case $\alpha >1$ we insert $V(x) \simeq V''(0)
x^2/2$ into (\ref{eq-effective_pot}) and find a distance between the
left and right maxima of $V_{\rm eff}(x)$ equal to $W_{\rm eff} = 2
(\hbar V''(0)/(2\alpha \gamma_0))^{1/(2\alpha-2)}$, these maxima
equaling $V''(0) (\hbar V''(0)/\gamma_0 )^{1/(\alpha-1)}$ up to
a factor of the order of unity.  As a result,
(\ref{eq-stability_cond}) implies the required stability of $V_{\rm
  eff}(x)$.  According to the discussion of
Sec.~\ref{sec-unstable_V},  the
object-pointer coupling can be switched off at time $t_{\rm int} \approx W_{\rm
  eff}/(\epsilon \ds) \approx ( \hbar V''(0) /\gamma_0
)^{1/(2\alpha-2)} (\epsilon \ds)^{-1}$.  This time must be chosen
small compared with $\TS$ and $\TP$ and large compared with the
entanglement time $\enttime= \Delta_{\rm th}/(\epsilon \ds)$, so as to
fulfill (\ref{timescales}) and (\ref{eq-stability_cond}).

\newparagraph

By comparing (\ref{eq-result-S}) and (\ref{eq-result-S_totaleq}) we
may conclude that the coherences of $\rhoPS(t)$ for $s \not= s'$ decay
to zero in same way for the two initial states (\ref{eq-init_state}),
at least in the early time regime when these coherences are not yet
very small. Furthermore, in view of (\ref{eq-R_tsimeqR_0}) the whole
discussion of Sec. \ref{sec-entanglementvsdecoherence} about the
emergence of classically discernible peaks remains qualitatively
valid.

\section{Decoherence times}\label{sec_dec_times}

Before presenting a derivation of our main results (\ref{eq-result-S})
and (\ref{eq-result-S_totaleq}) in the next section, we focus our
attention to the decoherence factor $\E^{-D_t}$.  It has been stressed
in Sec.~\ref{sec-decoherence} that the object-pointer matrix elements
\begin{equation} \label{eq-cohererencesPS_peak}
\rhoPS^{\rm peak} (t;s,s') 
 =
  \bra{s,x=\epsilon t s} \rhoPS(t) \ket{s',x'=\epsilon t s'}
   = \bra{s} \rhoS^{0} (t)  \ket{s'}\, R_t (0,0;s,s')\,
     \E^{-D_t^{\rm peak} (s,s') -\I \phi_t^{\rm peak}}  
\end{equation}
are of particular importance for decoherence in a quantum measurement.
Here $R_t (0,0;s,s')$ is equal to $\bra{0} \rhoP \ket{0}$ for the
initial state (\ref{parteqapp}), to $Z_{{\cal P},{\rm eff}}^{-1}$ for
the initial state (\ref{eqapp}) if $D_t^{\rm peak} (s,s') \lesssim 1$,
and to a more complicated function of $s$ and $s'$ for the initial
state (\ref{eqapp}) if $D_t^{\rm peak} (s,s') \gtrsim 1$.  The main
difference between (\ref{eq-cohererencesPS_peak}) and
(\ref{eq-no_decay_of_coherences}) lies in the presence of the damping
factor $\exp \{ -D_t^{\rm peak} (s,s') \}$ given by
\begin{equation} \label{eq-D_t_00}
D_t^{\rm peak} (s,s') = D_t(0,0;s,s') = 
 \frac{\epsilon^{2 \alpha}}{2 \hbar^2}  \bigl( {s'}^\alpha - s^\alpha \bigr)^2 
    \int_0^t \D \tau_1 \int_0^t \D \tau_2 \, \tau_1^\alpha
    \tau_2^\alpha h (\tau_1 - \tau_2) \;.
\end{equation}
%

\subsection{How does $D_t^{\rm peak}$ grow with time?}
\label{eq-D_t^peak}

The decoherence factor (\ref{eq-D_t_00}) is positive, vanishes for $s=s'$ 
(see (a) and (b) in Sec.~\ref{sec-parteqapp}), and satisfies the following properties:
\begin{itemize}
\item[(d)] $D_t^{\rm
  peak} (s,s')$ is an {\it increasing convex function} of time if $s^\alpha \not= s'^\alpha$. 
\item[(e)] $D_t^{\rm peak} (s,-s)=0$ if
$\alpha$ is even.
\item[(f)]
$D_t(x,x';s,s') = D_t^{\rm  peak} (s,s') \left( 1 + \oforder ((|x|+|x'|)(\epsilon
t |s-s'|)^{-1})\right)$  for  $|x|,|x'| \ll \epsilon t |s-s'|$.   
\end{itemize}  

\begin{center}
\begin{figure}
\centering
\psfrag{a}{$\alpha=1$}
\psfrag{b}{$\alpha=2$}
\psfrag{D}{$D^{\rm peak}_t$}
\psfrag{ln(T)}{$\ln \tau_{\rm dec}$}
\psfrag{ln(t)}{$\ln \tau_{\rm ent}$}  
\psfrag{t}{$\tau$}
\psfrag{tdec}{$\tau_{\rm dec}$}
\includegraphics[width=9.5cm]{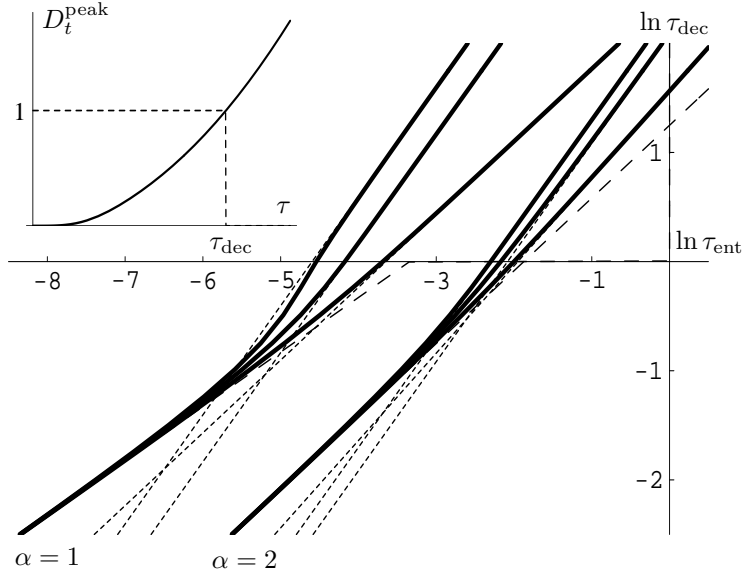}
\caption{\label{fig-3} 
Decoherence against entanglement times in units of $\TBmax=\hbar \beta$ 
 for the  harmonic oscillator bath 
considered in Sec.~\ref{sec-HObath} with $\eta=10^{-1}$, $c=1$,
and $w_D = 5$
(log-log scale).
Solid curves: exact results for 
$(\alpha,m)=(1,5)$, $(1,3)$, $(1,1)$, $(2,5)$, $(2,3)$, and $(2,1)$ (from left to right).
Broken lines: 
approximate expressions for $\tau_{\rm dec} \ll w_D^{-1}$ 
(dashed curves) and $\tau_{\rm dec} \gg 1$ (dotted
curves), see text.
Inset: decoherence exponent $D_t^{\rm peak}$ 
as a function of $\tau=t/\TBmax$ for $(\alpha,m)=(1,3)$.
}
\end{figure}
\end{center}

Property (d) means
that, quite generally, the graph of $D_t^{\rm peak}$ looks
qualitatively like in the inset in Fig.\ \ref{fig-3}.  To establish this result, we
take $x=x'=0$ in (\ref{eq-D(i)_with_FT}), differentiate both sides
with respect to $t$, and do the time integration by parts to get
\begin{equation}  \label{eq-derivative_D_t^peak}
\frac{\partial}{\partial t}  D_t^{\rm peak} (s,s')
 = 
   \frac{\epsilon^{2 \alpha}}{\hbar^2}  \bigl( {s'}^\alpha - s^\alpha \bigr)^2 
   \alpha \,t^{2\alpha}  \int_{0}^{\infty}    \frac{\D \omega}{\pi} 
    \frac{(\widehat{\Re h}) (\omega)}{\omega} 
    \int_0^1 \D u \, (1-u)^{\alpha-1} \sin (\omega t u)\;.
\end{equation}
Using the fact that the function $(1-u)^{\alpha-1}$ is positive and
decreasing between $0$ and $1$, it is easy to show that the integral
over $u$ in (\ref{eq-derivative_D_t^peak}) is positive for almost all
$\omega \geq 0$.  Bearing in mind that $\widehat{(\Re h )} (\omega )\geq 0$, this
establishes that $\partial D_t^{\rm peak}/\partial t >0$ for $t>0$. Hence
$D_t^{\rm peak}$ is an increasing function of $t$. By a similar
argument, $\partial^2 D_t^{\rm peak}/\partial t^2 >0$ and thus $D_t^{\rm peak}$ is
convex.  

\newparagraph

According to property (e), if $\alpha$ is even and the spectrum of $S$
is symmetric with respect to $s=0$, the coherences
(\ref{eq-cohererencesPS_peak}) for $s'=-s$ are not damped.  This comes
from the symmetry $x \leftrightarrow -x$ of the Hamiltonian $\HPB$ in
(\ref{eq-int_Hamiltonian}), which allows for the existence of
decoherence-free subspaces~\cite{Lidar03}.  Due to these long-living
coherences, $\Pc+\Sc$ fails to reach (at least within a time span $t
\ll \TP,\TS$) the statistical mixture required to be able to give a
classical result to the measurement.  We exclude that case from now
on.  More precisely, we assume that if $\alpha$ is even then $s/s'$ is
not close to $-1$ for all pairs $(s,s')$ of eigenvalues such that
$\bra{s} \rhoS \ket{s'} \not= 0$, i.e., $|
s'^\alpha-s^\alpha|/|s'-s|^\alpha$ is bounded below by a constant
$c_\alpha^{\rm min} >0$ of the order of unity.  With this restriction,
for $(|x|+|x'|)/ (\epsilon t |s-s'|)$ sufficiently small
the error term in property (f) is bounded by $(|x|+|x'|) (\epsilon t
|s-s'|)^{-1}$ times a constant independent of $x$, $x'$, $s$, $s'$, and
$t$~\cite{footnote-uniform_bound}.

\newparagraph

We are concerned in this section  with determining the time scale
$\dectime (s,s')$ characterizing the growth of $D_t^{\rm
  peak}(s,s')$ and the corresponding decay of the $(s\neq
s')$-coherences (\ref{eq-cohererencesPS_peak}).  This time, to be
called the decoherence time, is defined implicitly as
$D_{t=\dectime}^{\rm peak} (s,s') = 1$, i.e.,
\begin{equation} \label{t_dec_vs_t_ent}
\left( \frac{\enttime (s,s')}{\eta^{1/\alpha}} \right)^{2 \alpha}
= 
 \frac{c_\alpha (s,s')^2}{(\hbar \beta)^2}  
  \int_0^{\dectime(s,s')} \D \tau_1 \int_0^{\tau_1} \D \tau_2 \, \tau_1^\alpha
    \tau_2^\alpha\, \frac{\Re h (\tau_1 - \tau_2)}{\langle B^2 \rangle} 
\end{equation}
where 
\begin{equation} \label{eq-enttime}
\enttime(s,s') =
\frac{\Delta}{\epsilon |s'-s|}
\end{equation}
is the entanglement time (whose physical interpretation has been
illustrated in Sec.~\ref{sec-entanglement}), $\eta$ is the
(fluctuation of the) initial pointer-bath coupling energy in units of
$k_B T$,
\begin{equation} \label{eq-eta}
\eta = \langle B^2 \rangle^{1/2} \Delta^{\alpha} \beta \approx
\beta \bigl( \tr (\HPB^2 \,\rhoP \otimes \rhoB ) \bigr)^{1/2} \;,
\end{equation}
$c_\alpha(s,s')=1$ if $\alpha=1$, and
\begin{equation} \label{eq-min_s_part}
c_\alpha (s,s') 
 =  
    \frac{|s'^\alpha- s^\alpha|}{|s'-s|^\alpha} \quad {\rm if }\quad \alpha \geq 1\;. 
\end{equation}
For the initial state (\ref{eqapp}), one must set $\Delta=\Delta_{\rm th}$ in
(\ref{eq-enttime}) and (\ref{eq-eta}) and $\eta=\eta_{\rm th}$ must be
small enough, see (\ref{eq-stability_cond}). 
 By inspection of
(\ref{t_dec_vs_t_ent}), $\dectime(s,s')$ depends on the object-pointer
and pointer-bath coupling constants $\epsilon$ and $\eta$ through a single
parameter $\epsilon\, \eta^{1/\alpha}$.  Recalling that
(\ref{eq-result-S}-\ref{eq-D_t_00}) are valid with the proviso $t
\ll\TS,\TP$, the ``free''
evolutions of $S$ and $X$ must be slow compared to $\dectime (s,s')$, i.e.,
\begin{equation} \label{eq-short_time}
\dectime (s,s')  \ll \TS, \TP  \quad , \quad s\not= s'  \;.
\end{equation}
For given 
$s\not=s'$, if $\dectime (s,s') \geq \enttime(s,s')$ then at time $t \gg
\dectime (s,s')$ the peaks at $(x,x')=(\epsilon t s,\epsilon t s')$
of the pointer coherences in (\ref{eq-result-S})
and (\ref{eq-result-S_totaleq}) (second factors on the right-hand
sides) are  flatten down   
by decoherence (third factors), so that 
 $\bra{s,x} \rhoPS(t) \ket{s',x'} \simeq 0$ 
for all values of $(x,x')$.  This statement follows
from a similar argument as in Sec.~\ref{sec-decoherence} and
from property (f) (see the beginning of this section).  It is worth 
emphasizing that if, unlike in the situation just described, $\dectime
(s,s')$ is smaller than $\enttime (s,s')$ then the coherence $\bra{s
  ,x} \rhoPS(t) \ket{s',x'}$ may still be large at time $\dectime
(s,s')$ for some $(x,x') \simeq (\epsilon t s,\epsilon t s')$ 
with uncertainty $\Delta$.  In such a case the decoherence time must be defined as the
time $t$ at which the minimum of $D_{t} (x,x';s,s')$ over all values
of $(x,x')$ is equal to $1$. We postpone to a separate work the
determination of that decoherence time.

\newparagraph

The decoherence time $\dectime$ of
the measurement is the largest of the times $\dectime(s,s')$ for all
pairs of distinct eigenvalues $(s,s')$ such that $\bra{s} \rhoS
\ket{s'} \not= 0$ (with the proviso $\dectime \geq
\enttime=\Delta/(\epsilon\, \delta s)$ in light of the discussion
in the preceding paragraph).  This amounts to replacing 
$| s'^\alpha-s^\alpha|$ in (\ref{eq-D_t_00}) by its minimum value over
all such pairs $(s,s')$ (recall that $D_t^{\rm peak}$ is an increasing
function of time). For $\alpha=1$, this minimum value is by definition
equal to $\delta s$ (Sec.~\ref{sec-entanglement}); for $\alpha
\geq 2$, it depends on the spectrum of $S$ in a more subtle
way~\cite{footnote-upper_bound}.
At times $t \gg \dectime$, the object-pointer state $\rhoPS(t)$ is 
very close to the separable state
(\ref{eq-post-meas-1}). In other words, $\Sc$ and $\Pc$ are in the
statistical mixture (\ref{eq-post-meas-2}) with the probabilities
$p_s=\bra{s} \rhoS \ket{s}$ and with pointer states $\rhoP^{\,s}(t)$ given
by
\begin{equation} \label{eq-shifted-pointer_densities}
\bra{x} \rhoP^{\,s} (t) \ket{x'}
  = 
   R_t \bigl( x_{s}(t) , x_{s}'(t);s,s \bigr)
    \, \exp \left\{  - D_{-t}(x,x';s,s) - \I \phi_t \right\}
\end{equation}
with $R_t( x , x';s,s)$ equal to $\bra{x} \rhoP \ket{x'}$ for the
initial state (\ref{parteqapp}) and to the Gibbs-like density
(\ref{eq-R_tsimeqR_0}) for the initial state (\ref{eqapp}) when
$D_{-t}(x,x';s,s)\lesssim 1$.  We have used in
(\ref{eq-shifted-pointer_densities}) the time-invariance property (c),
see Sec.~\ref{sec-parteqapp}.  The initial superpositions of
eigenstates $\ket{s}$ have disappeared by {\it indirect decoherence 
  via the pointer}. The pointer is in a statistical mixture of
quasi-classical states having densities localized around $x=t \epsilon
s$ with uncertainty $\Delta$.  The essence of quantum measurements
lies in this loss of coherences: for indeed, as already pointed out in
Sec.~\ref{sec-summary_ent_dec} it is only when all
object-pointer coherences for $s \not= s'$ are vanishingly small that
a classical probability can be given for the result of the
measurement.

\newparagraph

The pointer matrix elements (\ref{eq-shifted-pointer_densities})
are also damped
by decoherence via the last exponential factor in (\ref{eq-shifted-pointer_densities}).
One can show, however, that for relevant values of $x$ and $x'$ satisfying
$|x-\epsilon t s| \leq \Delta$ and $|x'-\epsilon t s| \leq \Delta$, 
the corresponding damping time is much
larger than $\dectime$, at least in the two  limiting regimes
$\dectime \ll \TBmin$ and $\dectime \gg \TBmax$ studied below. The special case 
$\alpha=1$ will be discussed in the next subsection. 

\newparagraph

It is worthwhile mentioning here that one should expect that $\dectime \ll
t_{\rm class}$, save for extremely large 
object-pointer coupling constants $\epsilon$. Object and pointer are then never in a
Schr{\"o}dinger cat state as in (\ref{eq-Qsuperposition}),
because decoherence subdues linear superpositions 
(via the third factors in (\ref{eq-result-S}) and (\ref{eq-result-S_totaleq})) 
 faster than entanglement between $\Pc$
and $\Sc$ can produce them (second factors in
(\ref{eq-result-S}) and (\ref{eq-result-S_totaleq})).
Due to the simultaneous action of $\HPS$ and $\HPB$, the whole
measurement process directly produces the mixture of macroscopically
distinct pointer states $\rhoP^{\,s}(t)$, without allowing for the
intermediate appearance of macroscopic superpositions.
This is
one of the central results of the present paper.
In the situation described
in Secs.~\ref{sec-unstable_V} and \ref{sec-tot_eq_init_state},
i.e., if the (effective) pointer potential is unstable and the
object-pointer interaction is switched off at time $t_{\rm int}
\approx W/(\epsilon \delta s) \ll
\TP, t_{\rm class}$, 
even  mesoscopic superpositions do not appear at any stage of the measurement
when $\dectime \ll t_{\rm int}$.

\newparagraph

No assumption whatsoever was made on the bath correlation time $\TBmax$
to establish (\ref{eq-result-S}-\ref{eq-D_t_00}).  Our
results therefore go beyond the so-called Markovian limit which would
require $\TBmax \ll \dectime$.  This is an
important point, since for sufficiently large $\epsilon$ decoherence will
take place within the ``non-Markovian'' regime $t \lesssim \TBmax$.
Explicit  asymptotical results for $\dectime$ can now be
drawn from the foregoing expressions   for
both   $\dectime  \ll\TBmin$ and $\dectime  \gg\TBmax$.

\vspace*{-0.5cm}

\subsection{ Interaction dominated  regime  $\dectime \ll \TBmin$}
\label{sec-short-time}

In the (non-Markovian) regime $\dectime \ll \TBmin$, the dynamics is
dominated by the interactions $\HPS$ and $\HPB$.  For $t \ll \TBmin$,
one may approximate $h (\tau)$ in (\ref{eq-D_t_00}) by the thermal
variance $h(0)= \meanB{B^2}$ of the bath coupling agent. This yields
\begin{eqnarray} \label{D^peak_short_time}
D_t^{\rm peak} (s,s')
& = & 
\left(
\frac{t}{\dectime(s,s')} \right)^{2 \alpha+2}
\quad \text{(interaction-dominated regime),}
\\ \label{t_dec_short_time}
\dectime (s,s')
& = & 
  \left( 
   \frac{\sqrt{2} (\alpha+1)}{c_\alpha(s,s')} \right)^{\frac{1}{\alpha+1}}
   \left( \frac{\enttime (s,s')}{\hbar \beta\,\eta^{1/\alpha} } \right)^{\frac{\alpha}{\alpha+1}}
    \hbar \beta 
\end{eqnarray}
where we have used the entanglement time and dimensionless parameters
(\ref{eq-enttime}-\ref{eq-min_s_part}).  That result makes sense with
the proviso $\dectime (s,s') \ll \TBmin, \TS, \TP $.  The fact that
$D_t^{\rm peak} \propto t^{2\alpha+2}$ could be expected from 
(\ref{eq-ln_scalar_prod}): indeed, a contribution of
$t^2 ({x'}^\alpha-x^\alpha)^2$ to the decoherence exponent was found
for $\HPS=0$ and fixed $x$, $x'$; recalling that for $\HPS$ given by
(\ref{eq-int_Hamiltonian}) the positions of the peaks grow
proportionally with time, we recover the above-mentioned power law.
It should also be noted that $\dectime$ depends on the bath through the
single parameter $\eta$.  This comes from the fact that the bath
dynamics can be ignored when $\dectime \ll
\TBmin$~\cite{footnote-small_time}. 

\newparagraph

 Invoking (\ref{eq-D_t_00}) and $\widehat{(\Re h )}(\omega)
\geq 0$, it is easy to show that 
(\ref{D^peak_short_time}) gives an upper bound on $D_t^{\rm peak}
(s,s')$ for all times $t\geq 0$. By property (d) in
Sec.~\ref{eq-D_t^peak}, in the regime $\dectime \gtrsim \TBmin$
the decoherence time must be larger than the right-hand side of the
asymptotic formula (\ref{t_dec_short_time}).

\newparagraph

It has been stressed above that the
interpretation of $\dectime$ as the decoherence time of the
measurement relies on the assumption $\enttime \leq \dectime$. 
We now argue that this condition is fulfilled if the
pointer-bath coupling energy is of the order or smaller than 
$k_B T$ (i.e., $\eta \lesssim 1$) and $s'/s$ is not very close to unity.
In fact, under these assumptions one has even $\enttime (s,s') \ll \dectime (s,s')$.
This follows from (\ref{t_dec_short_time}), the
consistency condition $\dectime(s,s') \ll \TBmin$ and the inequality
$\TBmin \leq \hbar \beta$, which imply $\dectime (s,s') \ll \hbar
\beta$ and thus $\enttime (s,s') \ll \hbar \beta$.
In contrast, if $\alpha \geq 2$ and $|s'/s-1|\ll 1$
one sees by inspection of  (\ref{eq-min_s_part}) that 
$ c_\alpha (s,s') \simeq \alpha |1-s'/s|^{1-\alpha} \gg 1$.
Hence the
first factor on the right-hand side of (\ref{t_dec_short_time})
is small  and one may have 
$\enttime (s,s') \geq \dectime (s,s')$. 
This corresponds to an initial superposition of eigenstates $\ket{s}$
with closely lying eigenvalues, as discussed in Sec.~\ref{sec-summary_ent_dec}.
Similarly, one may have $\enttime (s,s') \geq \dectime (s,s')$ 
for a strong pointer-bath coupling, i.e., for $\eta \gg 1$.

\newparagraph

It is worthwhile comparing the strength of decoherence for different
values of the exponent $\alpha$ in the coupling Hamiltonian $\HPB$,
keeping its magnitude $\eta/\beta$  constant.  We
find that $\dectime$ is smaller in the nonlinear case $\alpha>1$ in
comparison with the linear case $\alpha=1$ by a factor
$\dectime^{(\alpha> 1)}/\dectime^{(\alpha= 1)}$ of the order of
$c_\alpha^{-1/(\alpha+1)} (\enttime /\dectime^{(\alpha=1)} )^{(\alpha-1)/(\alpha+1)} \ll 1$.
Interestingly, a linear pointer-bath coupling is much less efficient
than a nonlinear one in suppressing the coherences
(\ref{eq-cohererencesPS_peak}) for $s \not= s'$.  This has the
following important consequence: for a pointer-bath coupling of the
form $\HPB = B \,f(X)$ with $f(x)$ a smooth real function, a
dipole-like approximation consisting in linearizing $f(x)$ may lead to
an over-estimation of the decoherence time $\dectime$ even if $f''(0)
\Delta/f'(0)$ is small.  Actually, the quadratic coupling $B f''(0)
X^2/2$ gives a smaller decoherence time than the linear coupling $B
f'(0) X$ when $\enttime/\hbar \beta \lesssim \eta^{-1} (f''(0) \Delta
/f'(0))^2$ with $\eta= f'(0) \langle B^2\rangle^{1/2} \Delta\,\beta$.

\newparagraph 

We can now give an explicit condition ensuring that $\dectime$ is
smaller than the time $t_{\rm int} \approx W/(\epsilon \delta s)$
needed by object-pointer entanglement 
to produce superpositions of pointer positions
separated by the mesoscopic length $W$, 
$\Delta \ll W \ll \Delta_{\rm class}$: $\enttime$ 
must be large compared with $(\Delta/W)^{\alpha+1} \hbar \beta/ (\eta c_\alpha)$. 
In this limit decoherence is so fast that these mesoscopic superpositions
do not appear at any moment during the measurement.

\newparagraph

 As pointed out in Sec.~\ref{eq-D_t^peak}, it is appropriate to
demonstrate that the decay of the pointer matrix elements 
(\ref{eq-shifted-pointer_densities}) 
remains negligible for times $t$ until
well after the disappearance of the off-diagonal $(s\neq s')$ terms in
$\rhoPS(t)$.  Due to the peak structure of the pointer density (first
factor in the right-hand side of
(\ref{eq-shifted-pointer_densities})), the relevant values of $x,x'$
are such that $|x_s(t)|,|x_s'(t)| \leq \Delta$.   
Given $\enttime \ll \dectime$, such $x$ and $x'$ are separated by a distance
$|x-x'|\leq 2 \Delta$ much smaller than the interpeak distance
$\dectime  \epsilon \delta s$ relevant for the decay of the $(s,s'=s+\delta s)$ matrix
elements of $\rhoPS(t)$. 
  We restrict ourself to the
case $\alpha=1$ and consider the limit $t \ll \TBmin\leq \hbar \beta$.  Setting
$s=s'$, inverting the sign of the time $t$, and replacing
$h(\tau_1-\tau_2)$ by $\meanB{B^2}$ in (\ref{eq-D(i)}), one finds that
$D_{-t}(x,x';s,s)= \langle B^2 \rangle t^2 (x-x')^2/(2 \hbar^{2})$.  Note that
this decoherence exponent is the same as in (\ref{eq-post-meas-3}).  If
$\eta \lesssim 1$ and $x$, $x'$, and $t$ are in the range mentioned above then
$D_{-t}(x,x';s,s) \leq 2 \eta^2 t^2 (\hbar \beta)^{-2}\ll 1$.  Hence the
decoherence caused by the pointer-bath coupling has a small effect on
the pointer states $\rhoP^{\,s}(t)$ up to times $t \lesssim \dectime(s)$, 
the decoherence factor in
(\ref{eq-shifted-pointer_densities}) being still close to unity.  So
indeed, the bath does away with the ``off-diagonal'' ($s \not= s'$)
object-pointer  matrix
elements before the ``diagonal'' ones change noticeably.

\subsection{Markov regime $\dectime \gg \TBmax$}
\label{sec-Markov}

When $\dectime \gg \TBmin$ the off-diagonal matrix elements
(\ref{eq-cohererencesPS_peak}) have no time to decay between $t=0$ and
$\TBmin$.  Decoherence may then take place within the so-called Markov
regime $t \gg \TBmax$, also known in the mathematical literature as the
singular-coupling limit~\cite{Lieb73,Gorini76}.  Note that under our
condition $\dectime \ll \TS,\TP$ it is not appropriate to use a
rotating-wave approximation.  Decoherence
 is governed in the Markov regime by the small-frequency
behaviors of the Fourier transforms $(\widehat{\Re h})(\omega)$ and
$(\widehat{\Im h})(\omega)$ of the real and imaginary parts
of the bath correlator $h(t)$.  We shall make use of a few properties of
these Fourier transforms, which are explained in more detail in
Appendix~\ref{app-correl_functions}.  We assume that $(\widehat{\Im
  h})(\omega) \sim - \I \,\widehat{\gamma}\,\omega^{m}$ for $\omega \ll \TBmax^{-1}$,
$\widehat{\gamma}$ being a positive constant.  Bearing in mind that
$(\widehat{\Im h}) (\omega)$ is an odd function of $\omega$ and must be regular
enough (i.e., admit differentials of sufficiently high orders) in such
a way that $\Im h(t)$ decays rapidly to zero as $t \to \pm \infty$, we take
$m$ to be a positive odd integer.  By analogy with the case of a bath
of harmonic oscillators linearly coupled to $\Pc$, we speak of {\it
  Ohmic damping} when $m=1$ and of {\it super-Ohmic damping} when
$m>1$~\cite{Caldeira-Leggett83,Weiss}.  The behavior of $(\widehat{\Re
  h}) (\omega)$ at small frequencies can be deduced from that of
$(\widehat{\Im h}) (\omega)$ thanks to the Kubo-Martin-Schwinger (KMS) relation~(\ref{eq-KMS_FT}).
Such a relation holds because the average in
the correlation function $h(t)$ is  taken with respect to a bath Gibbs
state~\cite{Bratteli}.  It implies $(\widehat{\Re h}) (\omega) \sim 2
\,\widehat{\gamma} \, \omega^{m-1}/(\hbar \beta)$.

\newparagraph

Let us first discuss the super-Ohmic case $m \geq 3$.
The frequency integral in  (\ref{eq-derivative_D_t^peak}) can be rewritten
after an integration by parts as 
\begin{eqnarray}  \label{eq-derivative_D_t^peak_bis}
\nonumber
&  \displaystyle
\int_{0}^{\infty}    \frac{\D \omega}{\pi} 
    \frac{(\widehat{\Re h}) (\omega)}{\omega} 
    \int_0^1 \D u \, (1-u)^{\alpha-1} \sin (\omega t u)
 = 
t^{-1} \int_{0}^{\infty}    \frac{\D \omega}{\pi} 
    \frac{(\widehat{\Re h}) (\omega)}{\omega^2} 
    \Bigl( 1 
- \delta_{\alpha 1} \cos (\omega t) +
\\
& \displaystyle 
- (\alpha-1) \int_0^1 \D u \, (1-u)^{\alpha-2} \cos
      (\omega t u)
    \Bigr) 
\;\simeq\; 
t^{-1} \int_0^\infty \frac{\D \omega}{\pi} \frac{(\widehat{\Re h}) (\omega)}{\omega^{2}}
\end{eqnarray}
where we have neglected  in the last expression the  oscillatory 
integrals by invoking $t \gg \TBmax$. By inspection of
(\ref{eq-integral_a}) we conclude that 
 for $m \geq 3$ the frequency integral in (\ref {eq-derivative_D_t^peak})
can be approximated by 
$ t^{-1} | \int_0^\infty \D \tau \,\tau\,\Re h(\tau)|$.

\newparagraph

For an Ohmic bath $m=1$, the last 
integral in (\ref{eq-derivative_D_t^peak_bis})
diverges. 
We now argue that one can replace
$(\widehat{\Re h}) (\omega)$ by
$(\widehat{\Re h}) (0)=2\, \widehat{\gamma}\, (\hbar \beta)^{-1}$
 on the left-hand side of~(\ref{eq-derivative_D_t^peak_bis}), which becomes 
\begin{equation} \label{eq-approx_int_Markov}
 \int_0^\infty \frac{\D \omega}{\pi} 
    \frac{(\widehat{\Re h}) (0)}{\omega} 
    \int_0^1 \D u\, (1-u)^{\alpha-1} \sin (\omega t u)
    =  \frac{(\widehat{\Re h}) (0)}{2 \alpha}   
\end{equation}
in the limit $t \gg \TBmax$. 
(We have used $\int \D \omega \sin(\omega t u)/\omega = \pi$ for $t u >0$.)
Note that this 
amounts  to replacing $\Re h(t)$ by a white-noise correlator $2\, \widehat{\gamma} (\hbar
\beta)^{-1} \delta (t)$ in (\ref{eq-D_t_00}). 
Let us estimate the error  introduced  in the frequency integral 
in~(\ref{eq-derivative_D_t^peak}) by this substitution. This error 
is given by the left-hand side of
(\ref{eq-derivative_D_t^peak_bis}) modulo the replacement of
$(\widehat{\Re h})(\omega)$ by 
$(\widehat{\Re h})(\omega)-(\widehat{\Re h})(0)$.
Disregarding
oscillatory integrals as in the case $m\geq 3$, the error is 
equal  in the limit $t \gg \TBmax$ to 
$t^{-1} \int_0^\infty \D \omega\, ((\widehat{\Re h}) (\omega)- (\widehat{\Re h}) (0))
\, \omega^{-2}/\pi$. The latter integral converges 
since $(\widehat{\Re h})(\omega)-(\widehat{\Re h})(0)$ behaves like
$\omega^{2}$ for small $\omega$.
Comparing with (\ref{eq-approx_int_Markov}) (see also
(\ref{eq-integral_a_bis})), 
one concludes that 
the relative error introduced in (\ref{eq-derivative_D_t^peak})
by the substitution of $(\widehat{\Re h}) (\omega)$ by
its value for $\omega=0$  is small, 
of the order of $\TBmax/t$.
Hence, for $m = 1$ the frequency integral in (\ref {eq-derivative_D_t^peak})
can be approximated by 
$(\widehat{\Re h}) (0)/(2 \alpha)= \alpha^{-1} \int_0^\infty \D \tau \,\Re h(\tau)$.

\newparagraph

Collecting the above results and integrating
(\ref{eq-derivative_D_t^peak}) with respect to time,
we find in the Ohmic case $m=1$
\begin{eqnarray} \label{D_t-Markov-Ohmic}
D_t^{\rm peak} (s,s') & = & \left( \frac{t}{\dectime(s,s')} \right)^{2 \alpha+1} 
\quad  \quad \mbox{(Ohmic)}
\quad , 
\\  \label{t_dec-Markov}
\displaystyle
 \dectime (s,s')
& = & \left( 
   \frac{(2 \alpha+1)\, \langle B^2 \rangle \, \hbar \beta}
    {c_\alpha (s,s')^2 \int_{0}^\infty \D \tau\,\Re h (\tau) } 
   \right)^{\frac{1}{2 \alpha+1}} 
    \left( \frac{\enttime (s,s')}{\hbar \beta\,\eta^{1/\alpha} } \right)^{\frac{2\alpha}{2\alpha+1}}
     \hbar \beta 
\end{eqnarray}
and in the super-Ohmic case $m \geq 3$
\begin{eqnarray}  \label{D_t-Markov-sup_ohmic} 
D_t^{\rm peak} (s,s') & = & \left( \frac{t}{\dectime(s,s')} \right)^{2 \alpha} 
\quad \quad \mbox{(super-Ohmic)} \quad , 
 \quad 
\\ \label{t_dec-Markov_sup_ohmic}
 \dectime (s,s')
& = & \left( 
   \frac{2\, \langle B^2 \rangle\, \hbar^2 \beta^2}
    {c_\alpha (s,s')^2 | \int_{0}^\infty \D \tau\,\tau\, \Re h (\tau)| } 
   \right)^{\frac{1}{2 \alpha}} \frac{\enttime (s,s')}{\eta^{1/\alpha}}
\end{eqnarray}
with the proviso $\TBmax \ll \dectime (s,s')\ll
\TS,\TP$. 
We can interpret the growth of $D_t^{\rm peak}$ like $t^{2 \alpha+1}$ in the Ohmic case
by saying that  for fixed $x$ and $x'$, in the Markov regime 
$D_{t}$ must be proportional to $t (x'^{\alpha}-x^\alpha)^2$ 
(the fact that $D_t \propto t$ is well known~\cite{Giulini});
the indicated time behavior of $D_t^{\rm peak}$ then follows by replacing 
$(x,x')$ by $(\epsilon t s,\epsilon t s')$.   

\newparagraph

By using $|h(\tau)| \leq \langle B^2 \rangle$ and  $\Re h (\tau)
\simeq 0$ for $\tau \gg \TBmax$ (Sec.~\ref{sec-time scales}), 
one finds that the integrals 
$\int_0^\infty\D \tau\,\Re h (\tau)$ and  
$|\int_{0}^\infty \D \tau\,\tau\, \Re h (\tau)|$ are at most 
of the order of  
$\langle B^2 \rangle \TBmax$ and $\langle B^2 \rangle \TBmax^2$, 
respectively. 
If  $|s'/s-1|$ is
not close to unity (so that $c_\alpha (s,s')$ in
(\ref{eq-min_s_part}) is not very large)  
the factor inside the parenthesis in (\ref{t_dec-Markov_sup_ohmic})
is of the order of $(\hbar \beta/\TBmax)^2$ or larger. 
Thus, for coupling strength $\eta \lesssim \hbar \beta
/\TBmax$ the condition 
$\enttime(s,s') \leq \dectime(s,s')$ holds in the Markov regime for
super-Ohmic baths.  
The situation is different for Ohmic baths: then, by
(\ref{t_dec-Markov}), the condition in question is violated  
even for small $\eta$ if the entanglement time
$\enttime (s,s')$ is large enough compared with $\hbar \beta$.
More precisely, still
 assuming that $c_\alpha(s,s')$ is of the order of unity, 
$\dectime (s,s')$ becomes smaller than $\enttime (s,s')$ when 
\begin{equation} \label{eq-cond_tdec<tent}
\frac{\enttime (s,s')}{\hbar \beta} 
\gtrsim
\frac{\langle B^2 \rangle \hbar \beta}{\eta^{2} \int_0^\infty \D \tau \,\Re h (\tau)}\;.
\end{equation}

\newparagraph
 
For super-Ohmic baths, the decoherence time
(\ref{t_dec-Markov_sup_ohmic}) decreases by increasing $\alpha$ for $\eta \lesssim
\hbar \beta /\TBmax$ and $|s'/s-1|$ not close to unity, i.e., provided that
$\dectime (s,s') \geq \enttime (s,s')$. Then
$\dectime^{(\alpha>1)}/\dectime^{(\alpha=1)} \lesssim ( \eta \TBmax/(\hbar \beta))^{1-1/\alpha}
\leq 1$. 
Thus, for fixed weak enough coupling
strength $\eta$, nonlinear pointer-bath couplings always win over a
linear coupling in efficiency for decoherence. 
 This is in striking contrast with what happens in the
Ohmic case. Actually, for a Ohmic bath
nonlinear couplings become less efficient than a linear
coupling when  $\enttime(s,s')$ is large enough so as to fulfil  
(\ref{eq-cond_tdec<tent}). 
More precisely, we find by using $c_\alpha(s,s')\approx 1$ and (\ref{eq-cond_tdec<tent}) that the decoherence time
(\ref{t_dec-Markov}) is larger in
the nonlinear case $\alpha>1$ than in the linear case $\alpha=1$ by a factor
$\dectime^{(\alpha>1)}/\dectime^{(\alpha=1)}$ of the order of $(\enttime
/\dectime^{(\alpha=1)} )^{(2 \alpha-2)/(2 \alpha+1)} \geq 1$.

\newparagraph

Finally, it is worth mentioning that Ohmic baths win in efficiency
over super-Ohmic baths.  This can be shown by noting that
$(\dectime^{\rm Ohm}/\dectime^{\rm sup\,Ohm})^{2\alpha+1}$ is equal (up
to a numerical factor of the order of unity) to the product of $|
\int_0^\infty \D \tau \, \tau \,\Re h (\tau)| (\hbar \beta
\int_0^\infty \D \tau \,\Re h (\tau) )^{-1}$ by $\hbar \beta
/\dectime^{\rm sup\,Ohm}$. Since the last factor must be small
compared with $1$ for consistency (recall that $\hbar \beta \leq
\TBmax$), it follows that $\dectime^{\rm Ohm}(s,s')$ is smaller than
$\dectime^{\rm sup\,Ohm}(s,s')$.
  
\newparagraph

One may wonder if the results of this section 
could be strongly modified if a direct
coupling between the object $\Sc$ and bath $\Bc$ (which we do not admit in the present
model) was allowed for.  It is clear that one can answer this question
by the negative when the object-pointer coupling constant $\epsilon$ is large
enough, i.e., for small enough $\enttime$. In order to estimate 
how small must be $\enttime$, let us couple $\Sc$ and $\Bc$ via
the Hamiltonian $H_{\Sc \Bc}= \Delta^\alpha (S/\ds)^\alpha B$. This Hamiltonian  
has a magnitude comparable with the pointer-bath coupling 
(\ref{eq-int_Hamiltonian}) in the initial
state (\ref{parteqapp}).
We first consider Ohmic baths.   It is known
that the decay of the off-diagonal matrix elements $\bra{s} \rhoS (t)
\ket{s'}$ resulting from the coupling  $H_{\Sc \Bc}$ then goes  
like $\exp( - t/ T_{\rm dec}
(s,s'))$ in the Markov regime 
(we ignore  here the object-pointer coupling)~\cite{Giulini}. 
If  $T_{\text{dec}}(s,s') \ll \TS$, a condition fulfilled e.g. if 
$[S,\HS]=0$ (pure dephasing regime, $\TS=\infty$), 
the corresponding decoherence time is given by
$T_{\text{dec}}(s,s') = \hbar^{2} (\ds/\Delta)^{2 \alpha}
|{s'}^\alpha -s^\alpha|^{-2} / \int_0^\infty \D \tau \, \Re h
(\tau)$~\cite{Haake01}. 
The ratio between $T_{\text{dec}}(s,s')$  and the decoherence
time (\ref{t_dec-Markov}) 
for $s'=s+\ds$ is   
$ [ (c_\alpha \eta)^{-2} \enttime^{-1} ( \hbar
\beta)^2 \langle B^2 \rangle  /\int_0^\infty \D \tau\, \Re h (\tau)  
 ]^{2\alpha/(2 \alpha+1)}$  up to an irrelevant factor.
Taking into account that
$\int_0^\infty\D \tau\,\Re h (\tau) \lesssim \langle B^2 \rangle \TBmax$, we see that
it is well
justified to neglect the coupling of the object with all degrees of
freedom of the apparatus but the pointer 
provided that $\enttime/(\hbar \beta) \ll (\hbar \beta/\TBmax) (c_\alpha \eta)^{-2}$.
For a super-Ohmic bath, if $[S,\HS]=0$ then
the modulus of the off-diagonal matrix element $\bra{s} \rhoS (t)
\ket{s'}$ decays to a nonzero value under the coupling  $H_{\Sc \Bc}$
(for a discussion on this saturation of decoherence see e.g.~\cite{Ekert97}),
whereas indirect decoherence
via the pointer leads to a complete decay of the
object-pointer
coherences (this decay being given by the decoherence exponent (\ref{D_t-Markov-sup_ohmic})). 
It is also easy to show that 
$T_{\text{dec}}(s,s')$ is much larger than the
decoherence
time (\ref{t_dec_short_time})
provided that $t_{\rm ent}/(\hbar\beta) \ll (\hbar \beta/\TBmax)^{1+1/\alpha}
 (c_\alpha \eta)^{-2-1/\alpha}$.

\subsection{Bath of harmonic oscillators linearly coupled to ${\cal P}$}
\label{sec-HObath}

\begin{center}
\begin{figure}
\centering
\psfrag{a=1}{$\alpha=1$}
\psfrag{a=2}{$\alpha=2$}
\psfrag{a=3}{$\alpha=3$}
\psfrag{ln(e)}{$\ln\, \eta$}
\psfrag{ln(tdec)}{$\ln\, \tau_{\rm dec}$}
\includegraphics[width=10cm]{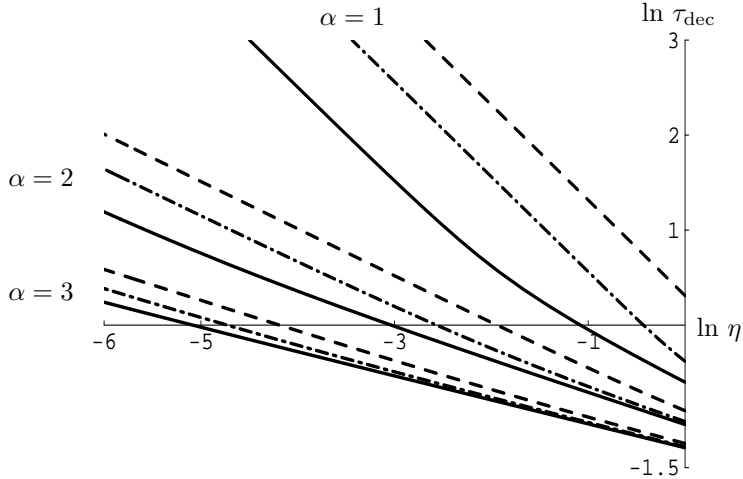}
\caption{\label{fig-5}
Decoherence time $\tau_{\rm dec}$ in units of $\TBmax=\hbar \beta$
as a function of the pointer-bath coupling strength $\eta$ 
for the
  same bath as in Fig.~\ref{fig-3} with
$\tau_{\rm ent}=0.1$, $m=3$, and $c=1$. 
Three distinct values of $w_D$ are shown: 
$w_D=2$ (solid curves),  $w_D=5$ (dotted-dashed curves), and 
$w_D=10$ (dashed curves). For each of these values, $\tau_{\rm dec}$ is shown
for $\alpha =1,2$, and $3$ (from top to bottom).
}
\end{figure}
\end{center}

To study the transition between the limiting time regimes discussed in
the two preceding subsections, let us consider a bath of $N \gg 1$
harmonic oscillators, $\HB = \sum_{\nu} \hbar \omega_\nu
(b_\nu^\dagger b_\nu+1/2)$, coupled to the pointer via a coupling
agent $B$ linear in each of its creation and annihilation operators
$b_\nu^\dagger$ and $b_\nu$, $B=\sum_{\nu} ( \kappa_\nu b_\nu^\dagger
+\kappa_\nu^\ast b_\nu)/\sqrt{N}$~\cite{Caldeira-Leggett83}.  Here
$\omega_\nu$ is the frequency and $\kappa_\nu$ the coupling constant
of the $\nu$th oscillator.  We shall take the following specific
choice for the power spectrum function:
\begin{equation} \label{eq-J(omega)}
J(\omega) 
= 
 \frac{\pi}{N} \sum_{\nu=1}^N |\kappa_\nu|^2 \delta ( \omega - \omega_\nu )
  = \widehat{\gamma}\, \omega^m 
   \E^{-\omega^2/\omega_D^2 } 
\end{equation}
wherein $m$ is an odd positive integer, $\widehat{\gamma}>0$, and
$\omega_D$ is a cutoff frequency.  We recall that the case $m=1$
corresponds to an Ohmic damping, whereas one speaks of
super-Ohmic damping for $m>1$. For instance, $m=d$ or $d+2$ for a
phonon bath in $d$ dimensions, depending on the underlying
symmetries~\cite{Weiss}.  As is
well known~\cite{Caldeira-Leggett83,Weiss}, the imaginary part of the
bath correlation function $h(t)$ is
temperature-independent, its Fourier transform being given by
$\I (\widehat{\Im h}) (\omega)= J(\omega)$ for $\omega\geq 0$.  By the
KMS property~(\ref{eq-KMS_FT}) this implies $ (\widehat{\Re h})
(\omega) = \coth ( \hbar \beta \omega / 2 ) J( | \omega|) $.
If $w_D = \hbar \omega_D \beta>1$, the thermal time $\TBmax=\hbar
\beta$ is the largest decay time of $\Re h(t)$. The other time scale
characterizing the variations of $\Re h(t)$ is the inverse cut-off
frequency $\TBmin = \omega_D^{-1} < \TBmax$.  By
(\ref{t_dec_vs_t_ent}), the decoherence and entanglement times in
units of $\TBmax$, $\tau_{\rm dec} = \dectime /\TBmax$ and
$\tau_{\rm ent} = \enttime/\TBmax$, are given by
\begin{equation} \label{eq-dectime_bath_HO}
\frac{\tau_{\rm ent}^{2 \alpha}}{c_\alpha^2 \,\eta^2}
 = 
  \frac{ \int_0^\infty d w\, \coth (w/2)\, w^m\,
     \E^{-w^2/w_D^2} \bigl| \int_0^{\tau_{\rm dec}} \D \tau\,
    \tau^\alpha\,\E^{-\I w \tau} \bigr|^2}
     {2 \int_0^\infty d w\, \coth (w/2) \, w^m\,
    \E^{-w^2/w_D^2}}
\end{equation}
where we have expressed $\Re h (t)$ in terms of its Fourier transform
and relied on (\ref{eq-J(omega)}).  We
did not write explicitly in (\ref{eq-dectime_bath_HO}) the dependence
of $\tau_{\rm ent}$, $\tau_{\rm dec}$, and $c_\alpha$ on $(s,s')$ .  The right-hand
side of (\ref{eq-dectime_bath_HO}) is shown in the inset 
in Fig.\ \ref{fig-3}.  We
have computed numerically the integrals appearing in this right-hand
side for various values of $\alpha$, $m$, and $w_D$, so as to obtain
$\tau_{\rm dec}$ as a function of $\tau_{\rm ent}$ and $\eta$. The main
results are shown in Figs.~\ref{fig-3} and~\ref{fig-5}.  
For fixed $\alpha$ and $\eta$, the plain curves representing
$\tau_{\rm dec}$ in Fig.\ \ref{fig-3} split by increasing $\tau_{\rm
  ent}$ into distinct branches corresponding to distinct $m$'s, as
predicted by (\ref{D_t-Markov-Ohmic}) and
(\ref{D_t-Markov-sup_ohmic}).  This splitting occurs when $\tau_{\rm
  dec}$ is in the transition region $w_D^{-1} \lesssim \tau_{\rm dec}
\lesssim 1$.  After this splitting $\tau_{\rm dec}$ is larger for
larger $m$. In particular, a Ohmic bath ($m=1$) has a smaller
decoherence time than a super-Ohmic bath ($m=3,5\ldots$) as stated
above.  For comparison, the power law behaviors found in
Sec.~\ref{sec-short-time} and~\ref{sec-Markov} in the small time
($\tau_{\rm dec} \ll w_D^{-1}$) and Markov ($\tau_{\rm dec} \gg 1$)
regimes are also shown in Fig.\ \ref{fig-3} (broken lines).  A
remarkably good agreement between the exact and asymptotic behaviors
of $\tau_{\rm dec}$ is obtained: the exact results are well
approximated by their small-time behaviors (\ref{t_dec_short_time}) up
to $\tau_{\rm dec} \leq w_D^{-1}$ and they are hardly distinguishable
from the Markov approximation as soon as $\tau_{\rm dec} \geq 1$.  
Our aforementioned statement that a nonlinear
pointer-bath coupling is more efficient for decoherence than a linear
one when $\tau_{\rm ent}$ is not too large (and even for arbitrarily
large $\tau_{\rm ent}$ if $m \geq 3$ and $\eta$ is small enough) is well
confirmed. Indeed, it is seen in Fig.\ \ref{fig-5} that for a
pointer-bath coupling strength $\eta \ll 1$,
$\tau_{\rm dec}$ becomes significantly smaller when the value of $\alpha$ is
increased from $\alpha=1$ to $\alpha=3$.  If the dotted lines 
 (Markovian results) in Fig.\ \ref{fig-3} were drawn
farther to the right, the two lines corresponding to
$(\alpha,m)=(1,1)$ and $(\alpha,m)=(2,1)$ would intersect; after this
intersection (not shown in the figure) the reverse situation of higher
values of $\alpha$ leading to higher values of $\tau_{\rm dec}$ occurs.
In contrast, for $m=3,5,\ldots$, the dotted lines associated with
$\alpha=1$ and $\alpha=2$ never intersect (their are parallel); hence $\tau_{\rm
  dec}$ decreases with $\alpha$ and $\tau_{\rm ent}\leq \tau_{\rm
  dec}$ for all values of $\tau_{\rm ent}$ (more precisely, 
$\tau_{\rm ent}\lesssim 10^{-4}\tau_{\rm
  dec}$ for $\alpha=1$ and 
$\tau_{\rm ent}\lesssim 10^{-2} \tau_{\rm  dec}$ for $\alpha=2$).  We also emphasize that
$\tau_{\rm dec}$ increases in~Fig.\ \ref{fig-5} with the cut-off
frequency $\omega_D$.  Even though the results in Figs.\ \ref{fig-3}
and~\ref{fig-5} correspond to the simplifying choice of a bath of
harmonic oscillators with power spectrum function (\ref{eq-J(omega)}),
for more general baths they should still give the correct
qualitative picture.

\vspace*{-0.5cm}

\subsection{Bath at very low temperature}
\label{sec-zero_T_bath}

\begin{center}
\begin{figure}
\centering
\includegraphics[width=10cm]{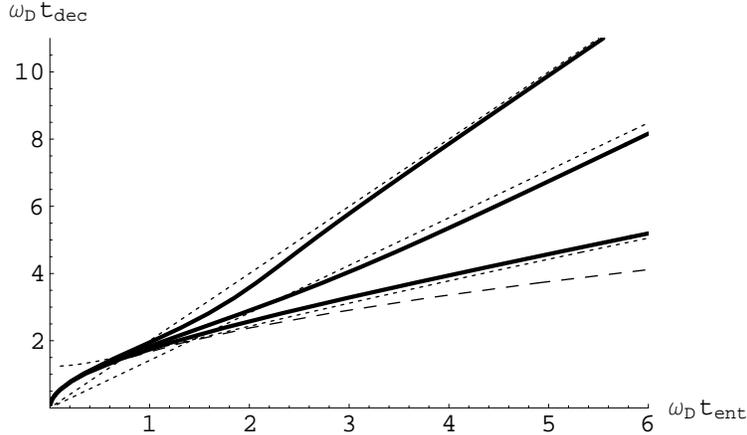}
\caption{\label{fig-6}
Decoherence against entanglement times in units of $\omega_D^{-1}$ 
for the 
bath at zero temperature of Sec.~\ref{sec-zero_T_bath}
with $\alpha=1$, $\eta_D=1$, and 
 $m=5,3,1$ (solid curves, from top to bottom). 
The approximations (\ref{t_dec_short_time}) and 
(\ref{eq-dectime_T=0_Markov}) for $\omega_D \dectime \ll 1$ and 
$\omega_D \dectime \gg 1$ are  shown as dashed lines.
}
\end{figure}
\end{center}

We have so far considered baths at finite temperature.
Motivated by experiments in solid state physics, we shall now 
discuss the case of a bath
initially in thermal equilibrium at very low temperature. Strictly
speaking, for the equilibrium apparatus initial state (\ref{eqapp})
extremely low temperatures have to be proscribed because of our
hypothesis $\hbar \beta \ll \TP$. However, taking e.g. $\TP=1$ s, this
separation of time scales holds even for the smallest temperatures
that can be achieved in experiments.  Furthermore, the stability
conditions (\ref{eq-stability_cond}) have a better chance to be met at
low temperature $T$ since $\Delta_{\rm th}$ decreases with $T$.  
To be specific, we consider the same bath of harmonic
oscillators as in the previous subsection, but now in the limit $w_D =
\hbar \omega_D \beta \gg 1$. Then only spontaneous emission plays a
role in the pointer-bath interaction. In other words, $(\widehat{\Re
  h} )(\omega) = \coth ( \hbar \beta \omega/2 ) J(|\omega|)$ can be
approximated by $J(|\omega| )$.  The zero-temperature variance of $B$
equals $\langle B^2 \rangle =\int \D \omega \,J(|\omega |)/(2 \pi)$.  For
our choice (\ref{eq-J(omega)}) of the power spectrum function, this
gives $2 \pi \langle B^2 \rangle =\widehat{\gamma} \, \omega_D^{m+1}
((m-1)/2 )!$.  The analog of (\ref{eq-dectime_bath_HO}) reads
\begin{equation} \label{eq-dectime_bath_HO_T=0}
\frac{(\omega_D \enttime )^{2 \alpha}}{c_\alpha^2 \,\eta^2_D}
 = 
  \frac{1}{ ((m-1)/2)!} 
   \int_0^\infty d v\, v^m\,\E^{-v^2} 
    \left| \int_0^{\omega_D \dectime} \D u\,
     u^\alpha\,\E^{-\I v u} \right|^2
 \end{equation}
 where $\eta_D$ is now the pointer-bath coupling strength in units of
 $\hbar \omega_D$, $\eta_D = \langle B \rangle^{1/2} \Delta^\alpha
 /(\hbar \omega_D)$.  Eq.~(\ref{eq-dectime_bath_HO_T=0}) holds
 provided that $\omega_D \dectime \ll w_D$, i.e., $\dectime \ll \hbar
 \beta$. This equation is the same as (\ref{eq-dectime_bath_HO}) apart
 from the substitutions  $\tau_{\rm ent} \to \omega_D\,\enttime$,
$\tau_{\rm dec} \to \omega_D\,\dectime$, 
$\eta \to \eta_D$ and $\coth(w/2)\,\E^{-(w/w_D)^2 }\to \E^{-v^2}$.
Explicit formulae for $\dectime$ can be given as before when
$\dectime$ is small or large compared with $\omega_D^{-1}$.  One reads the
small-time result directly on (\ref{t_dec_short_time}) by transforming
this expression according to the recipe mentioned above. This gives
$\omega_D \dectime \propto (c_\alpha^{-1/\alpha} \eta_D^{-1/\alpha} \omega_D \enttime )^{\alpha/(\alpha+1)}$ for
$\dectime \ll \omega_D^{-1}$.  Similarly, for a super-Ohmic bath $\dectime
\propto c_\alpha^{-1/\alpha} \eta_D^{-1/\alpha} \enttime $ 
when $\omega_D^{-1} \ll \dectime \ll \hbar \beta$.  To
find the proportionality factor, it is enough to realize that
$|\int_0^\infty \D \tau \, \tau \Re h (\tau)|=\int_0^\infty \D \omega\,(\widehat{\Re h}
)(\omega)\, \omega^{-2}/\pi$ has to be interpreted in
(\ref{t_dec-Markov_sup_ohmic}) as $\widehat{\gamma}\, \omega_D^{m-1} \int_0^\infty \D v \,v^{m-2}
\E^{-v^2}/\pi= 2\, \omega_D^{-2} \langle B^2\rangle/(m-1)$.  The Ohmic case requires
some extra work.  For indeed, replacing $(\widehat{\Re h} )(\omega)$ by
$J(|\omega|)$ leads to a vanishing second member in
(\ref{eq-approx_int_Markov}), even though the frequency integrals in
the second and third members in (\ref{eq-derivative_D_t^peak_bis}) are still divergent.  One actually finds
\begin{equation} \label{eq-approx-Ohmic_T=0}
\int_0^\infty \D v \, \E^{-v^2}  \int_0^1 \D u
\, ( 1-u)^{\alpha-1} \sin (\omega_D t\, u v) 
\sim \frac{\ln (\omega_D t)+k_\alpha}{\omega_D  t} 
\quad {\rm{ as}}  \quad \omega_D t \to \infty
\end{equation}
with $k_1\simeq 0.2886$ and $k_\alpha=k_1-1-1/2-\ldots-1/(\alpha-1)$
if $\alpha\geq 2$.  Substituting the frequency integral in
(\ref{eq-derivative_D_t^peak}) by the right-hand side of
(\ref{eq-approx-Ohmic_T=0}), one gets
\begin{eqnarray} \label{eq-dectime_T=0_Markov} 
\nonumber
\enttime(s,s')
& = &  
\Bigl( c_\alpha(s,s') \, \eta_D \Bigr)^{1/\alpha} \dectime(s,s') 
\Bigl( \ln (\omega_D \dectime(s,s'))+k_\alpha-(2\alpha)^{-1} 
\Bigr)^{\frac{1}{2\alpha}}
\quad   {\text{(Ohmic)}} 
\\
\enttime(s,s')
& = & 
\Bigl( c_\alpha(s,s')\, \eta_D/\sqrt{m-1}\Bigr)^{1/\alpha}
 \dectime (s,s')
\qquad \qquad  \qquad \qquad  {\text{(super-Ohmic).}} 
\end{eqnarray}
Instead of going through a proof of (\ref{eq-approx-Ohmic_T=0}), which
would lead us too far into technical details, let us compare the
formulae (\ref{eq-dectime_T=0_Markov}) to the exact results obtained
by numerical evaluations of the integrals in
(\ref{eq-dectime_bath_HO_T=0}).  It is seen in Fig.\ \ref{fig-6} that
the approximate values (\ref{eq-dectime_T=0_Markov}) closely follow
the exact curves when $\omega_D \dectime$ becomes large (in fact, even
for $\omega_D \dectime \simeq 2$ in the Ohmic case $m=1$).  Similar
pictures are found for higher $\alpha$'s.  Let us remark on
(\ref{eq-dectime_T=0_Markov}) that for a given $\enttime$, the ratio
between the decoherence times for Ohmic and super-Ohmic baths is
logarithmically small in the dimensionless time $\omega_D \dectime$.
Hence a Ohmic bath is not dramatically more efficient than a
super-Ohmic bath at very low temperature, in contrast with our
previous findings at ``high'' temperatures.

\section{Derivation of the result}
\label{sec_derivation}

We here fill in the derivation of the results presented in Sec.
\ref{simul_results}. An alternative derivation of (\ref{eq-result-S})
based on the time-dependent Redfield equation can be found
in~\cite{ProceedingQOIII}.  Our approach below is non-perturbative in
the pointer-bath coupling but makes use of the QCLT (Sec.~\ref{sec-time scales}) 
which holds due to the
additivity (\ref{eq-int_Hamiltonian}) of the bath coupling agent.

\subsection{Object-pointer dynamics}
\label{sec_dyn_small_t}

According to the results of Sec.~\ref{sec-sep_time_scales} we may
drop the Hamiltonians $\HS$ and $\HP$ in the full Hamiltonian $H$,
with the proviso that the object initial state $\rhoS$ is replaced by
$\rhoS^{0}(t)$ given by (\ref{eq-rhoS(0)}).  This means that at times $t \ll \TS, \TP$ the
exact evolution operator of $\Sc + \Pc +\Bc$ in (\ref{eq-rhoPS}) can
be approximated as $\E^{-{\I} t \HPSB/{\hbar}} \simeq \E^{-\I t
  \HB/\hbar} \,\widetilde{W} (t,0)\, \E^{- {\I} t ( \HS + \HP
  )/{\hbar}}$, where
\begin{equation} \label{eq-evol_op_approx}
\widetilde{W} (t,0) 
 = \E^{\I t\HB/\hbar}\, \E^{-\I t ( \HB + \HPS +
  \HPB)/\hbar}
=
\timeord \exp 
 \left\{ -\frac{\I}{\hbar}
     \int_0^t \D \tau 
      \bigr( \epsilon   S P +  X^\alpha  \widetilde{B}(\tau)
      \bigr) 
 \right\} 
\end{equation}
is the approximate evolution operator in the interaction picture.
Note that $\widetilde{B}(\tau)$ is
different from $B$ as soon as $\tau \gtrsim \TBmin$. Since we do not
assume here $t$ to be small compared with $\TBmin$ we keep the time
dependence of the bath coupling agent in (\ref{eq-evol_op_approx}). 
In view of the product structure $\rhoPSB (0)=\rhoS \otimes \rho_{\cal
  P B} $ of the initial state and by cyclic invariance of the trace,
the object-pointer state (\ref{eq-rhoPS}) becomes
\begin{equation} \label{eq-approx_rhoPS}
\rhoPS (t) 
 \simeq 
 \trB \Bigl(
   \widetilde{W} (t,0) \rhoS^{0}(t) \otimes \rho_{\cal P B} \,  
    \widetilde{W} (t,0)^\dagger
  \Bigr) 
\quad , \quad t \ll \TS, \TP  \;.  
\end{equation}
The pointer
Hamiltonian $\HP$ is absent in (\ref{eq-approx_rhoPS}) since  $\E^{-\I t \HP/\hbar} \,
\rho_{\cal P B} \,\E^{\I t \HP/\hbar} \simeq \rho_{\cal P B}$ at times $t
\ll \TP$ for the
initial states under study, see
Sec.~\ref{sec-sep_time_scales}.  

\newparagraph

The approximate evolution operator (\ref{eq-evol_op_approx}) can be
simplified by using the exact identity
\begin{equation} \label{eq-Generalized-Hausdorff-Campbell}
\widetilde{W} (t,0)
 = \E^{- {\I} t \epsilon  S P/{\hbar}}\,
   \timeord \exp
  \left\{ -\frac{\I}{\hbar}
     \int_0^t \D \tau  
      (X + \tau \epsilon   S)^\alpha   \widetilde{B} (\tau)
\right\}\;.
\end{equation}
We forego the proof of this (generalized Baker-Campbell-Haussdorff)
identity, which uses the role of
the momentum as generator of displacements,  $\E^{-\I t \epsilon S P/\hbar} (X+ t \epsilon S
)^\alpha\, \E^{\I t \epsilon S P/\hbar} = X^\alpha$.  Employing
(\ref{eq-Generalized-Hausdorff-Campbell}) in (\ref{eq-approx_rhoPS})
and setting $x_s(t) = x - t \epsilon s$ and $x_{s'} '(t) = x'
- t \epsilon s'$ as before we get
\begin{eqnarray} \label{eq-tilderhoPS}
\nonumber
\bra{s, x} \rhoPS (t) \ket{s',x'} 
& = &  
\bra{s}\rhoS^{0}(t)  \ket{s'}
 \bra{x_s (t)} 
\trB 
 \biggl(  
    \timeord \exp
  \left\{ -\frac{\I}{\hbar}
     \int_0^t \D \tau  \,
      x_s(t-\tau)^\alpha \widetilde{B} (\tau)
  \right\}
\rho_{\cal{PB}}   
 \\
& & 
  \left[ \timeord \exp
  \left\{ -\frac{\I}{\hbar}
     \int_0^t \D \tau  \,
       x_{s'}' (t-\tau)^\alpha \widetilde{B} (\tau)
  \right\}
  \right]^\dagger  
 \biggr)  
\ket{x_{s'}' (t)}\;.
\end{eqnarray}
The next step consists in evaluating the trace over the bath in
(\ref{eq-tilderhoPS}) by taking advantage of Wick's theorem
(\ref{eq-Wick}).  We discuss the two initial states (\ref{parteqapp})
and (\ref{eqapp}) separately.

\subsection{Partial-equilibrium initial state}
\label{sec-derivation-parteqapp}

For the partial-equilibrium initial state (\ref{parteqapp}) one has
$\rho_{\cal{PB}} = \rhoP \otimes \rhoB$ and the last matrix element in
(\ref{eq-tilderhoPS}) is the product of a pointer and a bath
expectation values,
\begin{equation} \label{eq-result-Sbis}
\bra{s, x} \rhoPS (t) \ket{s',x'} 
 =  
\bra{s}\rhoS^{0}(t)  \ket{s'}\,
   \bra{x_{s} (t)} \rhoP  \ket{x_{s'}' (t)}\,
K_t \bigl( x_s(t), x_{s'}'(t);s,s' \bigr)
\end{equation}
with
\begin{eqnarray} \label{eq-K}
\nonumber
K_t \bigl( x, x';s,s' \bigr)
& = &
\left\langle 
\left[ \timeord \exp
  \left\{ -\frac{\I}{\hbar}
     \int_0^t \D \tau  \,
       x_{s'}' (-\tau )^\alpha \widetilde{B} (\tau)
  \right\}
  \right]^\dagger
\right.
\\
& &
\left.
\timeord \exp
  \left\{ -\frac{\I}{\hbar}
     \int_0^t \D \tau  \,
      x_{s}(- \tau )^\alpha \widetilde{B} (\tau)
  \right\}
 \right\rangle \;. 
\end{eqnarray}
Here $\meanB{\cdot } = \ZB^{-1} \trB \bigl( \cdot \, e^{-\beta \HB}
\bigr)$ denotes the average with respect to the free bath thermal
state.  The QCLT and the additivity (\ref{eq-int_Hamiltonian}) of the
bath coupling agent imply
\begin{eqnarray} \label{eq-characteristic-functional}
& & 
\nonumber
F_{t,0} [ k,l] 
=
\left\langle 
 \left[ \timeord \exp 
  \biggl\{ 
    -\frac{\I}{\hbar} \int_0^t \D \tau \,k (\tau)\,\widetilde{B}(\tau) 
  \biggr\}
 \right]^\dagger
\timeord \exp
  \biggl\{ 
    -\frac{\I}{\hbar} \int_0^t \D \tau \,l (\tau)\,\widetilde{B}(\tau) 
  \biggr\}
\right\rangle
\\
&  & = 
\exp 
\biggl\{
- \frac{1}{\hbar^2}
 \int_0^t \D \tau_1 \int_0^{\tau_1} \D \tau_2 
  \bigl( {k (\tau_1)} - l (\tau_1) \bigr) 
 \bigl( 
{k (\tau_2)} h ( \tau_2, \tau_1)  -l(\tau_2) h (\tau_1,\tau_2)
    \bigr)
\biggr\}\,,
\end{eqnarray}
where $k (\tau)$ and $l(\tau)$ are two arbitrary real-valued functions
and $h(\tau_1,\tau_2)=h(\tau_1-\tau_2)$ is the two-point bath
correlator, see (\ref{eq-bath_correl_function}).  The identity
(\ref{eq-characteristic-functional}) is equivalent to Wick's theorem
(\ref{eq-Wick}). For ordered times $t> t_1> t_2>\cdots> t_n$ one
actually gets (\ref{eq-Wick}) from
(\ref{eq-characteristic-functional}) by setting $k=0$ in
(\ref{eq-characteristic-functional}) and taking the functional
derivative of both members with respect to $l(t_1),\ldots,
l(t_n)$ at $l=0$. The proof of converse statement is deferred to
Appendix~\ref{app-characteristic-functional}.  By using the parity
properties $\Re h (\tau)=\Re h (-\tau)$ and $\Im h (\tau)=-\Im h (-\tau)$
of the real and imaginary parts of $h$ and employing
(\ref{eq-characteristic-functional}) in (\ref{eq-K}) we get
\begin{equation} \label{eq-D(i)bis}
K_t \bigl( x, x';s,s' \bigr)
 =  
  \E^{- D_t (x,x';s,s') 
  - \I \phi_t(x,x';s,s')} 
\end{equation}
with a decoherence exponent $D_t$ and a phase $\phi_t$ given by 
\begin{eqnarray}
\nonumber
(D_t + \I \phi_t) (x,x';s,s') 
& = &
\frac{1}{\hbar^2} \int_0^t  \D \tau_1  \int_0^{\tau_1}  \D \tau_2  
    \bigl( x_{s'}' (- \tau_1 )^\alpha -x_{s} (- \tau_1)^\alpha \bigr)
   \biggl\{ 
     \bigl( x_{s'}' (- \tau_2 )^\alpha - x_{s}( - \tau_2 )^\alpha \bigr)
    \\
\nonumber
& & \times   \Re h (\tau_1-\tau_2)
 - \I   \bigl( x_{s'}' (- \tau_2 )^\alpha + x_{s}( - \tau_2 )^\alpha \bigr)
     \Im h (\tau_1 -\tau_2) \biggr\} \,.
\end{eqnarray}
Thus (\ref{eq-result-Sbis}) reduces to the result
(\ref{eq-result-S}) announced in Sec.~\ref{sec-parteqapp}.

\subsection{Equilibrium apparatus initial state}
\label{sec-derivation-eqapp}

Before deriving the expression corresponding to (\ref{eq-result-Sbis})
in the case $\rho_{\cal P B } = \rhoPB$, we determine the pertinent
initial density matrix of the pointer,
\begin{equation} \label{eq-new_label}
R_0 (x,x') 
 =
  \bra{x}  \trB ( \rhoPB ) \ket{x'} \;.
\end{equation}
It is convenient to introduce the $x$-dependent bath average
\begin{equation} \label{eq-bath_averagex}
\meanBx{ {\OB} }{x} =  
 \ZBx{x}^{-1} \trB  \bigl( {\OB} 
   \, \E^{-\beta (\HB +  x^\alpha B)} \bigr)
\quad , \quad 
\ZBx{x} = \trB  \bigl(  \E^{-\beta (\HB +  x^\alpha B)} \bigr)
\end{equation}
with $x$ a real number (the pointer position, for us). Note that
\begin{equation} \label{eq-Dyson_Euclidean}
 \E^{-\beta (\HB +  x^\alpha B)}
  = \E^{-\beta \HB} 
   \,\timeord 
    \exp
     \left\{
    - \frac{x^\alpha}{\hbar} \int_0^{\hbar \beta} \D z 
      \,\widetilde{B} (-\I z)
     \right\}
\end{equation}
with $\widetilde{B} ( -\I z) = \E^{z \HB/\hbar} B \E^{-z \HB / \hbar}$.
The normalization factor $Z_x$ can be determined by applying Wick's
identity (\ref{eq-characteristic-functional}) with $t=-\I \hbar \beta$,
$k(\tau)=0$, and $l(\tau) = x^\alpha$. This gives
\begin{equation} \label{eq-Zx}
\ZBx{x} 
 = 
\ZBx{0} \exp \left\{ \frac{x^{2\alpha} \beta \gamma_0}{\hbar} \right\}
\end{equation}
with 
\begin{equation}
\gamma_0 
 = 
  \frac{1}{\hbar\beta} \int_0^{\hbar \beta} \D z_1 \int_0^{z_1} \D z_2
  \, 
   h(-\I z_2 )\;.
\end{equation}
By using the analyticity and KMS
properties of the bath correlator $h(\tau)$, one can show that $0 \leq \gamma_0
\leq \hbar \beta \meanB{B^2}/2$ and that $\gamma_0=\gamma(0)$ coincides with the
following integral evaluated at $t=0$
\begin{equation}
\label{eq-gammat}
\gamma(t) = \int_{-\infty}^t \D \tau \,\Im h(\tau)\;.
\end{equation}
Details of this derivation are deferred to
Appendix~\ref{app-correl_functions}.

\newparagraph

Replacing (\ref{eq-Zx}) into the approximation
(\ref{eq-approx_total_equilibrium}) for the apparatus initial state and
inserting the high-temperature expression of $\bra{x} \E^{-\beta
  \HP/2} \ket{y}$ (see Sec.~\ref{sec-model}) yields
\begin{eqnarray} \label{eq-rhoP0}
\nonumber
R_0(x,x')
&  = & 
\ZPB^{-1} \int \D y \, Z_{y}\, \bra{x} \E^{- \beta \HP/2}  \ket{y} 
   \bra{y} \E^{- \beta \HP/2} \ket{x'}
\\
&   = &  
\ZBx{0} {\ZPB}^{-1} \int \D y \,\E^{\beta \gamma_0 \, y^{2\alpha}/\hbar}
\,\E^{-\beta ( V(x)+V(x')+2 V(y))/4}\,
 \E^{-4 \pi^2 ((x-y)^2+(x'-y)^2)/\lambda_{\rm th}^2} 
\end{eqnarray}
with $\lambda_{\rm th} = 2\pi \hbar ( \beta/M)^{1/2}$.  The stability
condition (\ref{eq-stability_cond}) does not guarantee that $V(x)$
compensates $-\gamma_0\, x^{2\alpha}/\hbar$ when $x \to \pm \infty$.
If this is not the case, i.e., if the effective potential $V_{\rm
  eff}(x)=V(x)-\gamma_0\, x^{2\alpha}/\hbar$ has the shape shown in
Fig.\ \ref{fig-0}, the integrals in (\ref{eq-rhoP0}) diverge. This
reflects the fact that the pointer interacting with the bath will
tunnel to infinity after a certain time.  Since we restrict our
attention to initial states describing a pointer initially localized
inside the potential well of $V_{\rm eff}(x)$, we shall disregard this
convergence problem by adding to $V(x)$ a positive potential vanishing
for $| x| \lesssim W_{\rm eff}$ and diverging exponentially for $x
\rightarrow \pm \infty$.  This regularization trick amounts to replace
$\rhoPB$ in (\ref{eq-new_label}) by the local thermal state of the
apparatus discussed in Sec.~\ref{sec-tot_eq_init_state}.  After
this regularization, the main contribution in the $y$-integral in
(\ref{eq-rhoP0}) comes from small values of $y$, $|y| \lesssim
\Delta_{\rm th}$.  In fact, the first exponential in this integral is
a slowly varying function on the scale $\lambda_{\rm th}$ since
$(\beta \gamma_0/\hbar)^{-1/(2\alpha)} \geq 2^{1/\alpha}\,\Delta_{\rm
  th} \gg \lambda_{\rm th}$, as follows from (\ref{eq-stability_cond})
and $\gamma_0 \leq \hbar \beta \langle B^2\rangle/2$.  Due to the
presence of the last exponential in (\ref{eq-rhoP0}), this first
exponential can be approximated by $\E^{\beta \gamma_0
  (x+x')^{2\alpha}/(2^{2\alpha} \hbar)}$ and taken out of the
integral.  Similarly, the second exponential varies noticeably on the
scale $\Delta_{\rm th} \gg \lambda_{\rm th}$ and can be approximated
by $\E^{-\beta (V(x)+V(x')+2 V(x/2+x'/2))/4}$ and taken out of
integral in (\ref{eq-rhoP0}).  Thus (\ref{eq-stability_cond}) and
$\Delta_{\rm th} \gg \lambda_{\rm th}$ entail
\begin{equation} \label{eq-noname}
R_0(x,x') 
 \simeq  Z_{{\cal P},{\rm eff}}^{-1}\,
  \E^{-\beta ( V_{\rm eff}( x) + V_{\rm eff}( x'))/2} \,
   \E^{-2 \pi^2 (x-x')^2/\lambda_{\rm th}^2} 
\end{equation}
with $V_{\rm eff} (x)$ given by (\ref{eq-effective_pot}) and $Z_{{\cal
    P},{\rm eff}} = \int \D x \, \E^{-\beta V_{\rm eff}( x)}$.  We
have used in (\ref{eq-noname}) the approximations $\beta V(x/2+x'/2)
\simeq \beta (V(x)+V(x'))/2$ and $\beta \gamma_0
(x+x')^{2\alpha}/(2^{2\alpha} \hbar) \simeq \beta \gamma_0 (
x^{2\alpha} +{x'}^{2\alpha})/(2 \hbar)$. This introduces an error
which is negligible against $(x-x')^2/\lambda_{\rm th}^2$ for
$|x|,|x'| \lesssim \Delta_{\rm th}$ and $\Delta_{\rm th} \gg
\lambda_{\rm th}$.  Hence the pointer is in a Gibbs-type state
with an effective potential $V_{\rm eff} (x)$, as
announced in (\ref{eq-R_tsimeqR_0}).

\newparagraph

We can now proceed to evaluating (\ref{eq-tilderhoPS}). 
Repeating the steps yielding to (\ref{eq-rhoP0}) and using 
the notation (\ref{eq-bath_averagex}), 
\begin{eqnarray} \label{eq-rhoPS_once_again}
\nonumber
& 
\displaystyle
\bra{s, x} \rhoPS (t) \ket{s',x'} 
 =   
\ZPB^{-1} \,
\bra{s}\rhoS^{0}(t)  \ket{s'}
\int \D y \, Z_{y}\, \bra{x_s (t)} \E^{- \beta \HP/2}  \ket{y} 
   \bra{y} \E^{- \beta \HP/2} \ket{x_{s'}' (t)} \times
\\
& \displaystyle 
\left\langle
 \left[ \timeord \exp
  \left\{ -\frac{\I}{\hbar}
     \int_0^t \D \tau  \,
       x_{s'}' (t-\tau)^\alpha \widetilde{B} (\tau)
  \right\}
  \right]^\dagger  
  \timeord \exp
  \left\{ -\frac{\I}{\hbar}
     \int_0^t \D \tau  \,
      x_s(t-\tau)^\alpha \widetilde{B} (\tau)
  \right\}
\right\rangle_{y}\;.
\end{eqnarray}
We set $\delta \widetilde{B}(\tau,y)=\widetilde{B}(\tau)
-\langle\widetilde{B}(\tau)\rangle_{y}$ and consider the 
(quantum) characteristic functional 
\begin{equation} \label{eq-characteristic-functionalbis}
F_{t,y} [ k,l] 
= \left\langle 
 \left[ \timeord \exp 
  \biggl\{ 
    - \frac{\I}{\hbar} \int_0^t \D \tau \,k
    (\tau)\,
    \delta \widetilde{B}(\tau,y) 
  \biggr\}
 \right]^\dagger
\timeord \exp
  \biggl\{ 
    - \frac{\I}{\hbar} \int_0^t \D \tau \,l (\tau)\,
    \delta  \widetilde{B}(\tau,y)  
  \biggr\}
\right\rangle_{y} \;.
\end{equation}
It is shown in Appendix~\ref{app-characteristic-functional} that 
all the correlation functions 
$\langle
\delta \widetilde{B}(\tau_1,y)
  \cdots 
\delta \widetilde{B}(\tau_n,y) 
\rangle_{\Bc,y}$
are independent of $y$, i.e., 
\begin{equation} \label{eq-characteristic-functional_y}
F_{t,y} [k,l] = F_{t,0} [ k,l]
\end{equation}
for any $y$, $t$, $k(\tau)$, and $l(\tau)$.  Wick's theorem
(\ref{eq-Wick}) also entails (see
Appendix~\ref{app-characteristic-functional})
\begin{equation} \label{eq-meanxB}
\meanBx{ \widetilde{B}(\tau) }{x}
   = -\frac{2 x^\alpha}{\hbar} \gamma(\tau)
\end{equation}
with $\gamma(\tau)$ given by (\ref{eq-gammat}). 
Formula (\ref{eq-meanxB}) is reminiscent of linear response theory since 
$\gamma (t) =\gamma_0 - \hbar\int \D \tau\,\chi(\tau) \,
\theta (t-\tau)/2$, 
where $\theta(\tau)$ is the Heaviside function 
 and $\chi(\tau)=-(2/\hbar)\, \theta (\tau)\, \Im h (\tau)$ the 
linear susceptibility. Let us point out, however, that (\ref{eq-meanxB})
is exact to all orders in $x^\alpha$. 
Collecting the above results one finds
\begin{equation} \label{eq-result-Sbis_totaleq}
\bra{s, x} \rhoPS (t) \ket{s',x'} 
 =  
\bra{s}  \rhoS^{0}(t)    \ket{s'}\,
R_t  \bigl( x_s(t), x_{s'}'(t);s,s' \bigr)\,
K_t \bigl( x_s(t), x_{s'}'(t);s,s' \bigr)
\end{equation}
where $K_t ( x, x';s,s' )$ is given by (\ref{eq-K}), 
\begin{eqnarray} \label{eq-R}
\nonumber
R_t (x,x';s,s')
& = & 
C \,
\E^{
   -\beta 
    ( V_{\rm eff}( x) + V_{\rm eff}(x' ) )/2} \, \E^{- 4 \pi^2 ( x^2 +
    {x'}^2)/\lambda_{\rm th}^2} \times \\
 & & 
\int \D \xi \, 
 \exp 
  \left\{
  - \xi^2 + 
\sqrt{8 \pi^2}\, \xi \, \frac{x+x'}{\lambda_{\rm th}}
    - 2 \I \, \xi^\alpha \, g_t(x,x';s,s') 
  \right\}\;,
\end{eqnarray}
$C$ is a time-independent normalization constant, and 
\begin{equation} \label{eq-g_t}
 g_t(x,x';s,s') 
 = 
  (8\pi^2)^{-\frac{\alpha}{2}} \, \frac{\lambda_{\rm th}^\alpha }{\hbar^2} 
    \int_0^t \D \tau \, \gamma (\tau) 
     \left(  x_{s'} ' (- \tau)^\alpha -  x_{s}(- \tau)^\alpha \right) \;.
\end{equation}
Therefore,  (\ref{eq-D(i)bis}) and
(\ref{eq-result-Sbis_totaleq}) account for (\ref{eq-result-S_totaleq}).
Moreover, by using
  (\ref{eqapp}), (\ref{eq-rhoS(0)}), (\ref{eq-new_label}), 
(\ref{eq-result-Sbis_totaleq}), and  $K_0=1$ 
one  easily establishes  that $R_0(x,x';s,s')=R_0(x,x')$. 
Taking $t=0$ in (\ref{eq-R}), 
 evaluating  the Gaussian
integral and comparing with (\ref{eq-noname}), one gets 
$C= \pi^{-\frac{1}{2}} Z_{{\cal P}, {\rm    eff}}^{-1}$.

\subsection{Justification of the approximation (\ref{eq-R_tsimeqR_0}) for $R_0$}
\label{sec-proof_on_g_t}

We here want to derive the inequality 
\begin{equation} \label{eq-exact_inequality_g_t}
g_t(x,x';s,s')^2 
 \leq
  (8\pi^2)^{-\alpha}
   \frac{\lambda_{\rm th}^{2 \alpha} \beta\, \gamma_0}{\hbar}\, 
   D_t (x,x';s,s') \;.
\end{equation}
Let us first point out that, putting together
(\ref{eq-exact_inequality_g_t}), the stability condition
(\ref{eq-stability_cond}), the separation of length scales
$\lambda_{\rm th} \ll \Delta_{\rm th}$, and the bound $2 \beta
\gamma_0/\hbar \leq \eta_{\rm th}^2/\Delta_{\rm th}^{2 \alpha}$, it
follows that $g_t (x,x';s,s')^2 \ll D_t (x,x';s,s')$ uniformly for all
$(x,x')$ and $(s,s')$.  This explains why the general expression
(\ref{eq-R}) reduces to (\ref{eq-R_tsimeqR_0}) for short times $t$
satisfying $D_t (x,x',s,s') \lesssim 1$; then $| g_t (x ,x' ;s,s')|
\ll 1$ and the phase factor inside the integral in (\ref{eq-R}) can be
neglected; by performing the resulting Gaussian integral, one gets
$R_t(x,x';s,s') \simeq R_0(s,s')$.  In the special case
$\alpha=1$, the integral in (\ref{eq-R}) can be evaluated exactly for
all times $t$. This leads to $R_t = R_0 \, \E^{-(g_t)^2-\I \phi_t'}$.
Here and in what follows $\phi_t'$, $\phi_t''$, etc, denote real
phases irrelevant for decoherence. Replacing the latter value of $R_t$
into (\ref{eq-result-S_totaleq}), the factor $e^{-(g_t)^2}$ can be dropped by
invoking $g_t^2 \ll D_t$ again. We are thus led to
\begin{equation} \label{eq-result-S_totaleq_alpha=1}
\E^{-D_t (x,x';s,s')} R_t ( x,x';s,s') 
\simeq 
\E^{-D_t (x,x';s,s')-\I \phi_t''} R_0 (x,x') 
\qquad (\alpha=1)
\end{equation}
which is now valid for all times $t \ll \TS, \TP$.  The integral
(\ref{eq-R}) can be evaluated exactly for $\alpha=2$ as well. By
(\ref{eq-exact_inequality_g_t}) and the same arguments as above, for
$\alpha>1$ the stronger condition $g_t (x,x';s,s')^2 \ll
\left(\lambda_{\rm th}/\Delta_{\rm th} \right)^2 D_t(x,x';s,s')$
holds. Using also the restriction $|x|, |x'| \lesssim \Delta_{\rm eff}
\approx \Delta_{\rm th}$ coming
from the factor in front of the integral in (\ref{eq-R}), one obtains
for all times $t \ll \TS,\TP$
\begin{equation} \label{eq-result-S_totaleq_alpha=2}
\E^{-D_t (x,x';s,s')} R_t ( x,x';s,s') 
\simeq 
\frac{\E^{-D_t (x,x';s,s')-\I \phi_t'''} R_{0}  
\bigl( x, x'  \bigr)}{[1+4 g_t^2 (x,x';s,s')]^{1/4}}
\qquad  (\alpha=2)\;.
\end{equation}
Notice that this equation is consistent with $R_t \simeq R_0$ at
times $t$ satisfying $D_t \lesssim 1$.

\newparagraph

Proceeding towards the inequality (\ref{eq-exact_inequality_g_t}) we
rewrite (\ref{eq-g_t}) as
\begin{equation} \label{eq-g_t_bis}
\frac{(8 \pi^2)^\alpha \hbar^4}{\lambda_{\rm th}^{2\alpha}}
\,  g_t(x,x';s,s')^2
 = 
 \left( 
\int_{-\infty}^\infty \frac{\D \omega}{2 \pi}\, \widehat{\gamma} (\omega) 
 \int_0^t \D \tau \, \cos (\omega \tau) 
   \bigl( x_{s'}'(-\tau)^\alpha -  x_{s}(-\tau)^\alpha \bigr)
\right)^2 
\end{equation}
where $\widehat{\gamma}(\omega)=\widehat{\gamma}(-\omega)\geq 0$
is the Fourier transform of $\gamma(t)$, see (\ref{eq-gammat}).
By using $\gamma_0=\gamma(0)=\int \D \omega \,
\widehat{\gamma} (\omega)/{2 \pi}$ and the Cauchy-Schwarz inequality,
\begin{equation} \label{eq-Cauchy-Schwarz}
\frac{(8 \pi^2)^\alpha \hbar^4}{\lambda_{\rm th}^{2\alpha}}
\,  g_t(x,x';s,s')^2
\leq
\gamma_0 
\int_{-\infty}^\infty \frac{\D \omega}{2 \pi} \, \widehat{\gamma} (\omega) 
\left( 
 \int_0^t \D \tau \, \cos (\omega \tau) 
   \Bigl( x_{s'}'(-\tau)^\alpha -  x_{s}(-\tau)^\alpha \Bigr)
\right)^2
\end{equation}
The integral over $\omega$ in the right-hand side of
(\ref{eq-Cauchy-Schwarz}) can
be bounded with the help of (\ref{eq-bound_Imh}) by
\begin{equation} \label{eq-upper_bound}
\frac{\hbar \beta}{2} 
\int_{-\infty}^\infty \frac{\D \omega}{2 \pi}\, (\widehat{\Re h}) (\omega)
\left[ \Re 
 \int_0^t \D \tau \, \E^{-\I \omega \tau} 
   \bigl( x_{s'}'(-\tau)^\alpha -  x_{s}(-\tau)^\alpha \bigr)
\right]^2 \;.
\end{equation}
Comparing (\ref{eq-upper_bound}) with (\ref{eq-D(i)_with_FT}), we
bound the last quantity by $\hbar^3\beta D_t(x,x',s,s')$ and have
thus established the inequality (\ref{eq-exact_inequality_g_t}).

\section{Conclusion}
\label{sec-conclusion}

Let us summarize the main results of this paper.  We have investigated
a model for a quantum measurement in which the entanglement produced
by the interaction between the measured quantum object and the pointer
is simultaneous with decoherence of distinct pointer readouts; the
apparatus (pointer and bath) is taken initially in a metastable local thermal
equilibrium, not correlated to the object.  Our model has four
parameters: the object-pointer coupling constant $\epsilon$, the
thermal variance $\langle B^2\rangle$ of the bath coupling agent, the
temperature $T=(k_B \beta)^{-1}$ of the bath, and the
exponent $\alpha$ in the pointer-bath
Hamiltonian (\ref{eq-int_Hamiltonian}).  One may construct out of the
first three parameters two relevant dimensionless constants.  The
first one is the entanglement time $\tau_{\rm ent} =\Delta
(\epsilon \,\delta s \, \hbar \beta)^{-1}$ in units of the thermal
time $\hbar \beta$. Here $\delta s$ is the separation between neighboring
eigenvalues of the measured observable and $\Delta$ the uncertainty in the initial pointer position. That
entanglement time $\tau_{\rm ent}$ describes the efficiency of the
pointer-bath interaction (a coupling is efficient if $\tau_{\rm ent}$
is small).  More precisely, $\tau_{\rm ent}$ is the time 
after which
pointer positions corresponding to distinct eigenvalues $s$ begin to
be resolved. The second dimensionless combination is the coupling
energy $\eta=\langle B^2\rangle^{1/2} \Delta^\alpha \beta$
in units of $k_B T$, which measures the strength of the pointer-bath
coupling. We have found that, after a certain time
$\dectime$, the object-pointer state is close to a statistical mixture
of separable states $\sum_s p_s \, \ketbra{s}{s} \otimes \rhoP^{\,s}$,
with $p_s= \bra{s} \rhoS \ket{s}$, $\rhoS$ the object initial state,
and $\rhoP^{\,s}$ a distinguished pointer state depending on $s$.
The decoherence time $\dectime = \hbar \beta \,\tau_{\rm dec}
\ll \TS, \TP$ is given by (we ignore here numerical factors, given
explicitly in (\ref{t_dec_short_time}), (\ref{t_dec-Markov}), and
(\ref{t_dec-Markov_sup_ohmic}))
\begin{equation} \label{eq-summary}
\tau_{\rm dec} \propto \left( \eta^{-1/\alpha} \tau_{\rm ent} \right)^{\gamma} 
\;\, , \;\; 
\gamma=
\left\{  
\begin{array}{lcll}
\frac{\alpha}{\alpha+1} 
& 
{\rm{if}}  
& \dectime \lesssim \TBmin
& \mbox{(interaction-dominated regime)}  
\\[3mm]
\frac{2\alpha}{2 \alpha+1}
& 
{\rm{if}} 
& \dectime \gtrsim \TBmax 
& \mbox{for an Ohmic bath (Markov)}
\\[3mm]
1
& 
{\rm{if}}
&  \dectime \gtrsim \TBmax  
& \mbox{for a super-Ohmic bath (Markov).}
\end{array}
\right.
\end{equation}
For  reasonably
strong pointer-bath
coupling and not too strong object-pointer coupling, 
the decoherence time $\dectime$ (needed
for the transformation of linear superpositions into statistical
mixtures) can be so small that the whole measurement is performed
without producing a Schr{\"o}dinger cat state as an intermediate step.
Two distinct regimes ought to be identified in (\ref{eq-summary}): in
the {\it interaction-dominated regime}, $\dectime$ is shorter than the
characteristic time $\TBmin$ after which the bath correlation function
$h(t)$ differs significantly from its value $\langle B^2
\rangle$ at $t=0$; in the opposite {\it Markov regime}, one must wait
more than the bath correlation time $\TBmax$, i.e., the largest decay
time of $h(t)$, to obtain the required statistical mixture.  While
$\dectime$ presents a universal behavior in the interaction-dominated
regime (it depends on the bath through the single parameter $\eta$),
in the Markov regime it is determined by the small-frequency behavior
of $\Im h(t)$,
$(\widehat{\Im h})(\omega) \sim -\I\, \widehat{\gamma}\, \omega^m$.
Larger values of $\dectime$ are found for larger $m$'s, with a
significant change of behavior between $m=1$ (Ohmic bath) and $m>1$
(super-Ohmic bath), see (\ref{eq-summary}).  In both time regimes,
$\dectime$ strongly depends  on the nonlinearity exponent
$\alpha$, as illustrated in Figs.\ \ref{fig-3} and \ref{fig-5}.  Smaller decoherence
times are obtained for larger $\alpha$'s save for the Markov regime if
$m>1$ and $\eta \gtrsim \hbar \beta/\TBmax$ or if $m=1$ and $\eta^2
\tau_{\rm ent} \gtrsim \hbar \beta/\TBmax$, where the reverse
statement holds.  The linearization of
the pointer-bath interaction with respect to the pointer position
(dipole approximation) may then
lead to an over-estimation of $\dectime$ in the interaction-dominated regime
or for super-Ohmic baths in the Markov regime.
For a bath at very low temperature, (\ref{eq-summary}) still
holds with $\tau_{\rm dec}$ and $\eta$ replaced by $\dectime/\TBmin$
and $\eta_D = \langle B^2\rangle^{1/2} \Delta^\alpha \TBmin/\hbar$,
save for the Ohmic case where $\dectime/\enttime$ becomes
logarithmically small in $\dectime/\TBmin$.

\newparagraph

Several generalizations of our results may be of interest.  The first
one concerns measurements of observables with continuous or dense
spectra. One must then allow for a finite resolution $\delta s$ in the
measurement result.  Unlike in the case of discrete
non-degenerate spectra studied in this work, the ``final''
object-pointer state will not be a separable state because coherences
for pairs $(s,s')$ of close eigenvalues ($|s-s'|\leq \delta s$) are
damped on a smaller time scale exceeding the time duration of the
measurement.  A second generalization concerns the bath, assumed in
this paper to consist of independent degrees of freedom. As stated in
Sec.~\ref{sec-model}, it can be
shown that the validity of the QCLT extends to baths of interacting
degrees of freedom if the correlator $\langle B_\mu B_\nu \rangle$
decays more rapidly than $1/|\mu-\nu|$ (see~\cite{Verbeure} for a
related version of the QCLT in this context).  This implies that our
results apply to a broad class of baths including certain interacting spin
chains.

\newparagraph

It would be interesting to investigate  
concrete models for the object and pointer
involving projective measurements in the ``no-cat'' regime
(decoherence fast compared with entanglement), in
connection with recent experiments in solid state physics. 
We should also mention that our results can be of interest
in a broader context. 
Actually, we have studied
quantitatively a
new scenario for
decoherence. In this  ``indirect decoherence'' scheme,
the decay of the quantum coherences of
the small system (object) does not result from a direct coupling 
to the many degrees of freedom of a bath, but rather
from a strong coupling  
to few degrees of freedom of the environment only (here, to the pointer).
These
few degrees of freedom are in turn coupled to all others bath
coordinates
and    serve as an
intermediate in the decoherence process.

\vspace{1cm}

\noindent \textbf{Acknowledgements:} We acknowledge support by the 
Deutsche Forschungsgemeinschaft  
(through the SFB/TR 12) 
and the Agence Nationale de la Recherche
(Project No. ANR-05-JCJC-0107-01). D.S. thanks M.  Guta 
for pointing out Ref.~\cite{QCTL} to him and 
D. Vion for his encouragement to study the low-temperature case.

\vspace{1cm}
\newpage

\appendix
\renewcommand{\theequation}{\Alph{section}\arabic{equation}}
\setcounter{equation}{0}


\section{Apparatus as a noninteracting infinite gas}
\label{app-noninteracting_gas}

Let us consider a gas of ${\cal N}=N+1$ noninteracting particles with mass
$m_\nu$, momentum $P_\nu$, and position $X_\nu$ ($\nu=0,\ldots , N$)
submitted to a slowly varying external potential $V(x)$.  To simplify
the discussion we restrict ourselves to a one-dimensional geometry.  Let
$M =\sum_\nu m_\nu$ and $P=\sum_\nu P_\nu$ be the total mass and
momentum and $X=\sum_\nu m_\nu X_\nu/M$ the center-of-mass
position; $R_{\nu 0}={\cal N}^{-1/2} (X_\nu-X_0)$ and $P_{\nu 0}$ are the
relative positions and their conjugate momenta.  Expanding $V(X_\nu)$
as $V(X) + (X_\nu-X) V'(X)$ and using $X_\nu-X=\sqrt{{\cal N}} (R_{\nu 0} -
\sum_\mu m_\mu R_{\mu 0}/M)$, the Hamiltonian of the gas reads 
\begin{equation} \label{eq-infinite_gas}
H_{\rm app} 
 = 
  \underbrace{\frac{P^2}{2 M} + {\cal N} V(X)}_{\HP} 
 + \underbrace{\sqrt{{\cal N}} \sum_{\nu=1}^N 
    \left( 1 - \frac{{\cal N} m_\nu}{M} \right) R_{\nu 0} \,V'(X)}_{\HPB}   
  + \underbrace{\sum_{\mu,\nu=1}^N 
     \frac{\ell_{\mu \nu}}{2 {\cal N} m_\nu}
      P_{\mu 0} P_{\nu 0}}_{\HB} 
\end{equation}
where $\ell$ is the $N\times N$ matrix with inverse 
$\ell^{-1} = (\delta_{\mu \nu}  - m_\nu/M)_{\mu,\nu=1}^N $. 
The pointer $\Pc$ is the center-of-mass degree of freedom. Its
Hamiltonian $\HP$ is given by the two first terms in
(\ref{eq-infinite_gas}).
The bath $\Bc$ is constituted by the $N$ relative degrees of freedom.
Its Hamiltonian $\HB$ is the last term in  (\ref{eq-infinite_gas}).
The third term in  (\ref{eq-infinite_gas}) describing the coupling 
between $\Pc$ and $\Bc$ has the form (\ref{eq-int_Hamiltonian})
if $V(x)= {\cal N}^{-1} (\alpha+1)^{-1} x^{\alpha+1}$ and $B$ is given by
(\ref{eq-int_Hamiltonian})
with $B_\nu = ( 1 - {\cal N} m_\nu/M ) R_{\nu 0}$.
If the measured system  is strongly coupled to the total momentum $P$
of the gas, one obtains a tripartite model of the kind discussed in
Sec.~\ref{sec-model} (although $\HB$ does not
satisfies all our hypothesis).

\setcounter{equation}{0}

\vspace{1cm}

\section{Approximation
 for the apparatus equilibrium state}
\label{app-total_equilibrium}

In this appendix we justify the approximation
(\ref{eq-approx_total_equilibrium}) for the initial Gibbs state
$\rhoPBeq$ of the apparatus.  Moreover, we show that $\rhoP^{0} (t)
= \E^{-\I t \HP/\hbar} \rhoP \, \E^{\I t \HP/\hbar} \simeq \rhoP$ when
$t \ll \TP$ for the quasi-classical pointer states $\rhoP$ considered
in Sec.~\ref{sec-model}.  A similar result holds for
$\rhoPBeq$.

\newparagraph

We recall that $\TP$ is defined by $\TP= ( M/V''(0))^{1/2}$.  Taking
$V(x)\simeq V''(0)\, x^2/2$ and invoking (\ref{eq-no_squeezing}),
(\ref{eq-rhoP}), and (\ref{eq-def_int_picture}) one easily finds that
$\tr_\Pc (\widetilde{X}(t)^2 \rhoP) - \Delta^2$ and $\tr_\Pc
(\widetilde{P}(t)^2 \rhoP ) - \Delta p^2$ are equal to lowest
order in time to $(-\Delta^2 \TP^{-2} + \Delta p^2 M^{-2} ) t^2
\approx \Delta^2 \TP^{-2} t^2 $ and $ ( - \Delta p^2 \TP^{-2} +
V''(0)^2 \Delta^2) t^2 \approx \Delta p^{2} \TP^{-2} t^2$,
respectively.  Hence $\TP$ can be identified with the time scale for
significant evolution of $\widetilde{X}(t)$ and $\widetilde{P}(t)$
when the pointer is in the quasi-classical state (\ref{eq-rhoP})
(Sec.~\ref{sec-time scales}).  One easily convinces oneself that
$\rhoP=\rhoP (X,P)$ is an operator-valued function of the position and
momentum operators $X$ and $P$. Letting $\rhoP$ evolve under the
Hamiltonian $\HP$ up to time $t$ amounts to substituting $X$ by
$\widetilde{X}(t)$ and $P$ by $\widetilde{P}(t)$, see
(\ref{eq-def_int_picture}).  This shows that $\rhoP^{0} (t) = \rho
(\widetilde{X}(t), \widetilde{P}(t)) \simeq \rhoP$ as
$\widetilde{X}(t) \simeq X$ and $\widetilde{P}(t) \simeq P$ for $t \ll
\TP$. 

In order to approximate $\rhoPBeq$
we shall rely on the
Baker-Campbell-Haussdorff formula  
\begin{equation} \label{eq-Baker-Campbell}
\E^{A} \E^{C} = \E^{A+C + [A,C]/2 + [A,[A,C]]/12 
+ [C,[C,A]]/12 + \cdots}
\end{equation}
wherein   $A$ and $C$ are  any two operators.
After a few transformations, (\ref{eq-Baker-Campbell}) becomes
\begin{equation} \label{eq-BCH}
\E^{A} \E^{2 C} \E^{A} 
= 
\E^{2A + 2C} \E^{ - [A,[A,C]]/3+2[C,[C,A]]/3+\cdots}\;.
\end{equation}
We write $(\widetilde{X}^\alpha )'(0)$ and $(\widetilde{X}^\alpha
)''(0)$ the two first time derivatives at $t=0$ of the free-evolved
observable $X^\alpha$, see (\ref{eq-def_int_picture}), and
similarly for $B$. Let us take $A=-\beta \HP/2$ and $C=-\beta
(\HB+\HPB)/2$. Then
\begin{eqnarray}
 \label{eq-commutators1}
\displaystyle
\bigl[ 
A,[ A, C ] 
\bigr] 
& = &  \frac{\hbar^2\beta^3}{8} 
(\widetilde{X}^\alpha)''(0)  B 
\\
 \label{eq-commutators2}  
\displaystyle
\bigl[ 
C , [ C,A ] 
\bigr]
& = & 
-  \frac{\hbar^2 \beta^3}{8} 
\left(
 (\widetilde{X}^\alpha) ' (0)  \,
 \widetilde{B}' (0)  
  - \frac{\alpha^2}{M} X^{2\alpha-2} B^2
\right) \;.
\end{eqnarray}
Each time derivative of $\widetilde{X}^\alpha$ (of $\widetilde{B}$)
gives a extra factor of $\TP^{-1}$ ($\TBmin^{-1}$).  Therefore, the
right-hand side of (\ref{eq-commutators1}) is smaller than $\beta
\HPB= \beta X^\alpha B$ by a factor of the order of $(\hbar
\beta/\TP)^2 \ll 1$.  By virtue of $\Delta_{\rm th} = (\beta V''(0)
)^{-1/2}$ and $\TP = (M/V''(0))^{1/2}$, the right-hand side of
(\ref{eq-commutators2}) is of the order of $(\hbar \beta)^2( \TP^{-1}
\TBmin^{-1} + \TP^{-2} \eta_{\rm th} ) \beta \HPB$ with $\eta_{\rm
  th}$ given by (\ref{eq-stability_cond}). Assuming that $\eta_{\rm
  th}$ is at most of order $1$ (this is the case in particular if
(\ref{eq-stability_cond}) holds true) this indicates that the double
commutators (\ref{eq-commutators1}-\ref{eq-commutators2}) are much
smaller than $C$ and can be neglected in (\ref{eq-BCH}) when $\hbar
\beta \ll \TP$ and $(\hbar \beta)^2 \ll \TP \,\TBmin$. 
Neglecting these commutators,  (\ref{eq-BCH}) reduces to
(\ref{eq-approx_total_equilibrium})  for 
the aforementioned choices of $A$ and $C$.  Notice that our approximation of
$\rhoPBeq$ is self-adjoint and is better than $\ZPB^{-1}\, \E^{-\beta
  \HP} \E^{-\beta (\HB +\HPB)}$ (the error is of one order smaller in
$\hbar \beta/\TP$). The approximation $\E^{-\I t \HP/\hbar} \rhoPB
\,\E^{\I t \HP /\hbar} \simeq \rhoPB$ for $t \ll \TP$ is obtained
similarly, by using (\ref{eq-Baker-Campbell}) with $A=-\I t \HP/\hbar$
and $C=-\beta ( \HP + \HB+ \HPB)$. One can
check explicitly  by means of similar arguments as in
Sec.~\ref{sec-derivation-eqapp}
that the relative errors are small.  Indeed, one can show that if one multiplies
(\ref{eq-commutators1}) or (\ref{eq-commutators2}) by the approximate
equilibrium state $\ZPB^{-1}\, \E^{- \beta \HP/2} \E^{-\beta ( \HB +
  \HPB )}\E^{- \beta \HP/2}$, traces out the bath variables and takes
the matrix elements between $\ket{x}$ and $\ket{x'}$, the matrix
elements so obtained are much smaller than (\ref{eq-noname}) if
(\ref{eq-stability_cond}) is satisfied and $\hbar \beta \ll \TP$.

\vspace{1cm}

\setcounter{equation}{0}
\section{Properties of the bath correlation function}
\label{app-correl_functions}

In this appendix, we establish some general properties of the bath
two-point autocorrelation function
\begin{equation} \label{eq-definition_h}
h(t_1,t_2) 
 = \meanB{ \widetilde{B}(t_1) \widetilde{B}(t_2) }
  = h(t_1-t_2) 
\end{equation}
and its Fourier transform
\begin{equation} \label{eq-FT}
\widehat{h} (\omega) = \int_{-\infty}^\infty \D t \, 
  h (t) \, \E^{\I \omega t}\;.
\end{equation}
Most  (but perhaps not all) of these properties are well known.
The average $\langle \cdot \rangle$ in (\ref{eq-definition_h}) is taken with respect to the
Gibbs state $\rhoB$, $\widetilde{B}(t)$ is the bath coupling agent in
the interaction picture, and $\meanB{B}=0$, see
(\ref{eq-def_int_picture}-\ref{eq-B=0}).  The
fact that $h(t_1,t_2)$ depends only on the time difference
$t_1-t_2$ is a consequence of the stationarity of
$\rhoB$~\cite{Cohen-Tannoudji}.

\vspace{3mm}

\noindent {\it Real and imaginary parts.} 
The real and imaginary parts of $h(t)$ are given by 
$\Re h(t)
= \meanB{\widetilde{B}(t) B + B \widetilde{B}(t)}/2 $ and $\Im h
(t) = - \I \meanB{ [ \widetilde{B}(t) , B ] }/2$. We write
$(\widehat{\Re h})(\omega)$ and $(\widehat{\Im h})(\omega)$ their Fourier
transforms. Then
\begin{equation} \label{eq-parity_h}
\begin{array}{ccc}
\Re h (t) = \Re h(-t) 
& \quad  , \quad &
\Im h (t) = - \Im h(-t) 
\\   
(\widehat{\Re h})(\omega) = (\widehat{\Re h})(- \omega)
& \quad  , \quad &
 (\widehat{\Im h})(\omega) = - (\widehat{\Im h})(- \omega)\;.
\end{array}
\end{equation}
The imaginary part $\Im h(t)$ is linked to the linear susceptibility
by $\chi(t)= - 2 \theta(t) \Im h(t)/\hbar$, where $\theta(t)$ denotes
the Heaviside function~\cite{Cohen-Tannoudji}.  Such a susceptibility
characterizes the response of the bath when its Hamiltonian is
perturbed by the time-dependent potential $V_\Bc(t) = - x^\alpha (t) B
$, where $x^\alpha(t)$ is a real-valued function of time. More
precisely, if $B(t)$ is the observable $B$ in the Heisenberg picture
(i.e., $\D B(t)/\D t =(\I/\hbar)[ \HB + V_{\cal B} (t), B(t)]$) then
$\meanB{B(t)}=\int \D \tau \,\chi (t) x^\alpha(t-\tau)$ up to terms of
order $x^{2\alpha}$.

\vspace{3mm}

\noindent {\it The function $h(t)$ is of positive type.} This means
that for any integer $n \geq 1$, complex numbers $c_1,\ldots,c_n$, and
times $t_1, \ldots , t_n$, one has
\begin{equation} \label{eq-h(t)>0type}
\sum_{i,j=1}^n c^\ast_i c_j h(t_i-t_j ) \geq 0\;.
\end{equation}
This property can be easily checked on (\ref{eq-definition_h}).  It is
equivalent to $\widehat{h} (\omega) \geq 0$ for any real $\omega$.
The real part of $h(t)$ is also of positive type, as $2 (\widehat{\Re
  h})(\omega)=\widehat{h}(\omega)+ \widehat{h}(-\omega) \geq 0$ for
any real $\omega$.  By using the Cauchy-Schwarz inequality for the
Hermitian  sesquilinear
form $(A,B) \mapsto \meanB{A^\dagger B}$, one shows that $| h(t)| \leq
h(0) = \langle B^2 \rangle$ for any time $t$.

\vspace{3mm}

\noindent {\it KMS property.} This property is specific to our choice
of the Gibbs state for the bath average. It says that $h(t)$ can be
extended to an analytic function in the strip $\{ z \in \complex ;
-\hbar \beta < z <0 \}$, continuous on $\{ z \in \complex ; -\hbar
\beta \leq z \leq 0 \}$, and such that~\cite{Bratteli}
\begin{equation} \label{eq-KMS}
h (t) = h( -t - \I \hbar \beta) \quad , \quad t \in \real  \;.
\end{equation}
Deforming the path of integration in (\ref{eq-FT}), one can show that
(\ref{eq-KMS}) is equivalent to $\widehat{h} (\omega) = \E^{\hbar
  \beta \omega} \widehat{h}(-\omega)$.  In view of
(\ref{eq-parity_h}), this means that
\begin{equation} \label{eq-KMS_FT}
(\widehat{\Re h}) ( \omega ) 
= 
\frac{\I  (\widehat{\Im h}) (\omega)}{ 
 \tanh ( \hbar \beta \omega/2)} \;.
\end{equation}
By
replacing in this equation $(\widehat{\Re h}) ( \omega )$ and
$(\widehat{\Im h}) ( \omega )$ by their Fourier integrals, expanding
and identifying each power of $\omega$, and using the parity
properties (\ref{eq-parity_h}), one finds relations between the
integrals $\int \D t \, t^a h(t)$ for even and odd $a$'s.
For instance, the identification of the zero-th power in $\omega$ in
(\ref{eq-KMS_FT}) yields
\begin{equation} \label{eq-I_1=beta_I_2}
\int_{-\infty}^{\infty} \D t \, t \, h(t) 
 =
  -\I \frac{\hbar \beta}{2} \int_{-\infty}^{\infty} \D t \,   
h(t) \;.
\end{equation}
We now assume that $\I (\widehat{\Im h}) (\omega) \sim
\widehat{\gamma} \, \omega^m$ as $\omega \to 0$ with $m$ a positive
odd integer and $\,\widehat{\gamma}>0$ (such a choice is motivated in
Sec.~\ref{sec-Markov}).  By (\ref{eq-KMS_FT}), this entails
$(\widehat{\Re h})(\omega)\sim 2 \,\widehat{\gamma}\,
\omega^{m-1}\,(\hbar \beta)^{-1}$.  Let $a$ be a nonnegative integer,
$a \leq m-2$. Then
\begin{equation} \label{eq-integral_a}
\int_{0}^\infty \D t\, t^a \,\Re h (t ) 
 = 
  \lim_{\varepsilon \to 0+} \int_{-\infty}^\infty \frac{\D \omega}{2 \pi} 
   (\widehat{\Re h}) (\omega) \int_0^\infty \D t \, t^a
   \,\E^{-\I (\omega-\I \varepsilon) t} 
   = - (-\I)^{a-1} a! \int_{-\infty}^\infty \frac{\D \omega}{2 \pi} 
   \frac{(\widehat{\Re h})  (\omega)}{\omega^{1+a}}\;.
\end{equation}
Note that the frequency integral converges for $a \leq m-2$, vanishes
for even $a$'s, $a < m-2$, and diverges for $a \geq m-1$.  For
$a=m=1$, a similar formula holds,
\begin{equation} \label{eq-integral_a_bis}
\int_{0}^\infty \D t\, t \,\Re h (t ) 
 = 
  - \int_{-\infty}^\infty \frac{\D \omega}{2 \pi} 
   \frac{(\widehat{\Re h})  (\omega)-(\widehat{\Re h})  (0) }{\omega^{2}}
\end{equation}
where the diverging frequency integral has been regularized by
subtracting $(\widehat{\Re h}) (0) $ to $(\widehat{\Re h}) (\omega)$.
This is equivalent to subtracting $(\widehat{\Re h}) (0) \,\delta(t)$ from
$\Re h(t)$ in (\ref{eq-integral_a}) and this does affect the left-hand
side of this equation.  The integral on the right-hand side of
(\ref{eq-integral_a_bis}) converges since $(\widehat{\Re h})
(\omega)-(\widehat{\Re h}) (0)$ behaves like $\omega^2$ as $\omega \to
0$.

\vspace{3mm} 

\noindent {\it Integration in the complex plane.} Let us denote by 
$\gamma(t)=\gamma(-t)$  the primitive 
of $\Im h (t)$ vanishing at $t\to \pm \infty$, 
\begin{equation} \label{eq-gammatbis}
\gamma(t) = \int_{-\infty}^t \D
\tau\, \Im h (\tau)
 = - \int_{t}^\infty \D
\tau\, \Im h (\tau)
 =
 \int_0^\infty \frac{\D \omega}{\pi} \widehat{\gamma}(\omega)
 \cos (\omega t )
\;.
\end{equation}
The KMS property (\ref{eq-KMS_FT}) and the bound $\tanh(u) \leq u$ for
$u \geq 0$ imply
\begin{equation} \label{eq-bound_Imh}
0 
 \leq 
  \widehat{\gamma}(\omega)
   = \frac{\I (\widehat{\Im h}) (\omega)}{\omega}
    \leq \frac{\hbar \beta}{2} (\widehat{\Re h})(\omega) \;.
\end{equation}
Substituting (\ref{eq-bound_Imh}) into (\ref{eq-gammatbis}) we obtain
the following 
inequalities for $\gamma_0 = \gamma(0)$
\begin{equation} \label{bounds}
0 \leq \gamma_0 \leq \frac{\hbar \beta}{2} h(0) 
\; .
\end{equation}
We can exploit the KMS property further to obtain the two identities
\begin{eqnarray} \label{eq-int_chi}
\int_0^{\hbar \beta} \D z \,h(-\I z - t ) 
& = &
2 \gamma(t)\;,
\\
\label{eq-int_chi2}
\int_0^{\hbar \beta} \D z_1 \int_{0}^{z_1} \D z_2 \, h (-\I z_2)
& = & \int_{0}^{\hbar \beta} \D z \,  h (-\I z)\,z  \;\; = \; \;
\hbar \beta \gamma_0
\;.
\end{eqnarray}
To show (\ref{eq-int_chi}), we deform the contour of integration in
the complex plane to get
\begin{equation}
\int_0^{-\I \hbar \beta} \D z \,h(z - t )
 = \int_0^\infty \D \tau \, \bigl( h ( \tau-t) - h( \tau-t - \I \hbar  \beta ) \bigr) 
\end{equation}
and then use (\ref{eq-KMS}), (\ref{eq-parity_h}), and (\ref{eq-gammatbis}). The second equality 
in (\ref{eq-int_chi2}) is established in a similar way, relying also on (\ref{eq-I_1=beta_I_2}). 
Finally, we note that
the left-hand side of (\ref{eq-int_chi2}) reads
\begin{equation}
\hbar \beta \int_0^{\hbar \beta} \D z_2\,h(-\I z_2  ) - \int_0^{\hbar \beta} \D z_2\, h(-\I z_2)\, z_2\;.
\end{equation}
Therefore, the first equality in (\ref{eq-int_chi2}) is a consequence
of the second one and of (\ref{eq-int_chi}).

\vspace{1cm}

\setcounter{equation}{0}
\section{Wick theorem}
\label{app-characteristic-functional}

We show in this appendix that  Wick's theorem (\ref{eq-Wick}) implies
formulas (\ref{eq-characteristic-functional}) and (\ref{eq-characteristic-functional_y}), i.e.,
\begin{eqnarray} \label{eq-F_t,y}
\nonumber
& & F_{t,y} [ k,l] 
=  \left\langle 
 \widetilde{U}_{t,0} [ k]^\dagger \, \widetilde{U}_{t,0} [ l] 
  \right\rangle_{y} 
  \exp 
  \left\{ -
     \frac{\I}{\hbar} \int_0^t \D \tau \,\bigl(  {k (\tau)} - l(\tau) \bigr)\,
    \meanBx{\widetilde{B}(\tau)}{y} 
  \right\}
\\
& & = 
\exp 
\biggl\{
- \frac{1}{\hbar^2}
 \int_0^t \D \tau_1 \int_0^{\tau_1} \D \tau_2 
  \bigl(  {k (\tau_1)} - l (\tau_1) \bigr) 
 \bigl( 
     {k (\tau_2)} h ( \tau_2, \tau_1)  -l(\tau_2) h (\tau_1,\tau_2)
    \bigr)
\biggr\}    
\end{eqnarray}
where $t$ and  $y$ are real numbers (time and position), $k$ and $l$ are 
(nice) real functions,  
$\meanBx{\cdot}{y}$ is the bath average (\ref{eq-meanxB}), 
$h(\tau_1,\tau_2)$ is the bath
function (\ref{eq-definition_h}), and
\begin{equation} \label{eq-appendix0}
\widetilde{U}_{t,0} [ k] 
 = 
  \timeord \exp
  \biggl\{ 
    - \frac{\I}{\hbar} \int_0^t \D \tau \,k (\tau)\,
    \widetilde{B}(\tau)  
  \biggr\} \;.
\end{equation}

\newparagraph

Let us first recall that Wick's theorem (\ref{eq-Wick}) 
can be rephrased as the following recursive relation for the bath 
$n$-point functions (\ref{eq-bath_correl_function}),  $n>2$, 
\begin{equation} \label{eq-Wick2}
h_n (t_1, \ldots , t_n )= \sum_{1 \leq j < i} 
h(t_j ,t_i  )  h_{n-2} ( t_1, \ldots, t_n )
   +  \sum_{i < j \leq n} 
h(t_i ,t_j  )  h_{n-2} ( t_1, \ldots, t_n )
\end{equation}
wherein $i$ is a fixed integer between $1$ and $n$ and the time arguments
of the $(n-2)$-point functions are the times appearing in the
$n$-point function on the left-hand side except for $t_i$ and $t_j$.
Wick's theorem (\ref{eq-Wick2}) holds for any value of $N$ if the bath
consists of $N$ harmonic oscillators linearly coupled to the pointer
(Sec.~\ref{sec-HObath}).  For the more general baths considered
in this work, its validity relies on the limit $N \gg 1$ and is a
consequence of the additivity of the bath coupling agent $B$ and
Hamiltonian $\HB$ in single-degree-of-freedom contributions 
and of the QCLT of Ref.~\cite{QCTL}. This
theorem provides a mapping between the correlation functions
$h_n(t_1,\ldots,t_n)$ and the correlation functions of a certain bath
of harmonic oscillators in thermal equilibrium. In such a mapping,
$B_\nu$ is identified with the position of the $\nu$th
oscillator~\cite{QCTL}.

\newparagraph

We now proceed to proving (\ref{eq-F_t,y}). Let us first consider the
case $y=0$.  To shorten the notation, we write $\D k(\tau)$ in place of
$k(\tau)\,\D \tau$.  The two members of (\ref{eq-F_t,y}) being equal
at time $t=0$, it is enough to prove that they satisfy the same
first-order time differential equation. Hence, we need to show that
$ F_{t,0}[k,l]= \langle \widetilde{U}_{t,0} [ k]^\dagger \,
\widetilde{U}_{t,0} [ l] \rangle$ satisfies
\begin{equation} \label{eq-equa_diff}
\frac{\partial F_{t,0}[k,l]}{\partial t} 
 = 
  - \frac{1}{\hbar^2} \bigl(  {k(t)} - l(t) \bigr) 
   \int_0^t  
    \bigl( \D  {k}(\tau)\, h (\tau,t) - \D l(\tau) \,h( t,\tau) \bigr)
     F_{t,0} [ k,l]
\;.
\end{equation}
But $\I \hbar \,\partial \widetilde{U}_{t,0} [k] /\partial t = k(t)
\widetilde{B} (t) \widetilde{U}_{t,0} [k]$, hence (\ref{eq-equa_diff})
is equivalent to
\begin{equation} \label{eq-appendix}
\left\langle 
    \widetilde{U}_{t,0} [ k]^\dagger \widetilde{B} (t) \widetilde{U}_{t,0} [ l] 
   \right\rangle_{0}
 = \frac{\I}{\hbar} \int_0^t  
    \bigl( \D  {k}(\tau) \, h (\tau,t) - \D l(\tau) \, h( t,\tau) \bigr)
   \left\langle 
    \widetilde{U}_{t,0} [ k]^\dagger \widetilde{U}_{t,0} [ l] 
   \right\rangle_{0}
\;.
\end{equation}
To show (\ref{eq-appendix}), let us expand the two time-ordered
exponentials on the left-hand side of (\ref{eq-appendix}).  Invoking also
(\ref{eq-Wick2}), $\langle \widetilde{B}(t) \rangle_0=0$, and setting
$h_0=1$, this left-hand side reads
\begin{eqnarray} \label{eq-appendix3}
\nonumber
& 
\displaystyle
\sum_{n+m\geq 1}^N
\frac{\I^n (-\I)^m}{\hbar^{n+m}} 
 \int_{0 \leq \tau_n \leq \cdots \leq \tau_1 \leq t}
 \D  {k}(\tau_1) \cdots \D  {k}(\tau_n) 
 \int_{0 \leq t_m \leq \cdots \leq t_1 \leq t}
  \D l (t_1) \cdots \D l(t_m) 
 \\
\nonumber
&  
\displaystyle
  \left(
    ( 1 - \delta_{n,0} ) \sum_{p=1}^n h (\tau_p,t) 
  h_{n+m-1} ( \tau_n, \ldots, \tau_{p+1},\tau_{p-1}, \ldots , \tau_1, t_1, \ldots , t_m ) +
  \right.
\\
& 
\displaystyle
  \left.
  ( 1 - \delta_{m,0} )
\sum_{q=1}^m h (t,t_q) 
   h_{n+m-1} ( \tau_n, \ldots,  \tau_1, t_1, \ldots , \tau_{q-1},\tau_{q+1}, \ldots , t_m )
\right) \;.  
\end{eqnarray}  
One may perform variable substitutions in the integrals in such a way
that the time arguments of the first $(n+m-1)$-point function become
$(\tau_{n-1}, \ldots, \tau_1,t_1,\ldots t_m)$ and those of the second
become $(\tau_{n}, \ldots, \tau_1,t_1,\ldots , t_{m-1})$.  Doing so and
resuming the series, we find that (\ref{eq-appendix3}) reduces to
(\ref{eq-appendix}).

\newparagraph

Turning to the case $y \not= 0$, we invoke
(\ref{eq-Dyson_Euclidean}) to write
\begin{equation} \label{eq-appendix4}
\left\langle
   \widetilde{U}_{t,0} [ k]^\dagger \, \widetilde{U}_{t,0} [ l]  
\right\rangle_{y} 
 =
 \frac{\ZBx{0}}{\ZBx{y}} 
  \left\langle
   \widetilde{U}_{-\I \hbar \beta,0} [ y^\alpha]\, 
   \widetilde{U}_{t,0} [ k]^\dagger \, \widetilde{U}_{t,0} [ l] 
  \right\rangle_{0} 
\end{equation}   
wherein $\widetilde{U}_{-\I \hbar \beta,0} [ y^\alpha]$ is obtained by choosing the complex time 
$-\I \hbar \beta$ and the constant function $y^\alpha$ in (\ref{eq-appendix0}).
By 
expanding the three time-ordered exponentials in 
the right-hand side of (\ref{eq-appendix4}) and using Wick's theorem (\ref{eq-Wick2}), one obtains
in a similar way as above
\begin{eqnarray}
\nonumber
\frac{\partial}{\partial t} 
\left\langle
   \widetilde{U}_{t,0} [ k]^\dagger \, \widetilde{U}_{t,0} [ l]  
\right\rangle_{y} 
& =  &
- \frac{1}{\hbar^2} \bigl(  {k(t)} - l(t) \bigr)
 \biggl(
   \int_0^t  
    \bigl( \D  {k}(\tau)\, h (\tau,t) - \D l(\tau) \,h( t,\tau) \bigr) +
\\    
&  &
   +\I y^\alpha \int_0^{\hbar \beta} \D z \, 
     h( -\I z, t)
 \biggr)
 \left\langle
   \widetilde{U}_{t,0} [ k]^\dagger \, \widetilde{U}_{t,0} [ l]  
\right\rangle_{y} \;.   
\end{eqnarray} 
Therefore, the functional
\begin{equation} \label{eq-G_t_y}
G_{t,y} [k,l] 
 = 
  \left\langle
   \widetilde{U}_{t,0} [ k]^\dagger \, \widetilde{U}_{t,0} [ l]  
  \right\rangle_{y} 
\exp
 \left\{
  \frac{\I y^\alpha}{\hbar^2} \int_0^t \D \tau \int_0^{\hbar \beta} \D z \,
  \bigl(  {k(\tau)} - l(\tau) \bigr) h(-\I z, \tau)
 \right\} 
\end{equation}
satisfies the same time differential equation (\ref{eq-equa_diff})
as $F_{t,0}[k,l]$. Moreover, it is equal to $1$ for $t=0$.
Thus $G_{t,y} [k,l]=F_{t,0}[k,l]$ for any $t$, $y$, $k$, and $l$.
Setting $k=0$ in this equation and differentiating with respect to $l(\tau)$ at $l=0$ yields
\begin{equation} \label{eq-meanBxB_y}
-\frac{\I y^\alpha}{\hbar^2} \int_0^{\hbar \beta} \D z \,h(-\I z, \tau )
 - \frac{\I}{\hbar} \meanBx{\widetilde{B}(\tau)}{y}
  =  - \frac{\I}{\hbar} \meanBx{\widetilde{B}(\tau)}{0} = 0 \;.
\end{equation}
This identity and (\ref{eq-int_chi}) imply (\ref{eq-meanxB}).  We may
now replace (\ref{eq-meanBxB_y}) into (\ref{eq-G_t_y}) to get
(\ref{eq-F_t,y}).  Let us stress that, although (\ref{eq-F_t,y}) and
(\ref{eq-meanBxB_y}) coincide with the lowest-order results of
perturbative expansions in $k$ and $l$, these formulas are in fact valid to
all orders in $k$ and $l$ in the limit $N\gg 1$, as a consequence of the
QCLT.

\newpage




\begin{thebibliography}{99}
\bibitem{footnote-generalization_f(X)} Our calculations can be 
generalized to a Hamiltonian $\HPB$ of the form $B f(X)$, where $f$
is a nice real function. In particular,  a cut-off
function at large distances can be introduced by choosing $f(x) \simeq 0$ for large
$x$'s and $f(x) \simeq x^\alpha$ otherwise, so that the pointer
coupling agent $f(X)$ remains bounded.
\bibitem{footnote-uniform_bound}
To show that the constant is uniform in time, it must 
be proven that for all $0\leq a,b \leq
\alpha$ one has $\sup_{t\geq 0} t^{2\alpha-a-b}
I_{a,b,t}/I_{\alpha,\alpha,t} < \infty$ with
$I_{a,b,t} = \int_0^t \D \tau_1 \int_0^t \D \tau_2
\,\tau_1^a \tau_2^b \,\Re h (\tau_1-\tau_2)$. 
This can be done by using the asymptotic analysis developed in 
Secs.~\ref{sec-short-time} and~\ref{sec-Markov}.
\bibitem{footnote-upper_bound} 
An upper bound $\dectime \leq \tilde{t}_{\rm dec}$ may be
obtained by substituting $\enttime(s,s')$ 
in (\ref{t_dec_vs_t_ent}) by
its maximum value $\enttime=\Delta/(\epsilon \delta s )$, $c_\alpha(s,s')$
by its minimum value $c_\alpha^{\rm min}$
and $\dectime (s,s')$ by $\tilde{t}_{\rm dec}$. Indeed, as the
right-hand side of (\ref{t_dec_vs_t_ent}) increases
with $\dectime (s,s')$ the solution $\tilde{t}_{\rm dec}$ obtained
after these substitutions is larger than $\dectime$.
\bibitem{footnote-small_time}
In fact, the small-time behavior (\ref{D^peak_short_time}) can be obtained directly,
without relying on our previous general results, by setting $\HPSB = \HPS + \HPB$ in
(\ref{eq-rhoPS}) and determining the dynamics in a similar way as
in~Ref.~\cite{Haake01}.
\end{thebibliography}
\end{document}